\documentclass[a4paper,11pt]{article}
\pdfoutput=1 

\usepackage{jheppub} 
\usepackage[T1]{fontenc}
\usepackage[italian,english]{babel}
\usepackage{hyperref}
\usepackage{ifpdf}
\usepackage{subfigure}
\usepackage{amssymb}
\usepackage{amsfonts}
\usepackage{epsf}
\usepackage{rotating}
\usepackage{graphicx}
\usepackage{amsmath}
\usepackage{fancyhdr}
\usepackage{lineno}
\usepackage{babel}
\usepackage{graphics}
\usepackage{pstricks}
\usepackage{color}
\usepackage{multirow}
\usepackage{nicefrac}
\usepackage{bm}
\usepackage{slashed}
\usepackage{float}
\usepackage{array}
\usepackage{boldline}

\usepackage{mathrsfs}

\newcolumntype{?}{!{\vrule width 3pt}}
\newcommand{\nn}{\nonumber}
\newcommand{\lsim}{\mathrel{\mathop{\kern 0pt \rlap
  {\raise.2ex\hbox{$<$}}}
  \lower.9ex\hbox{\kern-.190em $\sim$}}}
\newcommand{\gsim}{\mathrel{\mathop{\kern 0pt \rlap
  {\raise.2ex\hbox{$>$}}}
  \lower.9ex\hbox{\kern-.190em $\sim$}}}

\newcommand{\be}{\begin{equation}}
\newcommand{\ee}{\end{equation}}
\newcommand{\bea}{\begin{eqnarray}}
\newcommand{\eea}{\end{eqnarray}}

\def\ptmiss{\not\!\!{p_T}}

\title{\boldmath Distinguishing  Inert Higgs Doublet and Inert Triplet Scenarios}

\author[a]{Shilpa Jangid}
\author[b]{Priyotosh Bandyopadhyay}
\affiliation[a,b]{
Indian Institute of Technology Hyderabad, Kandi,  Sangareddy-502287, Telengana, India}
\emailAdd{ph19resch02006@iith.ac.in, bpriyo@phy.iith.ac.in}

\preprint{IITH-PH-0002/20 }

\abstract{In this article we consider a comparative study between Type-I 2HDM and $Y=0$, $SU(2)$ triplet extensions having one $Z_2$-odd doublet and triplet that render the desired dark matter(DM). For the inert doublet model (IDM) either a neutral scalar or pseudoscalar can be the DM, whereas for inert triplet model (ITM) it is a CP-even scalar. The bounds from perturbativity and vacuum stability are studied for both the scenarios by calculating the two-loop beta functions. While the quartic couplings are restricted to $0.1-0.2$ for a Planck scale perturbativity for IDM, these are much relaxed ($0.8$ ) for ITM. The RG-improved potentials by Coleman-Weinberg show the regions of stability, meta-stability and instability of the electroweak vacuum. The constraints coming from DM relic, the direct and indirect experiments like XENON1T, LUX and H.E.S.S., Fermi-LAT allow the DM mass $\gsim 700, \,1176$ GeV for IDM, ITM respectively. Though mass-splitting among $Z_2$-odd particles in IDM is a possibility for ITM we have to rely on loop-corrections. The phenomenological signatures at the LHC show that the mono-lepton plus missing energy with prompt and displaced decays in the case of IDM and ITM can distinguish such scenarios at the LHC along with other complementary modes.}


\begin{document}

\maketitle
\flushbottom

\section{Introduction} 
Higgs boson was the last key stone predicted by Standard Model (SM), which was discovered at the LHC ~\cite{Aad:2012tfa, Chatrchyan:2012xdj}. So far five decay modes of the SM Higgs boson are discovered  at the LHC \cite{Sirunyan:2018koj, ATLAS:2018doi} and they fall nearly by SM prediction. In spite of  immense success, SM cannot resolve many theoretical and experimental anomalies; like existence of dark matter (DM), explanation of very light neutrinos, Higgs mass hierarchy, vacuum stability, muon $g-2$, etc.  Though discovery of Higgs boson was a direct proof of the role of a scalar in electro-weak symmetry breaking (EWSB) the existence of other Higgs multiplets cannot be ruled out. Recent studies also show that SM stands in a metastable state  \cite{Isidori:2001bm} and need other scalar to make the electro-weak (EW) vacuum stable till Planck scale. This motivates to extend the SM by other Higgs multiplets. 

The simplest extension could be via a singlet \cite{singletex, HiggsDM1,BLscalar} but there could be a possibility of extension with another $SU(2)$ Higgs doublet, i.e.  two Higgs doublet model (2HDM) \cite{2HDMs, Honorez:2010re,HiggsDM, khan1, 2HDMpheno} or with a $SU(2)$ triplet \cite{Tripletex} which can enhance the vacuum stability.  The extensions of SM with fermions  motivated by Seesaw mechanisms often suffers from vacuum instability and one needs some extra scalar to compensate the negative effects \cite{exwfermion}-\cite{Garg:2017iva}.  Many of these extensions have a $Z_2$-odd particle, i.e. inert particle which is stable and being lightest among them, can be a dark matter candidate. 

Supersymmetric sector in its minimal framework has 2HMD of Type-II \cite{primer}. However, the minimal scenario is often challenged by fine-tuning of $\sim 125$ GeV light SM-like Higgs boson mass. One of the remedies of this problem is also to extend the Higgs sector beyond its minimal form. This can be achieved by extension by a SM gauge singlet \cite{NSSM,Bandyopadhyay:2015dio}, $SU(2)$ triplet \cite{TESSM}-\cite{Bandyopadhyay:2014tha} or via singlet and triplet superfields \cite{TNSSM}. In this case the DM particles is generated by $R$-parity and it is a supersymmetric particle with odd $R$-parity. The extended Higgs superfields mix at the superpotential level causing the mixing of Higgs bosons after EWSB among different representations, i.e. doublet-singlet, doublet-triplet, etc \cite{Bandyopadhyay:2015tva,dissusy,Bandyopadhyay:2015ifm, Bandyopadhyay:2017klv}. However, we see the situation is very different for non-SUSY Higgs extensions, especially for the inert models. There are no mixing among these extra Higgs states and the SM particles, making them more illusive to produce and detect at the colliders. Nevertheless, they can provide the much needed dark matter candidate and also make the EW vacuum more stable.  

In this article we consider two different extensions of SM to attain the dark sector.  In the first one we extend SM to  Type-I 2HDM  with  $Z_2$-odd $SU(2)$ doublet that constitutes the dark sector and the scenario is known as inert Higgs doublet (IDM). In the second case we consider the dark sector as $Y=0$ $SU(2)$ triplet which is again $Z_2$-odd and the scenario is known as inert Higgs Triplet scenario (ITM). Both the scenarios help in extending the vacuum stability \cite{2HDMs,Tripletex}; however, we will see that they differ in various constraints coming from perturbativity, vacuum stability, DM relic abundance, direct detection and collider searches. IDM has more scalar with relatively larger mass splitting among the $Z_2$-odd states whereas the ITM has only two $Z_2$-odd states mass degenerate at the tree-level.

Another aspect extended Higgs sector is the search for Higgs quartic coupling.  The SM Higgs quartic coupling is till to be measured precisely and only bounds are obtained from the di-Higgs production constraints at the LHC \cite{Sirunyan:2018two,ATLAS:2018otd}. Extended Higgs sectors have many such quartic couplings and they differ from IDM to ITM and are very crucial in determining the fate of the Higgs potential. One or few such quartic couplings can provide the much needed Higgs-DM coupling \cite{HiggsDM1,HiggsDM}.  In this case we focus our region where the DM mass is greater than discovered Higgs mass, i.e. $125.5$ GeV. Considering the bounds from vacuum stability, perturbativity, DM relic and direct DM searches we estimate the allowed parameter space and try to distinguish IDM and ITM at the LHC via the compressed spectrum and less number $Z_2$-odd states for the later.

Higgs sector dark matter  also has appeal as the quartic coupling between SM-like Higgs boson and dark sector is crucial in measuring such scenario experimentally as well as theoretically. There have been lots of work done in measuring Higgs-DM coupling \cite{Honorez:2010re, HiggsDM1,HiggsDM,Arina:2009um, Araki:2010zz}; nevertheless a comprehensive study including bounds from vacuum stability, perturbativity, DM relic and direct DM is expected and which is the topic of this article.

This article is arranged as follows. In section~\ref{IDM} and section~\ref{ITM} we discuss the IDM and ITM  briefly along with electro-weak symmetry breaking conditions and  the tree-level Higgs boson masses. The comparative study of tree-level mass spectra between IDM and ITM is detailed in section~\ref{mass}. The perturbativity and vacuum stability  bounds are discussed in section~\ref{perturb} and section~\ref{stability} respectively. The DM relic and direct dark matter constraints are calculated in  section~\ref{DMrelic}  and section~\ref{DMdirect} respectively. Indirect bounds are discussed in section~\ref{HESS}. In section~\ref{scale} we dispense the parameter space verses the validity scale and  in section~\ref{pheno}  we discuss the LHC phenomenology briefly. Finally we conclude in section~\ref{concl}.

\section{Inert Doublet Model (IDM)}\label{IDM}
The inert 2HDM is a minimalist (apart from SM singlet) extension of the SM with a second $SU(2)$ Higgs doublet $\Phi_2$ with the same quantum numbers as the SM Higgs doublet $\Phi_1$. The Lagrangian is invariant under the $Z_2$ parity transformation where $\Phi_2 \rightarrow -\Phi_2$, $\Phi_1 \rightarrow \Phi_1$ and all the SM fields are even under this symmetry. Such discrete symmetry guarantees the absence of Yukawa couplings between  fermions and the inert doublet $\Phi_2$ and prohibits any tree-level flavor changing neutral currents. The most general renormalizable, CP conserving potential for inert doublet model \cite{Honorez:2010re,Gustafsson:2010zz}-\cite{LopezHonorez:2006gr} is given by
 \begin{eqnarray}\nonumber
\rm  V_{scalar} &&= m_{11}^2\Phi_1^\dagger \Phi_1 + m_{22}^2\Phi_2^\dagger\Phi_2 + \lambda_1(\Phi_1^\dagger \Phi_1)^2 + \lambda_2(\Phi_2^\dagger \Phi_2)^2 +
 \lambda_3(\Phi_1^\dagger \Phi_1)(\Phi_2^\dagger \Phi_2) \\ &&+ \lambda_4(\Phi_1^\dagger \Phi_2)(\Phi_2^\dagger \Phi_1) + [\lambda_5((\Phi_1^\dagger \Phi_2)^2) + h.c],  \label{Eq:1.1}
 \end{eqnarray}
  where,
  \begin{eqnarray}\nonumber
  \Phi_1=\begin{pmatrix}
  \phi^+_1\\
  \phi^0_1
  \end{pmatrix},  \quad \quad  \Phi_2=\begin{pmatrix}
  \phi^+_2\\
  \phi^0_2
  \end{pmatrix}
  \end{eqnarray}
 and $m_{11}^2$, $m_{22}^2$ and $\lambda_{1-5}$ are real parameters.
Electro-weak symmetry breaking is achieved by giving real vev to the first Higgs doublet i.e. $\Phi_1$ and the second Higgs doublet does not take part in EWSB. At EW minima, 
\begin{equation}
\langle \Phi _1 \rangle \ = \ \frac{1}{\sqrt 2}\left(
\begin{array}{c}
0 \\
v \\
\end{array}
\right),
\end{equation}
with $v\simeq 246$ GeV, whereas the second Higgs doublet, being $Z_2$-odd, does not take part in symmetry breaking; hence the name  is`inert 2HDM'.

Using minimization conditions, we express the mass parameter $m_{11}^2$ in terms of other parameters as follows: 
\begin{align}
m_{11}^2 \ = \ -\lambda _1v^2.
\end{align}
Except for the SM Higgs boson, $h$, four new physical scalar states are present: one charged Higgs boson pair $H^{\pm}$, one CP-even neutral Higgs boson $H_0$ and one CP-odd neutral Higgs boson $A$. Lightest of the the two neutral Higgs bosons can be a candidate of cold dark matter that would be discussed later. After  electroweak symmetry breaking, the masses of the scalar particles are given by:
\begin{eqnarray}\label{mass1}
\rm M_{h}^2 &= &2\lambda_1 v^2 \nn\\
\rm M_{H_0}^2 &=& \frac{1}{2}(2m_{22}^2+v^2(\lambda_3+ \lambda_4+2\lambda_5))\nn\\
\rm M_{A}^2 &=& \frac{1}{2}(2m_{22}^2+v^2(\lambda_3+\lambda_4-2\lambda_5))\nn\\
\rm M_{H^\pm}^2 &=& m_{22}^2+\frac{1}{2}v^2
\lambda_3.
\end{eqnarray} 
Since, $\Phi_2$ is inert, there is no mixing between $\Phi_1$ and $\Phi_2$ and the gauges eigenstates are same as the mass eigenstates for the Higgs bosons. The $Z_2$ symmetry prevents any such mass mixing through Higgs portal and it also prevents the second  Higgs doublet to couple to fermions. In this case we get two CP-even neutral Higgs $h$ and $H_0$, where $h$ is likely to be the discovered Higgs boson around 125 GeV at the LHC~\cite{Aad:2012tfa, Chatrchyan:2012xdj} and the other is yet to be found out. Similarly we are also looking for the pseudoscalar Higgs boson $A$ and the charged Higgs boson $H^\pm$ at the collider. It can be seen from Eq.~\ref{mass1} that $H_0$, $A$ and $H^\pm$ are nearly degenerate. Depending upon the sign of $\lambda_5$ one of scalar between $H_0$ and $A$ can be lighter and a cold dark matter candidate \cite{Gustafsson:2010zz}-\cite{LopezHonorez:2006gr}. Unlike \cite{khan1,2HDMpheno} here we concentrate of $M_{H_0}, M_A > m_h$ and the corresponding couplings. 


\section{ Inert Triplet Model (ITM)} \label{ITM}
In completing SM with a dark sector we can have DM in the $SU(2)$ triplet representation
which does not take part in the EWSB.  This can be simply achieved by adding a $SU(2)$ real triplet scalar with $Y=0$ hypercharge and again making it $Z_2$-odd to provide to take part in EWSB \cite{Tripletex}.  Here we introduce in addition to SM Higgs doublet i.e. $\Phi$, another  $SU(2)_L$ triplet scalar with Y=0, i.e. $T$ and due to $Z_2$-odd nature, the triplet field does not take part in EWSB, i.e. the vev of the triplet, $\rm v_T=0$.
\begin{center}
	$	\Phi
	= \left(\begin{array}{c}
	\phi^+   \\
	\phi^0   \end{array}\right) $ , 
	$T =\frac{1}{2} \left(
	\begin{array}{cc}
	T_0 & \sqrt{2} T^+ \\
	\sqrt{2} T^- & -T_0 \\
	\end{array}
	\right)$.
\end{center}
The Higgs Lagrangian for ITM case can be written as,
\be
{\cal L}_k=(D^\mu \Phi)^\dagger (D_\mu \Phi)+Tr[(D^\mu T)^\dagger (D_\mu T)]-V(\Phi,T), 
\ee
where the covariant derivatives involving the gauge-fields are given by,
\bea\label{Tripeq}
D_\mu \Phi&=&(\partial_\mu -i \frac{g}{2}\tau^a W_\mu^a-i\frac{g^\prime }{2}B_\mu Y)\Phi,  \\ 
D_\mu T&=&(\partial_\mu -i \frac{g}{2}\tau^a W_\mu^a)T.
\eea

Now we impose an additional $Z_2$ symmetry under which triplet is assigned to be odd and other fields are even. The Lagrangian is invariant under the $Z_2$ parity transformation where $T \rightarrow -T$ and all the SM fields are even. A $Z_2$ symmetric potential for ITM  can be written as:
\be
V=m_h^2\Phi^\dagger\Phi+m_T^2Tr(T^\dagger T)+\lambda_1|\Phi^\dagger\Phi|^2+\lambda_t(Tr|T^\dagger T|)^2+\lambda_{ht}\Phi^\dagger\Phi Tr(T^\dagger T). \label{Eq:2.4}
\ee

In ITM the triplet field does not get vev i.e., $\rm v_T=0$ and only doublet gets vev as given by,
\begin{center}
	$	\Phi
	= \left(\begin{array}{c}
	G^+   \\
	\frac{1}{\sqrt{2}}(v_h+h)+i G^0  \end{array}\right) $, \qquad \qquad
	$T =\frac{1}{2} \left(
	\begin{array}{cc}
	T_0 & \sqrt{2} T^+ \\
	\sqrt{2} T^- & -T_0 \\
	\end{array}
	\right).$
\end{center}\vskip 0.5cm
Here $v=246$ GeV and the model in known as `inert triplet model'. In minimization conditions, we express the mass parameter $m_{h}^2$ in terms of other parameters as follows: 
\be
m_h^2= -\lambda _1 v_h^2.
\ee
Triplet field  does not contribute to mass of any of the SM particle and the gauge bososn masses solely get contribution from $\Phi$ as shown below:
\be
M_w^2 = {g^2\over 2} v_h^2,  \quad 
M_z^2 = {(g^2+g'^2)\over 4}v_h^2.
\ee
Thus in this case $\rho = {M_w^2\over cos^2\theta_w M_z^2}$ stays in SM value at the tree-level. Except for the SM Higgs boson, $h$, three new physical scalar particle states are present: one charged Higgs boson pair $T^{\pm}$ and one CP-even neutral Higgs boson $T_{0}$. After EWSB the physical Higgs boson masses can be read as:
\begin{eqnarray}\label{mass2}
M_h^2 &=& {2 \lambda_1 v_h^2}\nn\\
M_{T_0}^2&=&\frac{1}{2} v_h^2 \lambda _{\text{ht}}+m^{2}_T \nn\\
M^2_{T^\pm}&=&\frac{1}{2} v_h^2 \lambda _{\text{ht}}+m^{2}_T,
\end{eqnarray}
where $m_T$ and $\lambda_{ht}$ are the parameters as shown in the Higgs potential Eq.~\ref{Eq:2.4}. Note that at the tree-level from Eq.\ref{mass2}, masses of neutral and charged components are the same, but loop corrections tend to make the charged components, $T^{\pm}$ slightly heavier than the neutral one $T_0$ with a mass gap of $\delta M( M_T^{\pm}, M_{T_0}) = 166$ MeV \cite{Cirelli:2005uq}. Hence, $T_0$  turns out to be lightest component of triplet scalar and a suitable DM candidate.

Next we compare both the models after EWSB by their physical mass eigenstates, mass spectrum and perturbativity, stability bounds.  We mentioned earlier that for IDM we have one extra excitation as CP-odd Higgs boson i.e. $A$ which can be a DM candidate. Whereas in case of ITM the DM is always a purely CP-even scalar. In sections below we categorically address the issues regarding the mass spectrum, bounds from perturbativity and vacuum stability, DM relic and  direct dark matter detection.

\section{Mass spectrum of IDM and ITM}\label{mass}
\begin{figure}[H]
	\hspace*{+0.5cm}
	{\mbox{\subfigure[]{\includegraphics[width=0.47\linewidth,angle=-0]{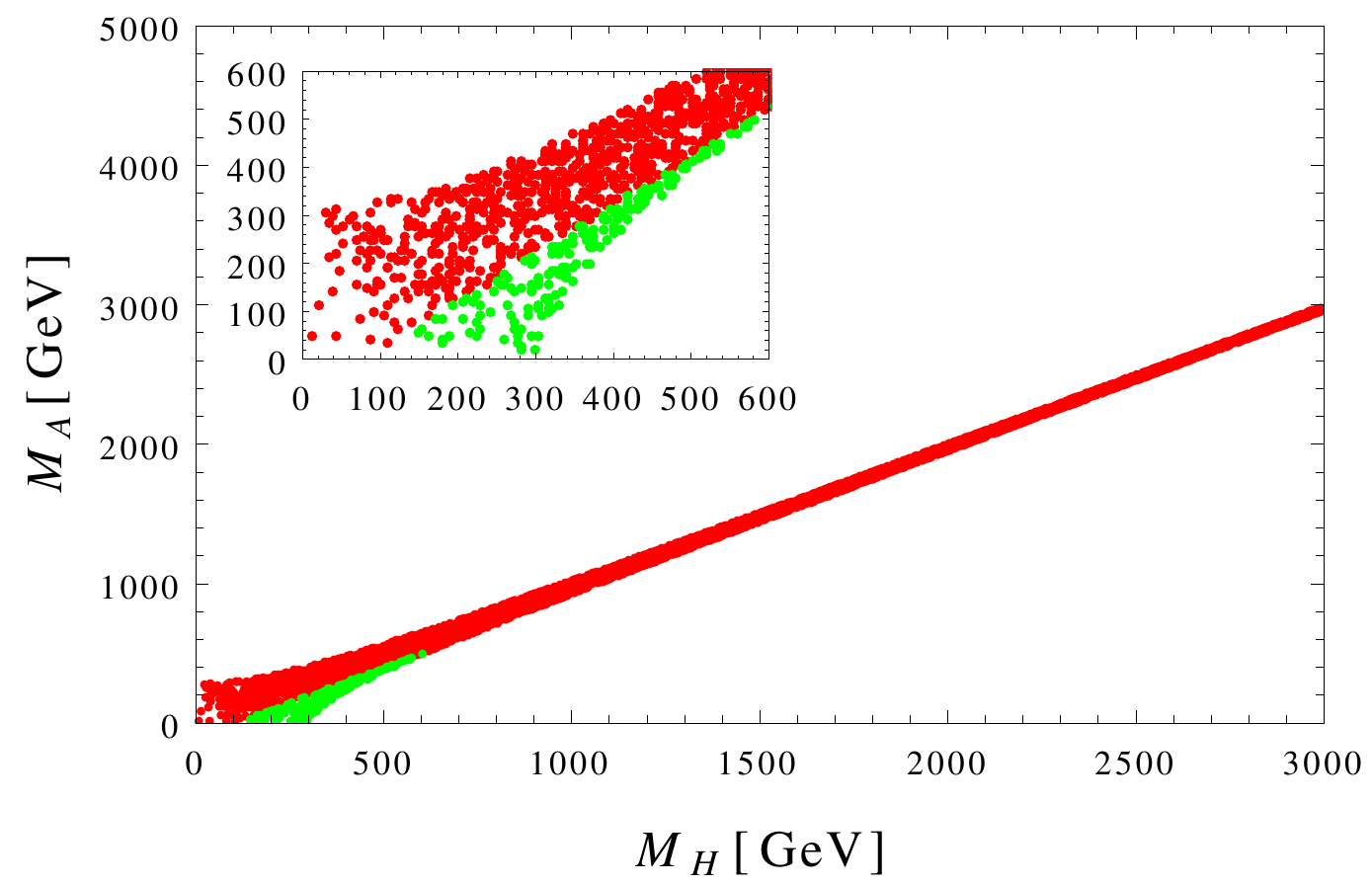}\label{f1}}
			\subfigure[]{\includegraphics[width=0.47\linewidth,angle=-0]{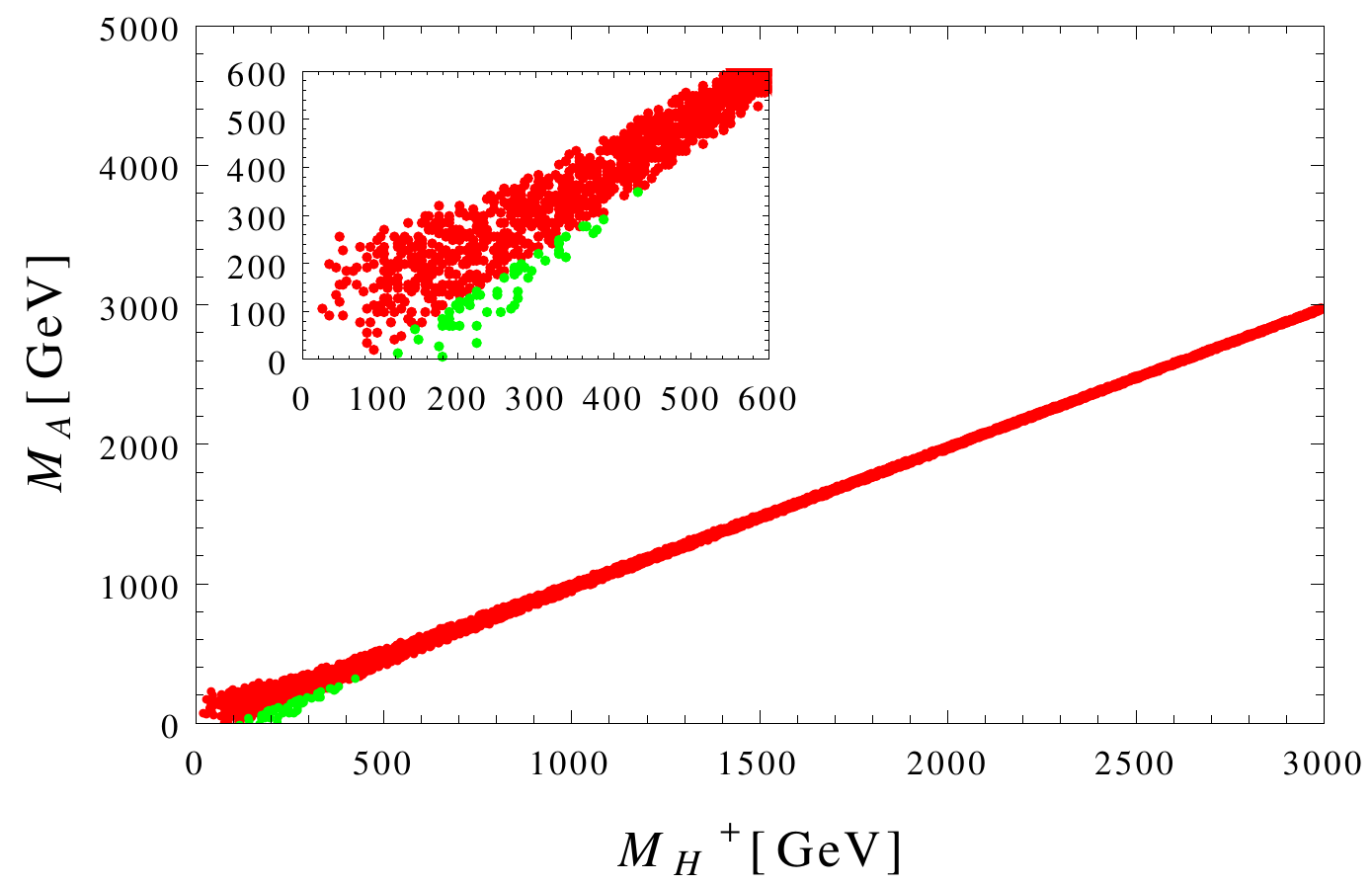}\label{f2}}}
		\caption{\ref{f1}: Mass correlation between $\rm M_{H_0}$ and $\rm M_A$ in GeV; \ref{f2}: Mass correlation between $M_H^{+}$ and $M_A$ in GeV. Green color corresponds to mass splitting greater than $\rm M_W$  and red color corresponds to mass splitting less than $\rm M_W$. The mass splitting between $M_H^{+}$ and $M_A$ is lower than the corresponding splitting between $M_{H_0}$ and $M_A$.}\label{fig1}}
\end{figure}

\begin{figure}[H]
	\hspace*{+0.5cm}
	{\mbox{\subfigure[]{\includegraphics[width=0.45\linewidth,angle=-0]{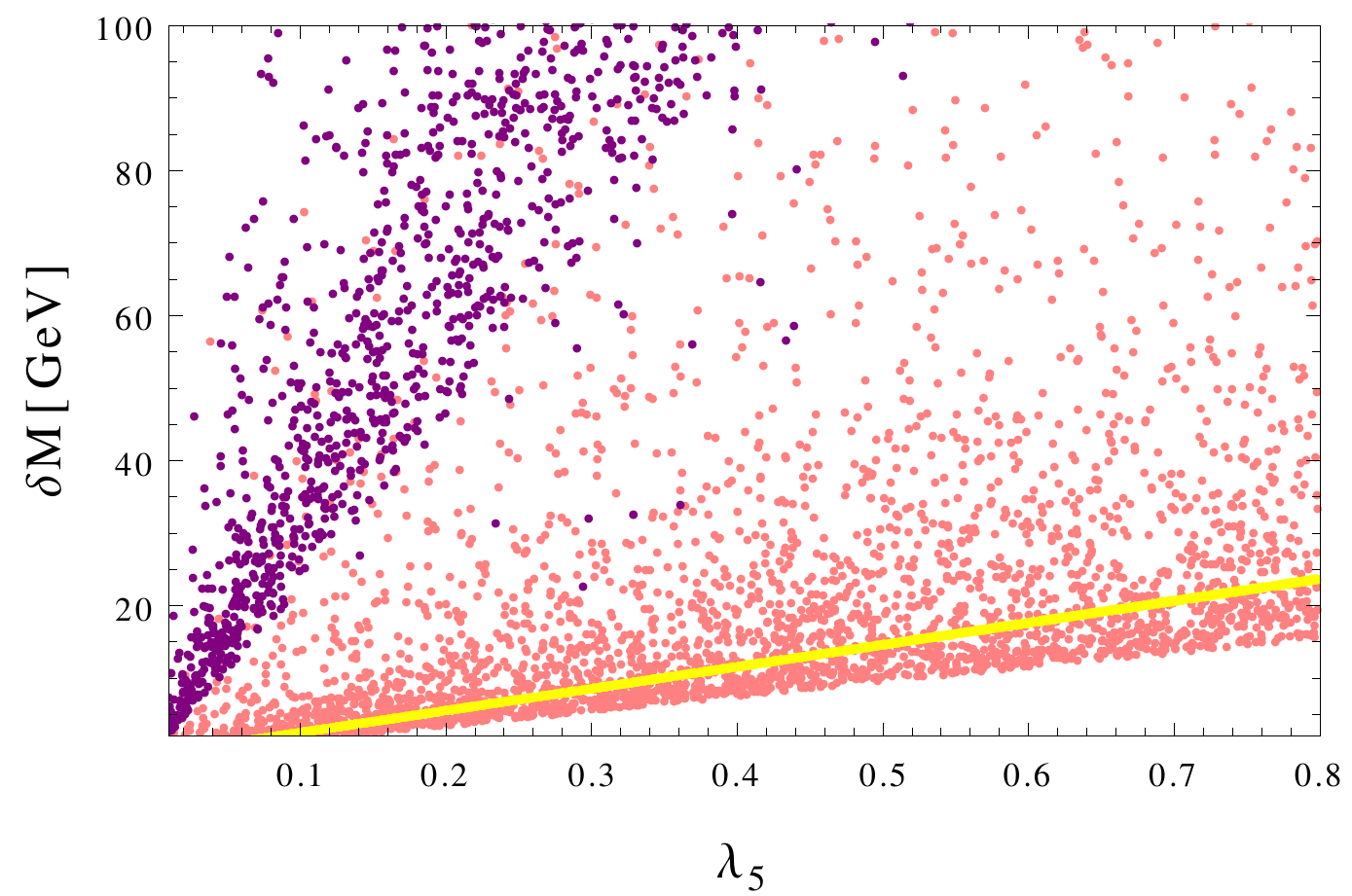}\label{f3}}
			\subfigure[]{\includegraphics[width=0.45\linewidth,angle=-0]{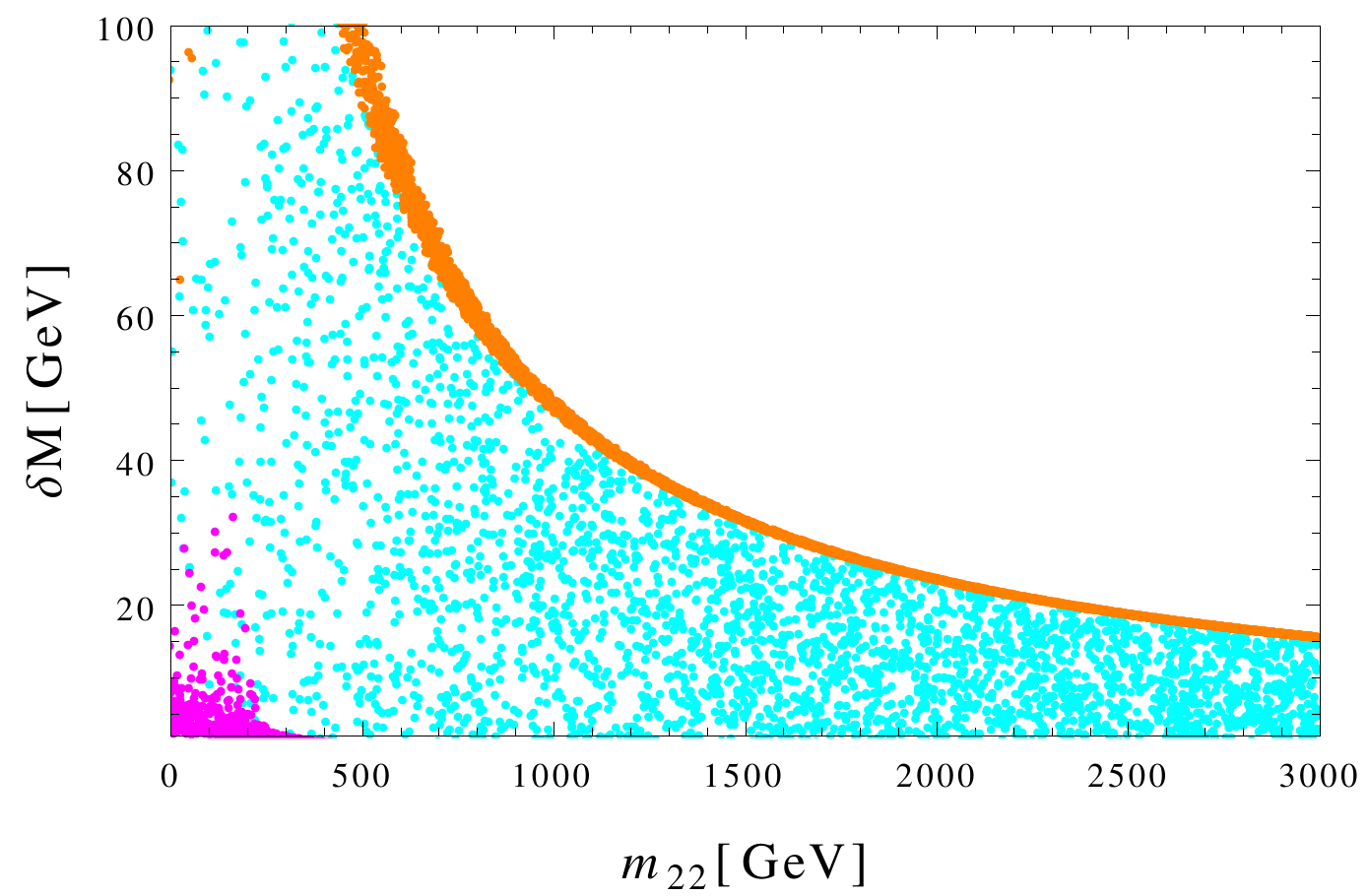}\label{f4}}}
		\caption{\ref{f3}: Variation of $\lambda_5$ verses $\delta M(M_{H_0}-M_A)$  in GeV  for different values of $m_{22}$. Purple, yellow and pink colours describe the variation for $m_{22}$=150, 2000 GeV and for $100-3000$ GeV respectively; \ref{f4}: Variation of $m_{22}$ verses $\delta M(M_{H_0}-M_A)$  in GeV for different values of $\lambda_5$.  Here the magenta and orange colours correspond to $\lambda_5=0.01, 0.8$ respectively and the cyan region corresponds to $\lambda_5=0.01-0.80 $. For lower values of $m_{22}$, mass splitting can be greater than $\sim$ 100 GeV and it comes down to $\sim$ 20 GeV for higher values $m_{22} \sim 3000$ GeV depending on the allowed parameter space.}\label{fig2l}}
\end{figure}

In Figure~\ref{fig1} we describe the mass correlations among the heavier Higgs states for both IDM. Figure~\ref{f1} depicts the mass correlation between $\rm M_{H_0}$ and $\rm M_A$ in GeV and the green colour corresponds to the mass-splitting greater than $M_W$ and red colour describes the mass-splitting less than $M_W$. In this case the tree-level mass splitting is generated by the $\lambda_5$ term. Such mass splitting is greater in the lower mass range and  as the mass spectrum increases, $m_{22}$ term dominates over the $\lambda_{3-5}$ which makes  all $Z_2$ odd states almost degenerate. We find that the mass splitting between $\rm M_H$ and $\rm M_A$ is greater than $W$ boson mass till $\rm M_{H_0}=600$GeV. This mass-splitting between $M_H^{+}$ and $M_A$ keeps below $M_W$ for  $M_{H^\pm} \lesssim 400$ GeV as can be seen from Figure~\ref{f2}. We also note that the mass splitting between $M_H^{+}$ and $M_A$ is lower than the corresponding splitting between $M_{H_0}$ and $M_A$.
Figure~\ref{f3} shows the variation of $\lambda_5$ and $\delta M(M_{H_0}-M_A)$ for different values of $m_{22}$. Purple, yellow and pink colours describe the variation for $m_{22}$=150, 2000 GeV and for $100-3000$ GeV respectively. As the value of $m_{22}$ is increasing it dominates the splitting effect of quartic couplings and the mass-splitting becomes lower and lower. Figure~\ref{f4} depicts the mass splitting  $\delta M $ $(M_{H_0}-M_A)$ with  $m_{22}$ for different values of $\lambda_5$. Here the magenta and orange colours correspond to $\lambda_5=0.01, 0.8$ respectively and the cyan region corresponds to $\lambda_5=0.01-0.80 $. For lower values of $m_{22}$, mass splitting can be greater than $\sim$ 100 GeV and it comes down to $\sim$ 20 GeV for higher values $m_{22} \sim 3000$ GeV depending on the allowed parameter space.
\begin{figure}[H]
	\begin{center}
		\includegraphics[width=0.5\linewidth,angle=-0]{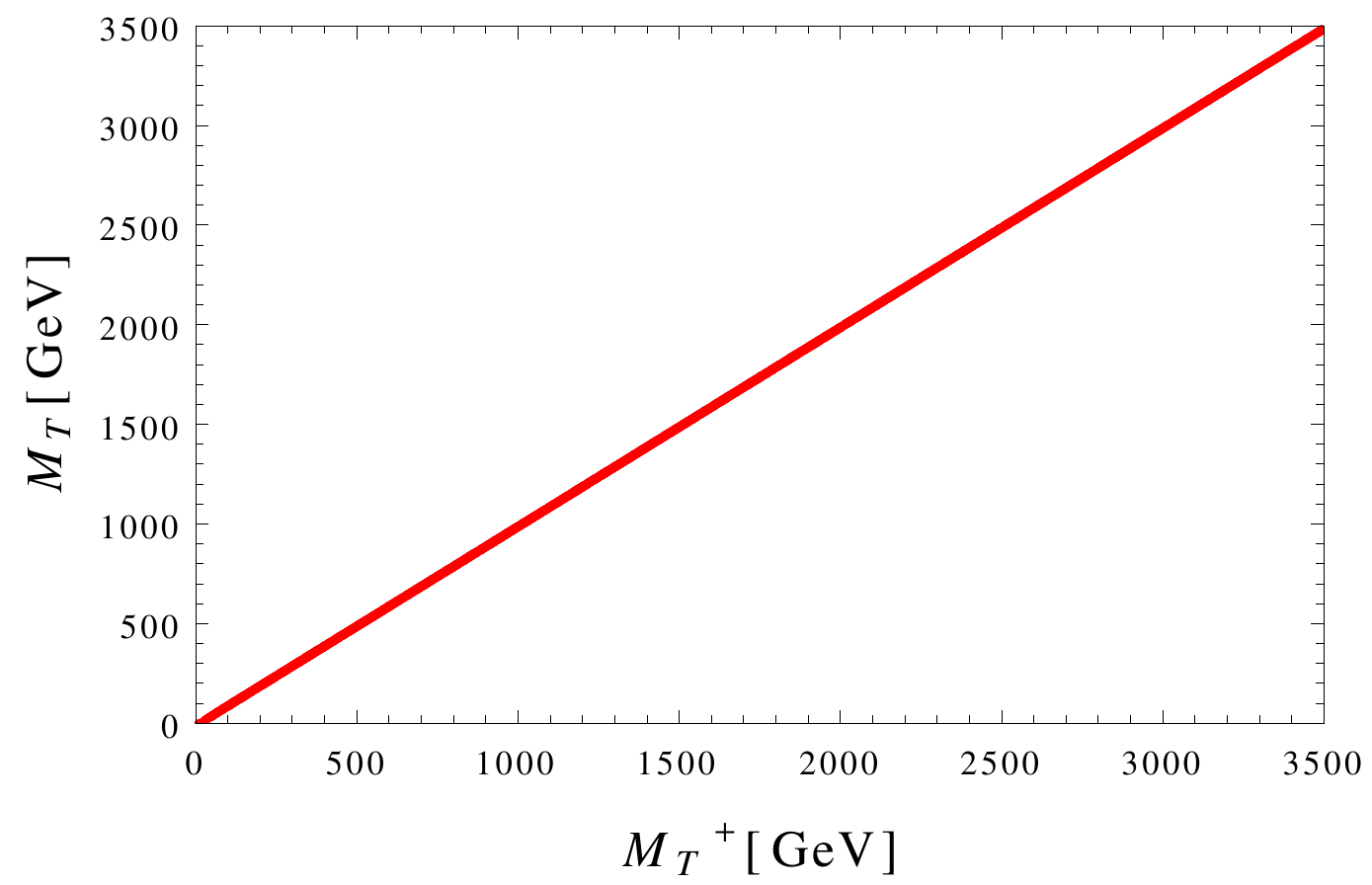}
		\caption{Variation of $M_T^{\pm}$ vs $M_{T_0}$ in GeV. At the tree-level there is no mass-splitting and the width describes the different solutions allowed by parameter space. }\label{fig3l}
	\end{center}
\end{figure}
Figure~\ref{fig3l} describes the variation of $M_T^{\pm}$ vs $M_{T_0}$ in GeV at tree-level. We see that at the tree-level there is no mass-splitting between triplet states. One has to rely on the loop-contributions for $\mathcal{O}(166)$ MeV mass splitting between $T^\pm$ and $T_0$ which will be crucial for the phenomenological studies \cite{Cirelli:2005uq}.

\section{Perturbativity Bound}\label{perturb}
To emulate the theoretical bounds from  perturbative unitarity of the dimensionless couplings, we impose that all dimensionless couplings of the model must remain perturbative for a given value of the energy scale $\mu$, i.e. the couplings must satisfy the following constraints:
\begin{align}
\left|\lambda_i\right|  \ \leq \ 4 \pi, \qquad
\left|g_j\right| \ \leq \ 4 \pi, \qquad \left|Y_k\right|  \ \leq \ \sqrt{4\pi} \, ,
\end{align}
where $\lambda_i$ with $i=1,2,3,4,5$ are the scalar quartic couplings; $g_j$ with $j=1,2$ are EW gauge couplings;\footnote{The running of the strong coupling $g_3$ is same as in the SM, so we do not show it here.}  and $Y_k$ with $k=u,d,\ell$ are all Yukawa couplings for the up, down types quarks and leptons respectively.  The two-loop beta functions generated by  {\tt SARAH 4.13.0}~\cite{Staub:2013tta}, given in Appendix~\ref{betaf1} and Appendix~\ref{betaf2} are used to check the variations of the dimensionless couplings with the scale of the variation ($\mu$ in GeV).
\begin{figure}[H] 
	\begin{center}
		
		\mbox
		{\subfigure[]{\includegraphics[width=0.5\linewidth,angle=-0]{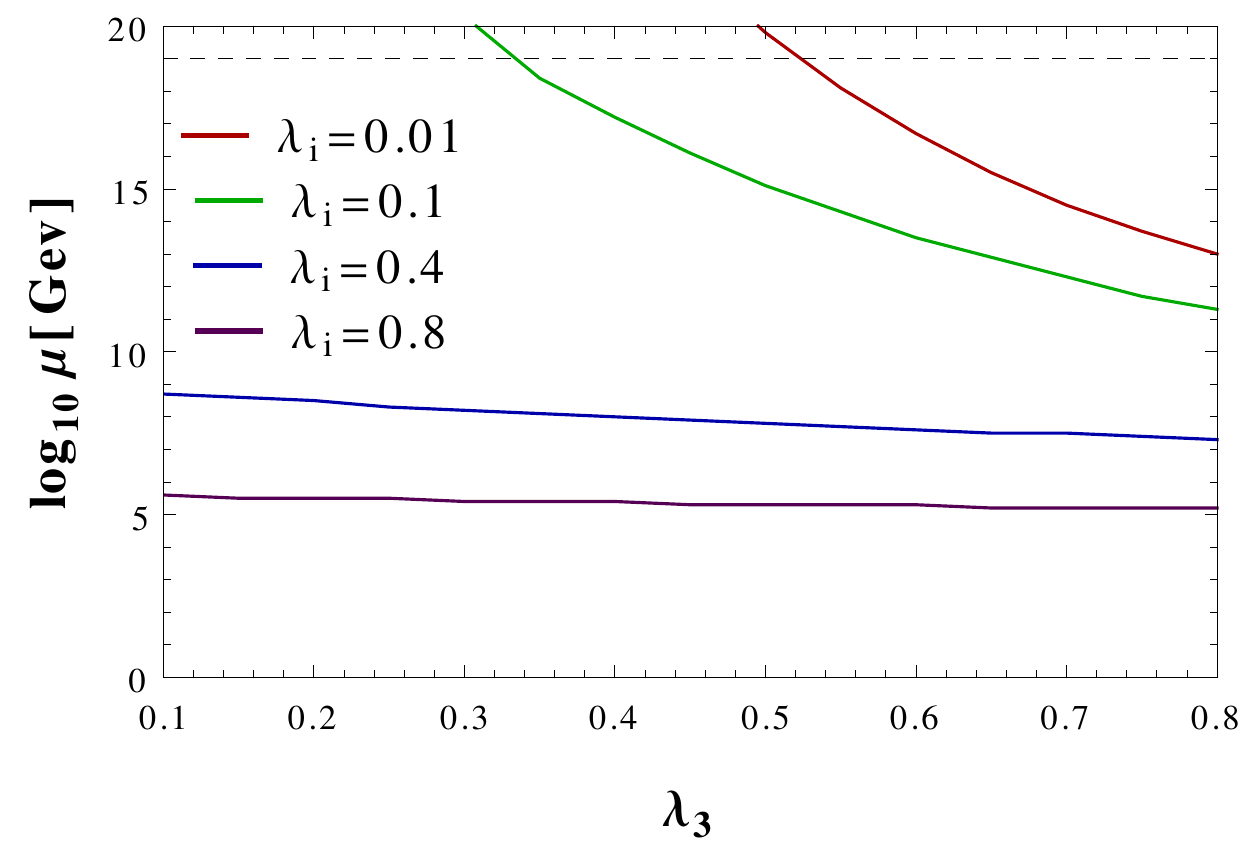}\label{f7}}	\subfigure[]{
				\includegraphics[width=0.5\linewidth,angle=-0]{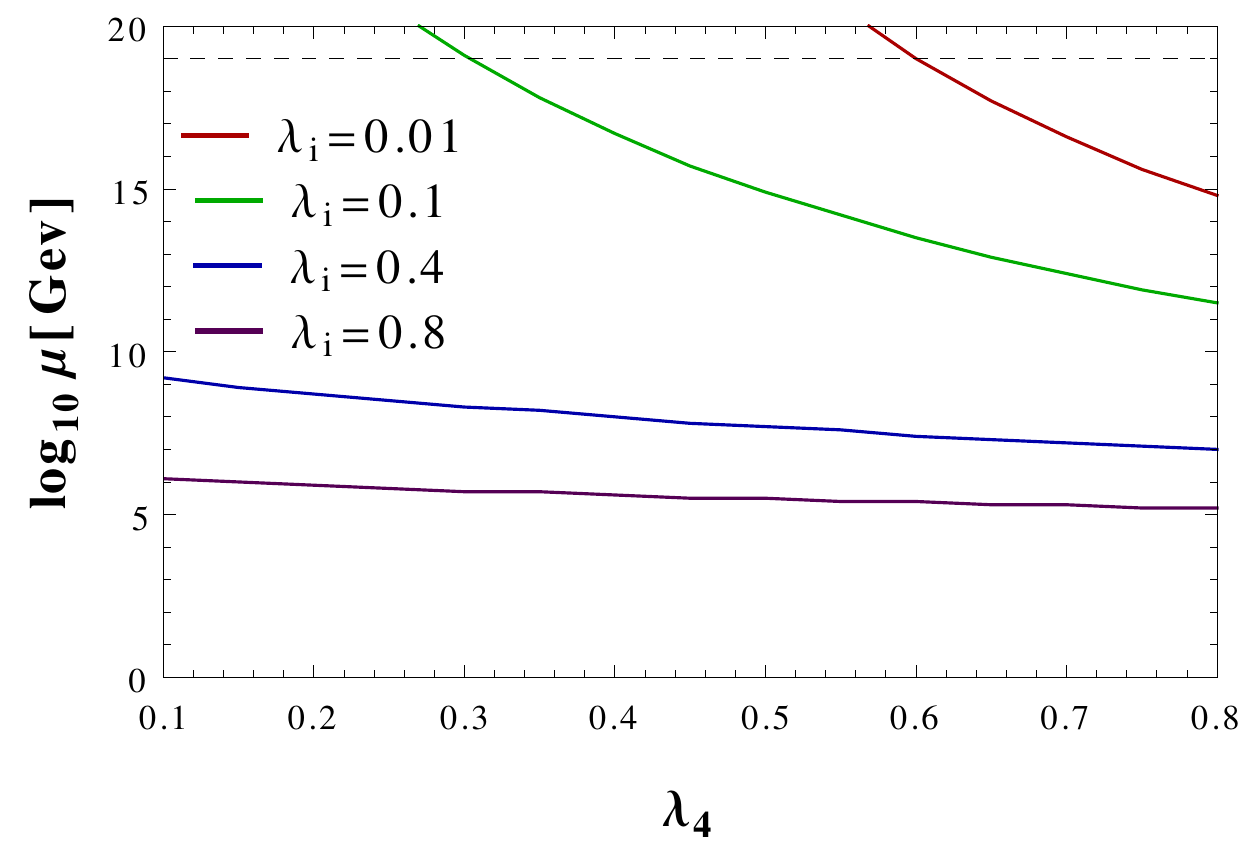}\label{f8}
		}}
		\mbox{\subfigure[]
			{
				\includegraphics[width=0.5\linewidth,angle=-0]{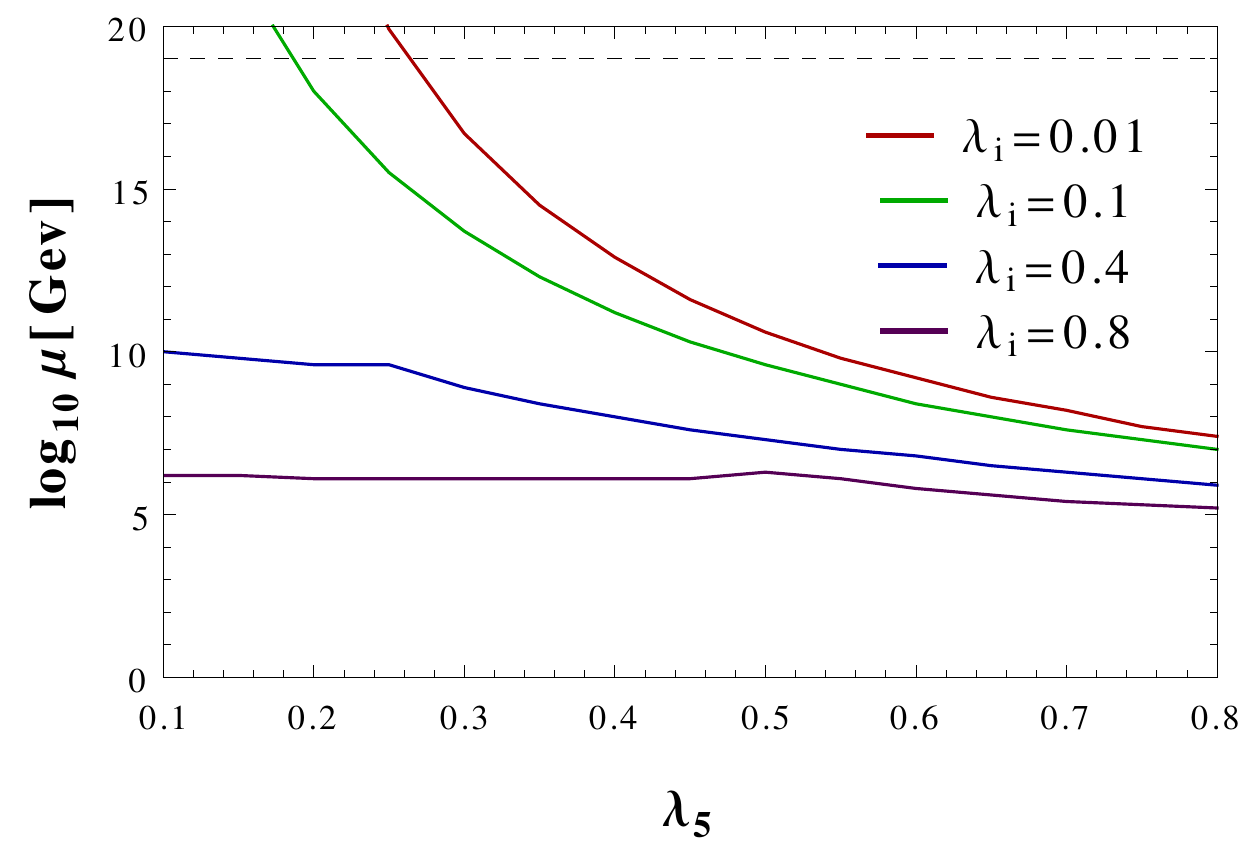}\label{f9}}
		}
		\caption{Two-loop running of the scalar quartic couplings $\lambda_3$, $\lambda_4$ and $\lambda_5$ as a function of perturbative scale. Here red, green, blue and purple curves in each plot correspond to different initial conditions for $\lambda_i$ (with $i=2,3,4,5$) at the EW scale, representative of very weak ($\lambda_i = 0.01$), weak ($\lambda_i = 0.1$), moderate ($\lambda_i = 0.4$) and strong ($\lambda_i = 0.8$) coupling limits respectively.}\label{fig4l}
	\end{center}
\end{figure}

 The perturbativity behaviour of the scalar quartic couplings $\lambda_{3,4,5}$ is studied in Figure~\ref{f7}-\ref{f9} respectively where the other quartic couplings $\lambda_{(i=2,3,4,5)}$ are fixed at some values. Here red, green, blue and purple curves in each plot correspond to different initial conditions for other $\lambda_i$ at the EW scale, representative of very weak ($\lambda_i = 0.01$), weak ($\lambda_i = 0.10$), moderate ($\lambda_i = 0.40$) and strong ($\lambda_i = 0.80$) coupling limits respectively. The dashed black line corresponds to Planck scale ($10^{19}$ GeV). Higgs quartic coupling $\lambda_3$ remains perturbative till Planck scale for $\lambda_3 \lesssim 0.51, 0.32$ for $\lambda_i(\rm{EW}) =0.01, 0.10$ respectively as shown in Figure~\ref{f7}. For $\lambda_i (\rm{EW})=0.40, 0.80$ theory becomes non-perturbative at much lower scale $\sim$ $10^{8.9}, \, 10^{5.6}$ GeV respectively for almost all initial values of $\lambda_3$.

 Figure~\ref{f8} shows similar behaviour for $\lambda_4$ and here for the choice of $\lambda_i(\rm{EW})=0.01, 0.10$ the perturbative limits remain valid till Planck scale  for $\lambda_4 \lesssim 0.60, \, 0.30$ respectively. For higher values of $\lambda_i(\rm{EW})$ the perturbative bounds remain similar to Figure~\ref{f7}.  Figure~\ref{f9} depicts the behaviour for $\lambda_5$ for the chosen other  $\lambda_i(\rm{EW})$.  Here for $\lambda_i(\rm{EW})=0.01, 0.10$ the perturbative limit till Planck scale   is valid for $\lambda_5$  $\lesssim 0.28, \, 0.19$ respectively.  In general when $\lambda_i\simeq 0.1-0.2$ at the EW scale, all the quartic couplings remains perturbative till Planck scale for IDM.

\begin{figure}[H]
	\begin{center}
		\includegraphics[width=0.65\linewidth,angle=-0]{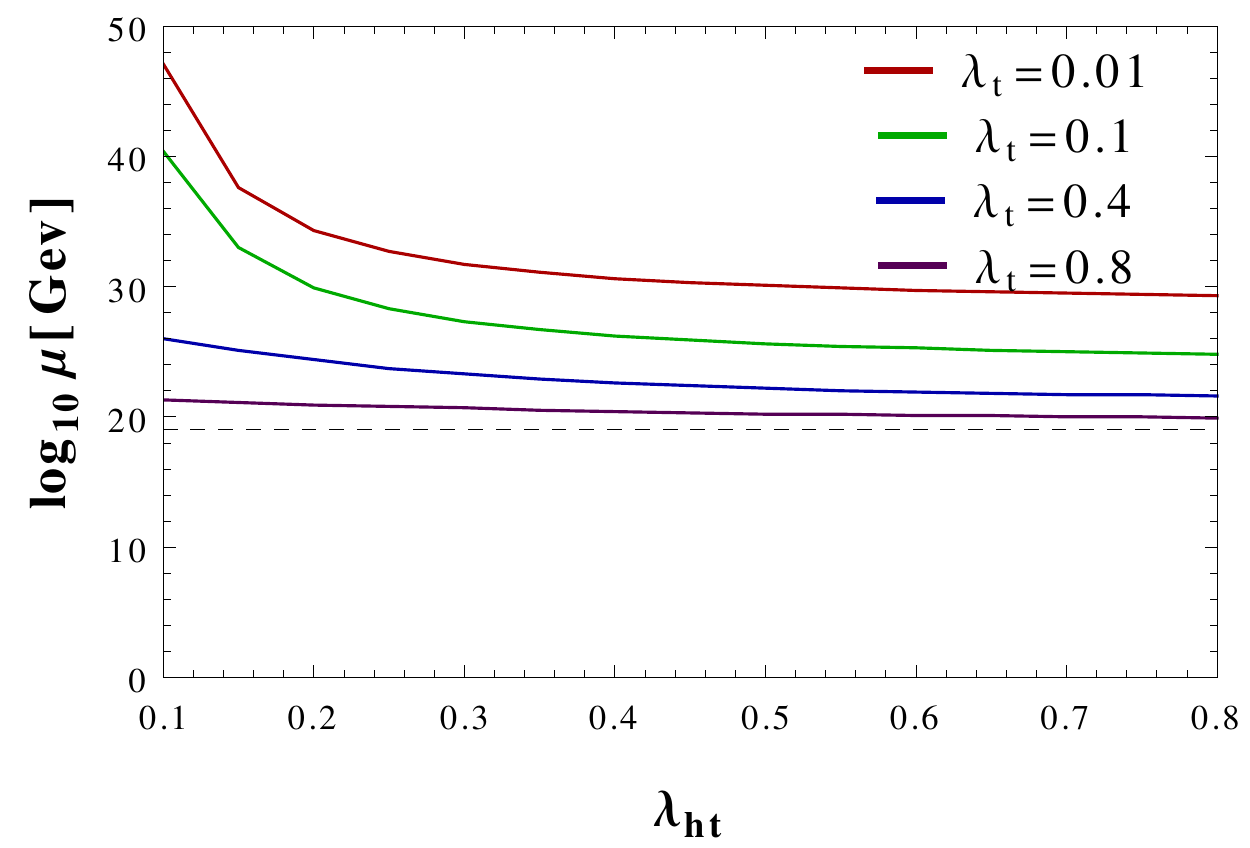}
		\caption{Two-loop running of the scalar quartic coupling $\lambda_{ht}$ as a function of perturbative scale. Here red, green, blue and purple curves in each plot correspond to different initial conditions for $\lambda_t$ and $\lambda_{ht}$  at the EW scale, representative of very weak ($\lambda_t = 0.01$), weak ($\lambda_t = 0.1$), moderate ($\lambda_t = 0.4$) and strong ($\lambda_t = 0.8$) coupling limits respectively. }\label{fig5l}
	\end{center}
\end{figure}

Figure~\ref{fig5l} shows the variation of quartic coupling $\lambda_{ht}$ which describes the interaction between SM doublet and  $Y=0$ Higgs triplet.  The dashed black line corresponds to the Planck scale. Due to the existence of lesser number of quartic couplings compared to the 2HDM, the theory stays perturbative till Planck scale for much higher values of quartic couplings $\lambda_t, \lambda_{ht}$. For choice of $\lambda_t(\rm EW)$=0.01, 0.1, 0.4 and 0.8 the perturbative limits remains valid till Planck scale for $\lambda_{ht}=0.1-0.8$ at EW scale. Perturbativity puts  upper bound on Higgs quartic coupling  $\lambda_{ht}\lesssim 0.50$ for $\lambda_t=1.3$ at the EW scale. For ITM case the SM-like Higgs quartic coupling only takes part in the EWSB breaking and its values at two-loop level is fixed at $0.1264$ for the SM-like Higgs boson mass  at $125.50$ GeV.

 \section{	Stability Bound} \label{stability}
 In this section we discuss the stability of Higgs potential via two different approaches. Firstly via calculating two-loop scalar quartic couplings and checking if the SM-like  Higgs quartic coupling $\lambda_h$ is getting negative at some scale. In this case $\lambda_h=\lambda_1$ at tree-level but at one-loop and two-loop levels $\lambda_h$ gets contribution from SM fields as well as the BSM scalars as we describe in the subsection~\ref{loops}. For the simplicity in subsection~\ref{loops} we give the expressions of the corresponding beta functions at one-loop level and in the Appendix~\ref{betaf1}, \ref{betaf2} the two-loop beta functions are given. 
 
\subsection{RG Evolution of the Scalar Quartic Couplings }\label{loops}
To study the evaluations of dimensionless couplings we implemented both the IDM and the ITM scenarios in {\tt SARAH 4.13.0}~\cite{Staub:2013tta}  and the corresponding $\beta$-functions for various gauge, quartic and Yukawa couplings are calculated at one- and two-loop levels. The explicit expressions for the two-loop $\beta$-functions can be found in Appendix~\ref{betaf1}, \ref{betaf2} and they are used in our numerical analysis of vacuum stability in this section. To illustrate the effect of the Yukawa and additional scalar quartic couplings on the RG evolution of the SM-like Higgs quartic coupling $\lambda_1$ in the scalar potential~\eqref{Eq:1.1} and \eqref{Eq:2.4}, let us first look at the one-loop $\beta$-functions. $\lambda_h=\lambda_1$ at tree-level and at the one-loop level, the $\beta$-function for the SM Higgs quartic coupling in this model receives two different contributions: one from the SM gauge, Yukawa, quartic interactions and the second from the inert scalar sectors of IDM/ITM as shown below: 
\be\label{lfull}
\beta_{\lambda_1 }   =  \beta_{\lambda _1}^{\rm SM} + \beta_{\lambda _1}^{\rm IDM/ITM},
\ee
where, 
\begin{eqnarray}\label{b1}
\beta_{\lambda_1}^{\rm SM} & =  & \frac{1}{16\pi^2}\Bigg[
\frac{27}{200} g_{1}^{4} +\frac{9}{20} g_{1}^{2} g_{2}^{2} +\frac{9}{8} g_{2}^{4} -\frac{9}{5} g_{1}^{2} \lambda_1 -9 g_{2}^{2} \lambda_1 +24 \lambda_1^{2}\nonumber \\
&&  +12 \lambda_1 \mbox{Tr}\Big({Y_u  Y_{u}^{\dagger}}\Big) +12 \lambda_1 {\rm Tr}\Big({Y_d  Y_{d}^{\dagger}}\Big) +4 \lambda_1 \mbox{Tr}\Big({Y_e  Y_{e}^{\dagger}}\Big)  \nonumber\\
&&-6 \mbox{Tr}\Big({Y_u  Y_{u}^{\dagger}  Y_u  Y_{u}^{\dagger}}\Big)-6 \mbox{Tr}\Big({Y_d  Y_{d}^{\dagger}  Y_d  Y_{d}^{\dagger}}\Big) -2 \mbox{Tr}\Big({Y_\ell  Y_{\ell}^{\dagger}  Y_\ell  Y_{\ell}^{\dagger}}\Big) \Bigg], \label{eq:7.2} \\
\beta_{\lambda_1}^{\rm IDM} & =  & \frac{1}{16 \pi^2}\Big[ 2 \lambda_{3}^{2} +2 \lambda_3 \lambda_4 +\lambda_{4}^{2}+4 \lambda_{5}^{2}\Big]. \label{eq:7.3}\\
\beta_{\lambda_1}^{\rm ITM} & =  & \frac{1}{16 \pi^2}\Big[ 8 \lambda_{ht}^{2}\Big]. \label{eq:7.4}
\end{eqnarray}		
Here $g_1,g_2,g_3$ are respectively the $U(1)_Y$, $SU(2)_L$ and $SU(3)_c$ gauge couplings, and $Y_u, Y_d, Y_\ell$ are respectively the up, down and lepton-Yukawa coupling matrices of SM. We use the SM input values for these parameters at the EW scale: $\lambda_1=0.1264$, $g_1=0.3583$, $g_2=0.6478$, $y_t=0.9369$ and other Yukawa couplings are neglected~\cite{Degrassi:2012ry, Buttazzo:2013uya}.

\begin{figure}
	\begin{center}
		
		\mbox
		{\subfigure[]{\includegraphics[width=0.5\linewidth,angle=-0]{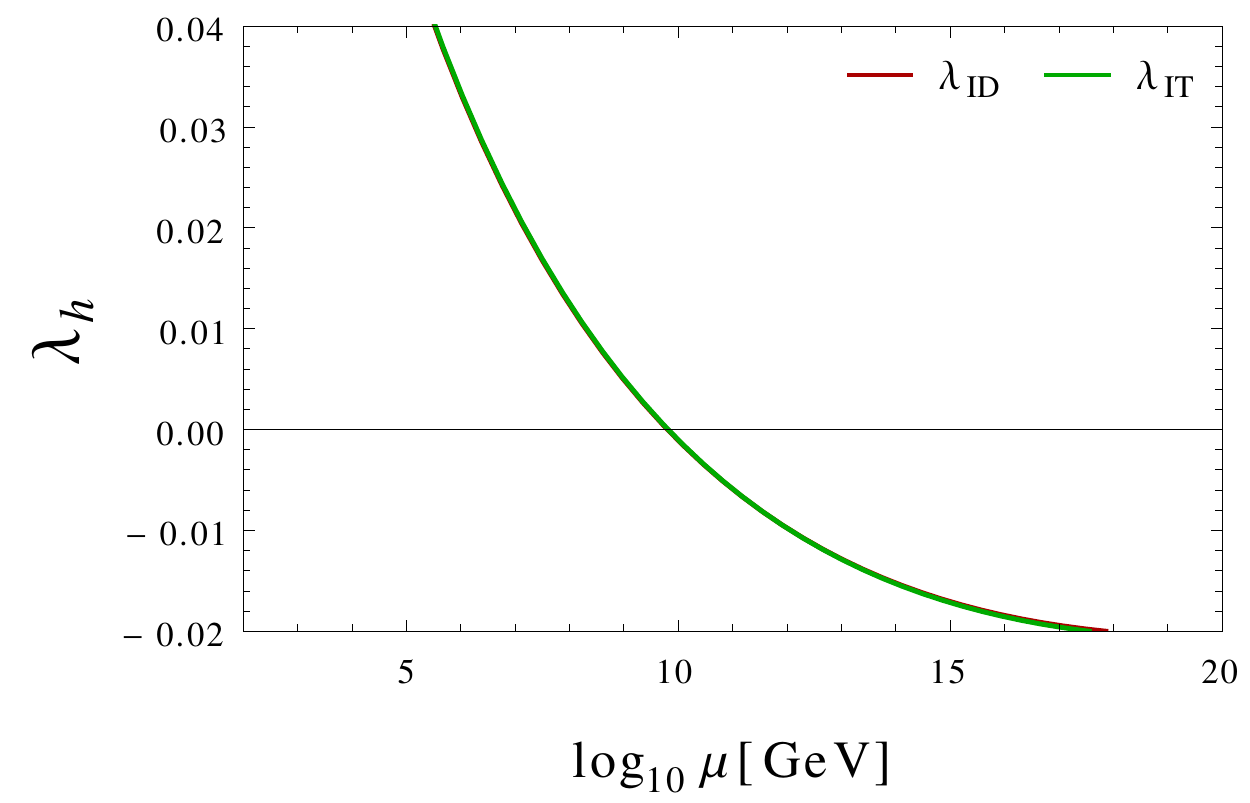}\label{f10}}	\subfigure[]{
				\includegraphics[width=0.5\linewidth,angle=-0]{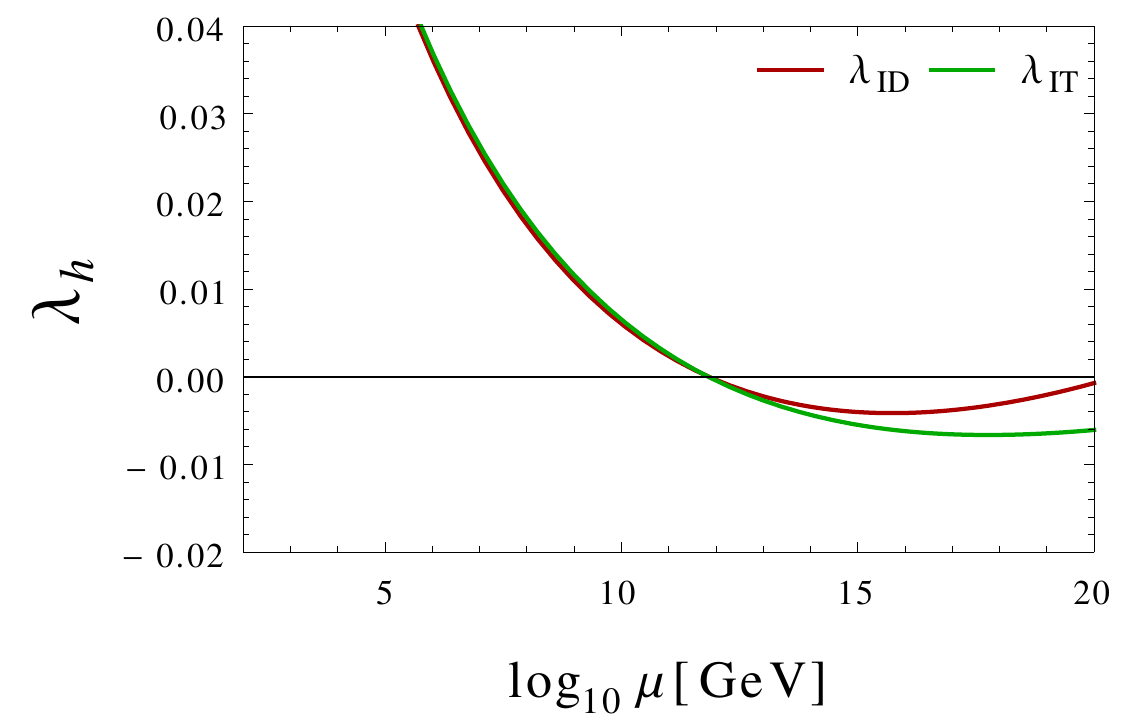}\label{f11}
		}}
		
		\mbox
		{\subfigure[]{\includegraphics[width=0.5\linewidth,angle=-0]{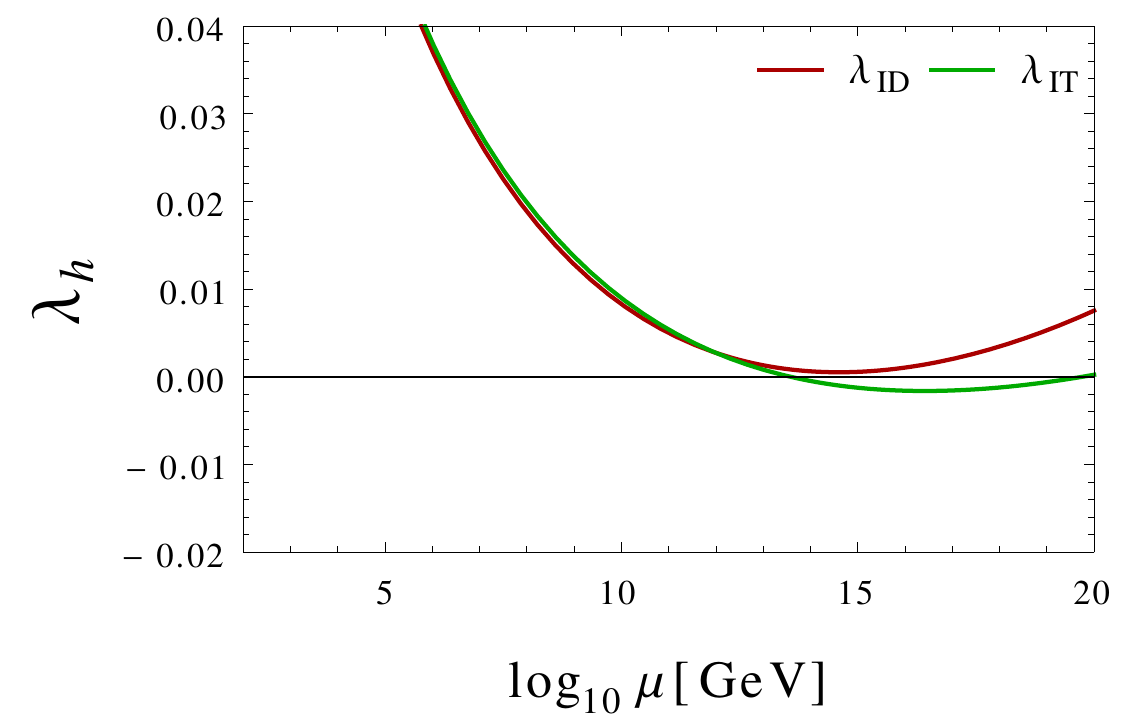}\label{f12}}	\subfigure[]{
				\includegraphics[width=0.5\linewidth,angle=-0]{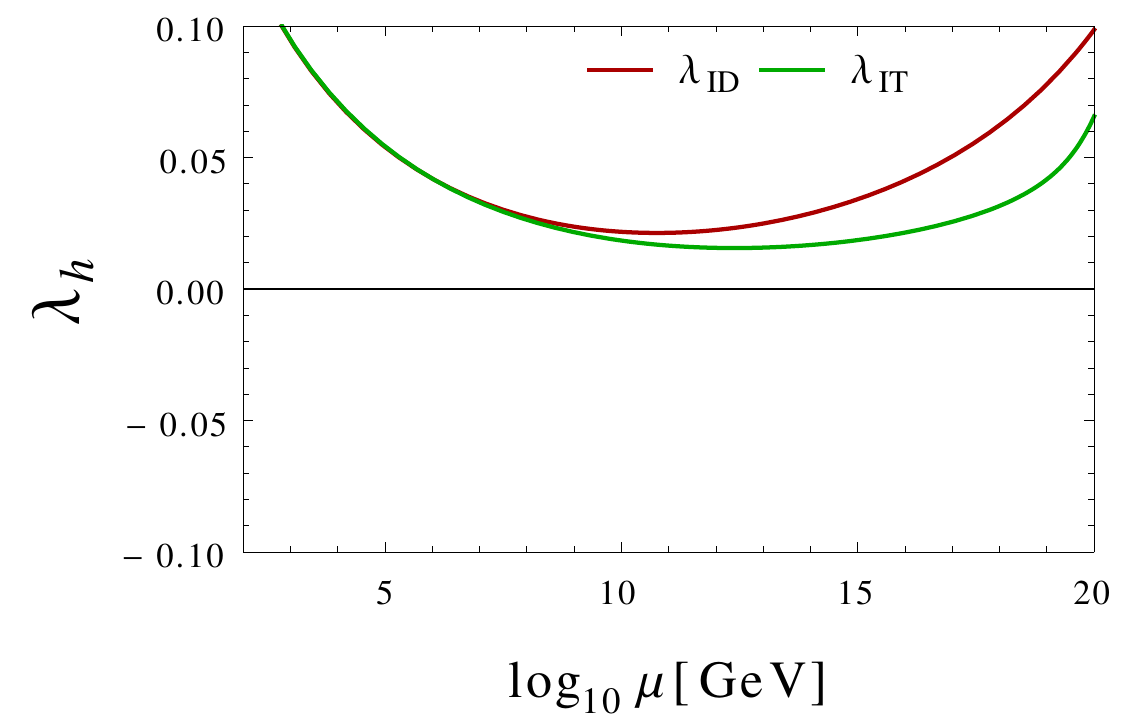}\label{f13}
		}}
		\caption{Two loop running of the SM like Higgs quartic coupling in IDM and ITM for four benchmark values of the Higgs quartic couplings $(\lambda_{1=2,3,4,5})$ in IDM and $(\lambda_{ht,t})$ in ITM to be  0.010, 0.060, 0.068 and  0.100 respectively. Here the red curve corresponds to the inert 2HDM and the green curve corresponds to the inert triplet Model.}\label{fig6l}
	\end{center}
\end{figure}

Figure~\ref{fig6l}  depicts the running of SM-like Higgs quartic coupling at two-loop level  for four benchmark points with $(\lambda_{2,3,4,5})$ for IDM and $(\lambda_{ht,t})$ for ITM to be  0.010, 0.060, 0.068 and  0.100 respectively. For both the cases $\lambda_1=0.1264$ is kept at two-loop level for the SM-like Higgs boson mass at $125.5$ GeV. Here the red curve corresponds to the IDM and the green curve corresponds to the ITM. For $\lambda_i \rm (EW)$=0.010, in Figure~\ref{f10}, the  effect of scalars on stability is less and both IDM and ITM becomes unstable at same scale $\sim 10^{9.7}$. In Figure~\ref{f11} for $\lambda_i (\rm{EW})=0.060$ we see that the $\lambda_h$ becomes negative around $10^{12}$ GeV but $\lambda_h$ turns upward at $10^{16}$ GeV  and touches zero value for $10^{20}$ GeV in the case of IDM while for ITM it still stays negative. As $\lambda_i (\rm{EW})$ enhances to 0.068 in Figure~\ref{f12}, the stability scale increases to $\sim 10^{13.5}$ in ITM while IDM becomes completely stable. Since, there are more number of scalars in IDM than ITM, the theory becomes stable at much lower values of $\lambda_i$. Further enhancement of $\lambda_i (EW)$ to 0.100, Figure~\ref{f13} makes both IDM and ITM stable till Planck scale.

\subsection{Vacuum Stability from RG-improved potential Approach }\label{vstability}
In this section, we investigate the vacuum stability via RG-improved effective potential approach by Coleman and Weinberg~\cite{Coleman:1973jx}, and calculate the effective potential at one-loop for IDM/ITM. The parameter space of the models are then scanned for the stability, metastability and instability of the potential by calculating the effective Higgs quartic coupling and implementing the constraints as discussed in the paragraph follows. 

Before going to quantum corrected potential  lets look at the stability conditions of the tree-level potential of IDM/ITM. The tree-level potential of IDM is given in Eq.~\eqref{Eq:1.1} and  the potential is bounded from below in all the directions is ensured by the tree-level stability conditions given by~\cite{Branco:2011iw}
\begin{align} \label{stabTHDM1}
\lambda_1 \ \geq \ 0 \, , \quad \lambda_2 \ \geq \ 0 \, , \quad 
\lambda_3 \ \geq \ -\sqrt{\lambda_1 \lambda_2} \, ,\quad \lambda_3+\lambda_4- |\lambda_5| \ \geq \ -\sqrt{\lambda_1 \lambda_2}  \, .
\end{align}
Similarly, the tree-level potential of ITM is given in Eq.~\eqref{Eq:2.4} and  the corresponding tree-level stability conditions are given by~\cite{Araki:2010zz}
\begin{align} \label{stabITM1}
\lambda_h \ \geq \ 0 \, , \quad \lambda_t \ \geq \ 0 \, , \quad 
 |\lambda_{ht}|  \geq -2\sqrt{\lambda_h \lambda_t}.
\end{align}
Considering the running of couplings with the energy scale in the SM, we know that the  Higgs quartic coupling $\lambda_h$ gets a negative contribution from top Yukawa coupling $y_t$, which makes it negative around $10^{9-10}$ GeV \cite{Buttazzo:2013uya,EliasMiro:2011aa} and we expect a second deeper minimum for the high field values. Since, the other minimum exists at much higher scale than the EW minimum, we can safely consider the effective potential in the $h$-direction to be 
\begin{align}
V_{\rm eff}(h,\mu) \ \simeq \ \lambda_{\rm eff}(h,\mu)\frac{h^4}{4},\quad {\rm with}~h\gg v \, ,
\label{eq:4.2}
\end{align}
where $\lambda_{\rm eff}(h,\mu)$ is the effective quartic coupling which can be calculated from the RG-improved potential. The stability of the vacuum can then be guaranteed at a given scale $\mu$ by demanding that $\lambda_{\rm eff}(h,\mu)\geq 0$. We follow the same strategy as in the SM in order to calculate $\lambda_{\rm eff}(h,\mu)$ in our model, as described below. 

The one-loop RG-improved effective potential in our model can be written as 
\begin{align}
V_{\rm eff} \ = \ V_0+V_1^{\rm SM}+V_{1}^{\rm IDM/ITM} \, ,
\label{eq:4.3}
\end{align}
where $V_0$ is the tree-level potential given by Eq.~\eqref{Eq:1.1} for IDM and Eq.~\eqref{Eq:2.4} for ITM.  $V_1^{\rm SM}$ is the effective Coleman-Weinberg potential of the SM that contains all the one-loop corrections involving the SM particles at zero temperature with vanishing momenta. $V_{1}^{\rm IDM}$ and $V_{1}^{\rm ITM}$ are the corresponding one-loop effective potential terms from the IDM and the ITM loops. In general, $V_1$ can be written as 
\begin{align}\label{qc}
V_1(h, \mu) \ = \ \frac{1}{64\pi^2}\sum_{i} (-1)^F n_iM_i^4(h) \Bigg[\log\frac{M_i^2(h)}{\mu^2}-c_i\Bigg],
\end{align}
where the sum runs over all the particles that couple to the $h$-field, $F=1\, \rm{and} \,0$ for fermions  and the bosons in the loop, $n_i$ is the number of degrees of freedom of each particle, $M_i^2$ are the tree-level field-dependent masses given by 
\begin{align}
M_i^2(h) \ = \ \kappa_i h^2-\kappa'_i,
\label{eq:4.5}
\end{align}
with the coefficients given in Table~\ref{table:1} and $m^2$ corresponds to Higgs mass parameter. Note that the massless particles do not contribute to Eq.~\eqref{eq:4.5}, and so to Eq.~\eqref{qc}. Therefore, for the SM fermions, we only include the dominant contribution from top quarks, and neglect the other quarks. We take $h=\mu$ for the numerical analysis as at that scale the potential remains scale invariant \cite{Casas:1994us}.

\begin{table}[h!]
	\begin{center}
		\begin{tabular}{||c|c|c|c|c|c|c||}\hline\hline
			Particles & $i$ & $F$ & $n_i$ & $c_i$ & $\kappa_i$ & $\kappa'_i$ \\ \hline\hline
			& $W^\pm$ & 0 & 6 & 5/6 & $g_2^2/4$ & 0\\
			& $Z$ & 0 & 3 & 5/6 & $(g_1^2+g_2^2)/4$ & 0\\
			SM & $t$ & 1 & 12 & 3/2 & $Y_t^2$ & 0\\
			& $h$ & 0 & 1 & 3/2 & $\lambda_h$ & $m^2$\\
			& $G^\pm$ & 0 & 2 & 3/2 & $\lambda_h$ & $m^2$\\
			& $G^0$ & 0 & 1 & 3/2 & $\lambda_h$ & $m^2$\\ \hline
			& $H^\pm$ & 0 & 2 & 3/2 & $\lambda_3/2$ & 0\\
			IDM & $H$ & 0 & 1 & 3/2 & $(\lambda_3+\lambda_4+2\lambda_5)/2$ & 0\\
			& $A$ & 0 & 1 & 3/2 & $(\lambda_3+\lambda_4-2\lambda_5)/2$ & 0\\ \hline
				ITM &$T^\pm$ & 0 & 2 & 3/2 & $\lambda_{ht}/2$ & 0\\
				& $T$ & 0 & 1 & 3/2 & $\lambda_{ht}/2$ & 0\\ \hline \hline
		\end{tabular}
	\end{center}
	\caption{Coefficients entering in the Coleman-Weinberg effective potential, cf.~Eq.~\eqref{qc}.}	\label{table:1}
\end{table}

Using Eq.~\eqref{qc} for the one-loop potentials, the full effective potential in Eq.~\eqref{eq:4.3} can be written in terms of an effective quartic coupling as in Eq.~\eqref{eq:4.2}. This effective coupling can be written as follows: 
	{\allowdisplaybreaks  \begin{align} \label{totalL}
		\lambda_{\rm eff}\left(h,\mu\right) & \ \simeq \  \underbrace{\lambda_h\left(\mu\right)}_{\text{tree-level}}+\underbrace{\frac{1}{16\pi^2}\sum_{\substack{i=W^\pm, Z, t, \\ h, G^\pm, G^0}} n_i\kappa_i^2 \Big[\log\frac{\kappa_i h^2}{\mu^2}-c_i\Big]}_{\text{Contribution from SM}}   
		+\underbrace{\frac{1}{16\pi^2}\sum_{i = H,A,H^\pm}n_i\kappa_i^2 \Big[\log\frac{\kappa_i h^2}{\mu^2}-c_i\Big]}_{\text{ Contribution from IDM/ITM }},
		\end{align}}
	where the corresponding coefficients for all the required fields are given in the Table~\ref{table:1}. The nature of $\lambda_{\rm eff}$ in the models thus guides us to identify the possible instability and metastability regions, as discussed below.

\subsection{Stable, Metastable and Unstable Regions}
The parameter space where $\lambda_{\rm eff }>0$ is termed as the {\it stable} region, since the EW vacuum is the global minimum in this region. For $\lambda_{\rm eff}<0$, there exists a second minimum deeper than the EW vacuum. In this case, the EW vacuum could be either unstable or metastable, depending on the tunnelling probability from the EW vacuum to the true vacuum. The parameter space with $\lambda_{\rm eff}<0$, but with the tunnelling lifetime longer than the age of the universe is termed as the {\it metastable} region. The expression for the tunnelling probability to the deeper vacuum at zero temperature is given by 
\begin{equation}
\rm P \ = \rm \ T_0^4 {\mu}^4 e^{\frac{-8 {\pi}^2}{3 \lambda_{\rm eff}(\mu)}} \, ,
\label{eq:P}
\end{equation}
where $T_0$ is the age of the universe and $\mu$ denotes the scale where the probability is maximized, i.e. $\frac{\partial P}{\partial \mu}=0$. This gives us a relation between the $\lambda$ values at different scales: 
\begin{align}
\lambda_{\rm eff}(\mu) \ = \ \frac{\lambda_{\rm eff}(v)}{1-\frac{3}{2\pi^2}\log\left(\frac{v}{\mu}\right)\lambda_{\rm eff}(v)} \, ,
\label{eq:lamb}
\end{align}
where $v\simeq 246$ GeV is the EW VEV. Setting $P=1$, $T= 10^{10}$ years and $\mu=v$ in Eq.~\eqref{eq:P}, we find $\lambda_{\rm eff}(v)$ =0.0623. The condition $P< 1$, for a universe about  $T= 10^{10}$ years old is equivalent to the requirement that the tunnelling lifetime from the EW vacuum to the deeper one is larger than $T_0$ and we obtain the following condition for metastability  \cite{Isidori:2001bm}: 
\begin{align}\label{meta}
0 \ > \ \lambda_{\rm eff}(\mu) \ \gtrsim \ \frac{-0.065}{1-0.01 \log\left(\frac{v}{\mu}\right)}.
\end{align}
The remaining parameter space with $\lambda_{\rm eff}<0$, where the condition~\eqref{meta} is not satisfied is termed as the {\it unstable} region. As can be seen from Eq.~\eqref{totalL}, these regions depend on the energy scale $\mu$, as well as the model parameters, including the  gauge, scalar quartic and Yukawa couplings.

\begin{figure}[H] 
	\begin{center}
				\mbox{\subfigure[]
			{
				\includegraphics[width=0.5\linewidth,angle=-0]{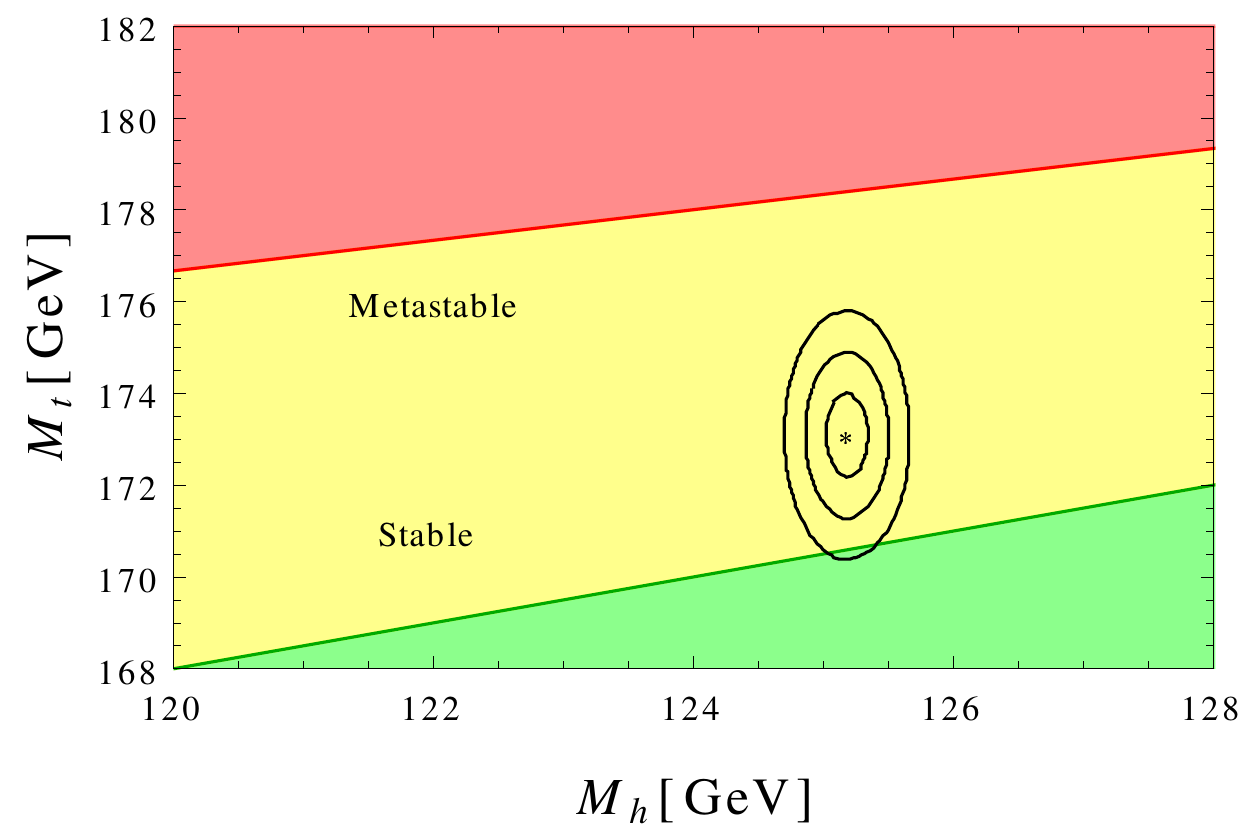}\label{f14}}
		}
		\mbox
		{\subfigure[]{\includegraphics[width=0.5\linewidth,angle=-0]{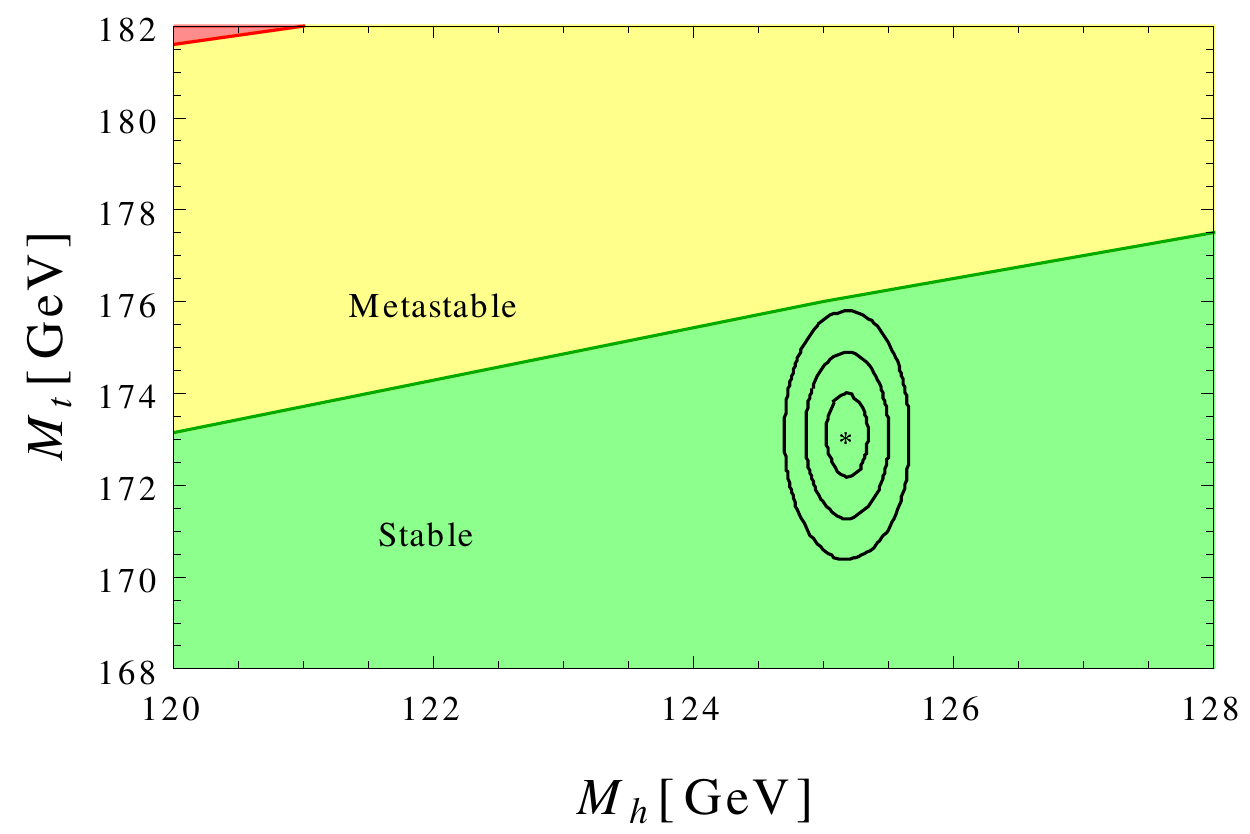}\label{f15}}	\subfigure[]{
				\includegraphics[width=0.5\linewidth,angle=-0]{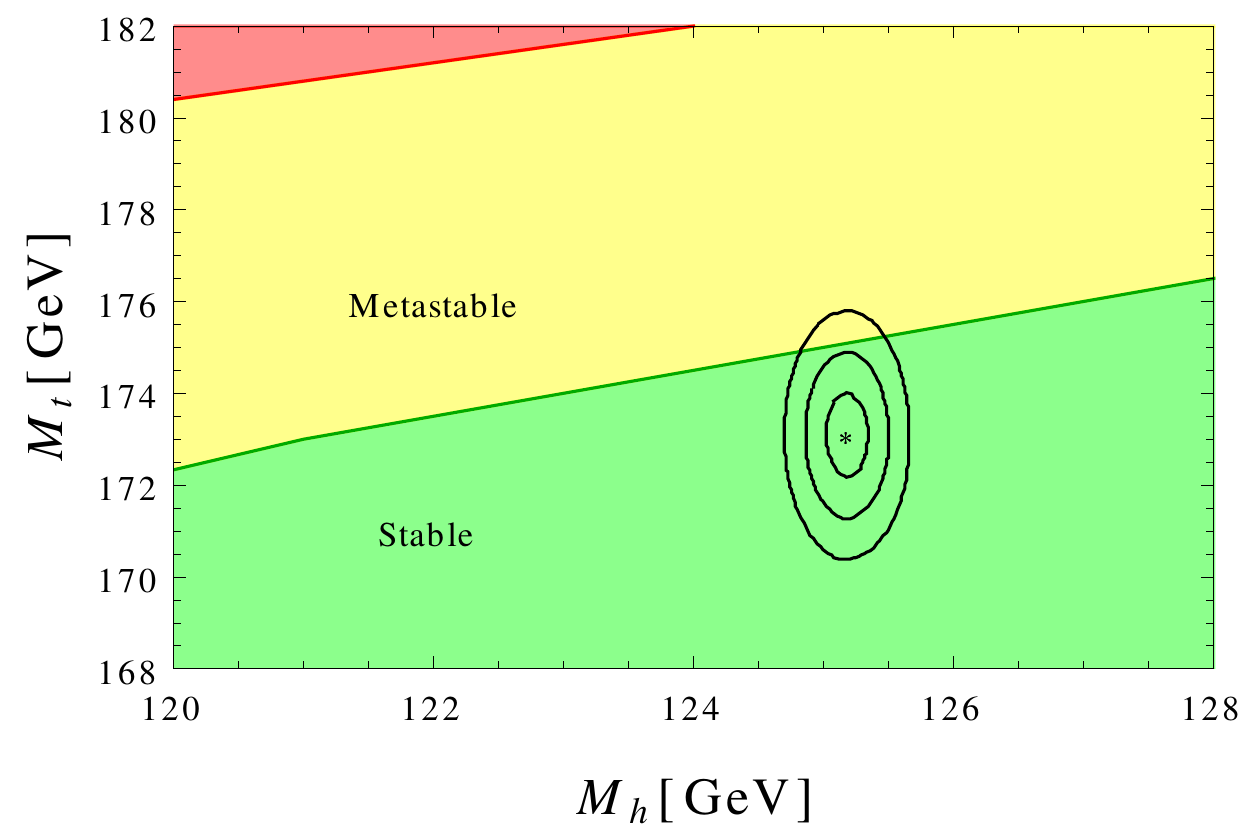}\label{f16}
			}}
			\caption{Phase diagram in terms of Higgs and top pole masses in GeV for \ref{f14}: SM like scenario, \ref{f15}: Inert Higgs Doublet Model and \ref{f16}: Inert Higgs Triplet Model. The red colour corresponds to the unstable region, yellow color corresponds to the metastable region and the green colour corresponds to the stable region. The contours and the dot show the current experimental $1\sigma,2\sigma,3\sigma$ regions and central value in the $(M_h,M_t)$ plane.
				}\label{fig7l}
		\end{center}
	\end{figure}

Figure~\ref{fig7l} represents the phase diagram in terms of Higgs and top pole masses in GeV. The red, yellow and green regions correspond to the unstable, metastable and stable regions respectively. The contours and the dot show the current experimental $1\sigma,2\sigma,3\sigma$ regions and central value in the $(M_h,M_t)$ plane \cite{Buttazzo:2013uya,Masina:2012tz}. 
To obtain the regions we vary all the $\lambda_i(\rm EW)=0.01-0.80$ for random values maintaining the Planck scale perturbativity  and also maintain the $m_h$ and $m_t$ within limits shown in Figure~\ref{fig7l}. Figure~\ref{f14} shows the scenario where $\lambda_{1}\neq 0$ and all other $\lambda_i =0$ and clearly the region is in metastbale state as expected for SM \cite{Buttazzo:2013uya}. Introduction of inert doublet adds more scalars to the effective potential so the $\lambda_{\rm{eff}}$ becomes more positive and the region is fully in the stable region as can be seen from Figure~\ref{f15}. In Figure~\ref{f16} we depicts the scenario for ITM, where such extra scalar degrees of freedoms are lesser than IDM but more than SM, so the $3\sigma$ contour in $m_h-m_t$ plane includes some region of metastability. In this context we also want to mention that the extra scalars are necessary and come as saviour for the models with right-handed neutrino with $\mathcal{O}(1)$ neutrino Yukawa coupling \cite{exwfermion}-\cite{Garg:2017iva}. 

\section{Calculation of Relic Density in freeze out scenario for IDM and ITM}\label{DMrelic}
\begin{figure}
		\hspace*{-1.0 cm}
		\includegraphics[width=0.8\linewidth,angle=-90]{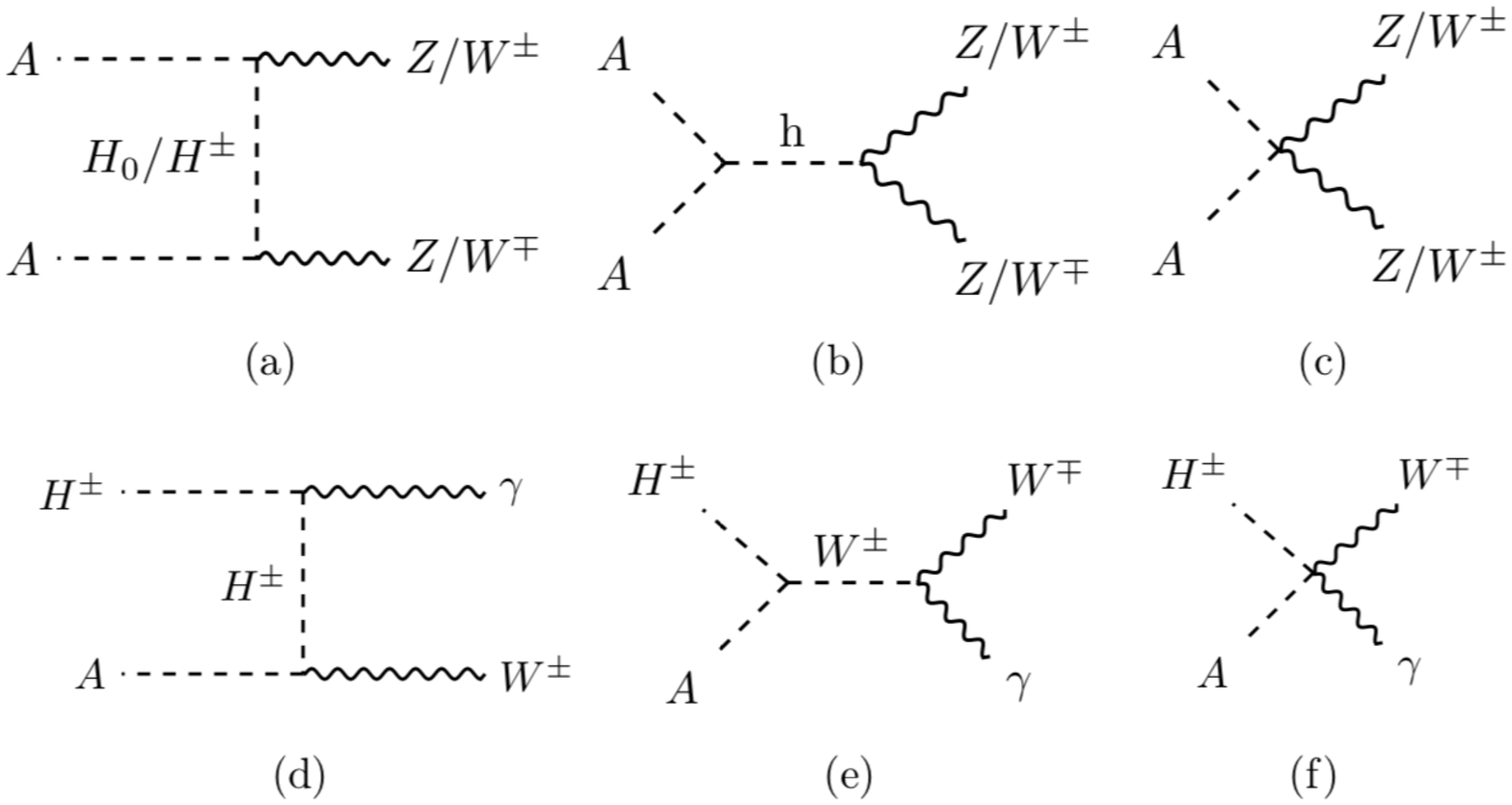}
		\vspace*{-2.7cm}
	\caption{Dominant annihilation and co-annihilation modes that contribute to DM relic for IDM.}\label{fig8l}
\end{figure}

\begin{figure}
\vspace*{-1.0cm}
\hspace*{-1.0 cm}
		\includegraphics[width=0.8\linewidth,angle=-90]{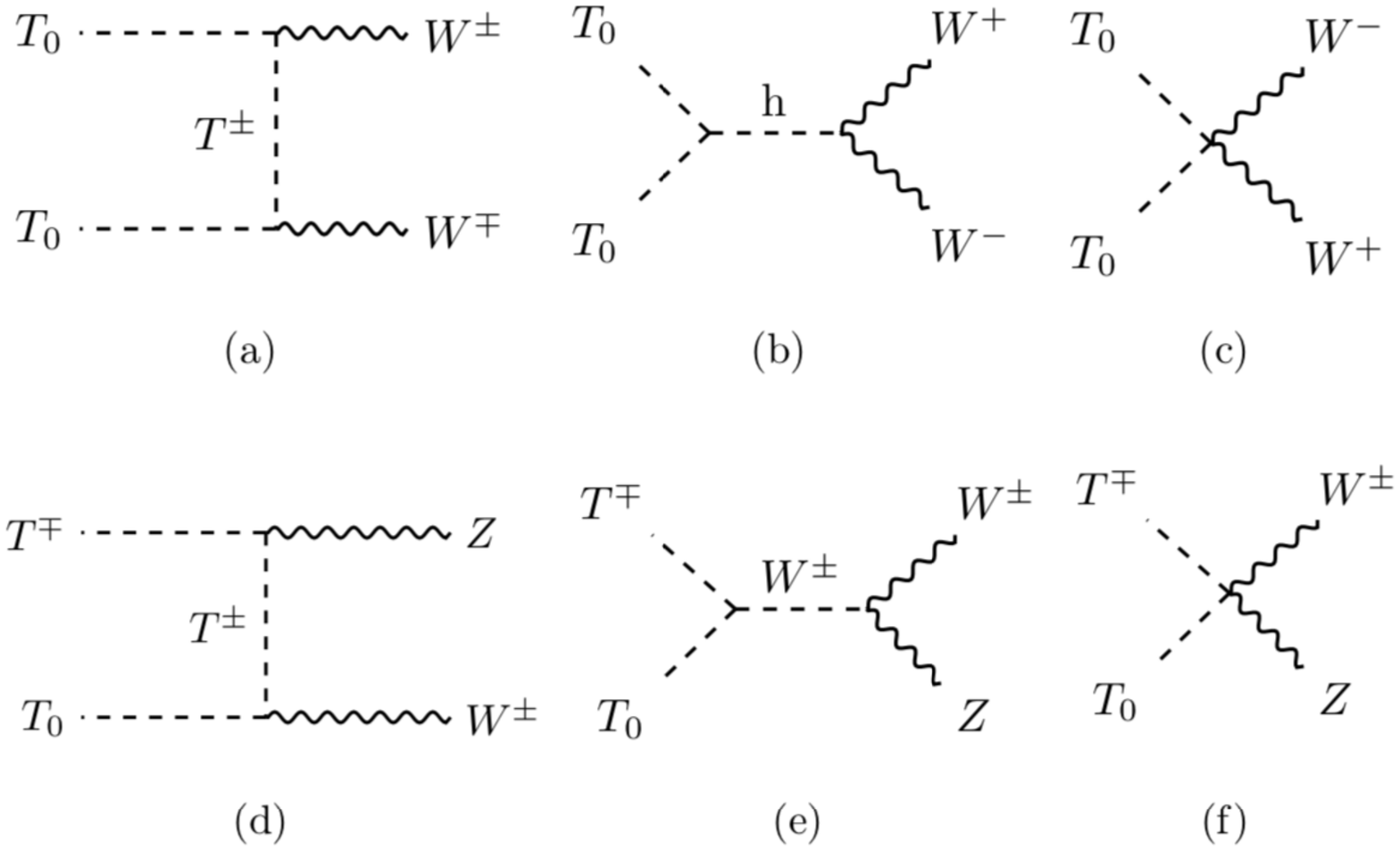}
			\vspace*{-2.7cm}
	\caption{Dominant annihilation and co-annihilation modes that contribute to DM relic for ITM.}\label{fig9l}
\end{figure}

After the theoretical constraints from perturbativity and vacuum stability we focus on the constraints coming from the dark sector. In case of IDM lightest of the $A$ and $H_0$ can be dark matter candidate being $Z_2$-odd. For our study we focus on the parameter space for which $A$ is the lightest and serves as the DM. However, for the ITM case we have only one $Z_2$-odd  neutral scalar, i.e. $T_0$ which serves as the DM. The different possible annihilation and co-annihilation diagrams are shown in Figure~\ref{fig8l} for IDM and in Figure~\ref{fig9l} for ITM respectively. Both of these DM candidates $A/T_0$ are charged under $SU(2)$ and thus the dominant mode of annihilation is $W^\pm W^\mp$. Being $Y=0$ triplet in the case of ITM there is no direct annihilation to $Z Z$ via contact or $t$-channel, which exist in the case of IDM. However, a $s$-channel annihilation via SM Higgs boson is possible. Apart from the annihilation channels,  $T^\pm T_0$ and $H^\pm A$ co-annihilation to $W^\pm Z/\gamma $ channels exist in both the scenarios which are secondary contributors. In both the cases DM annihilation channels to $hh$ is subdominant one and annihilation to fermion pair is negligible. 

The matrix element squared for the dominant DM annihilation and co-annihilation channels, i.e. to $W^\pm W^\mp, \,ZZ$ and $ZW^\pm $ are given in the Appendix~\ref{annihi}.  Once we  have the matrix element squared  we  calculate the $\big<\sigma v \big>$ in the non-relativistic limit following Eq.~\ref{relicv}

\be\label{relicv}
\big<\sigma v \big>=\frac{1}{k_f! 16\pi s }\frac{\sqrt{s- 4 m^2_f}}{m_i}{\overline{|\mathcal{M}|^2}}
\ee
where $s=E^2_{cm}$, $v$ stands for the relative velocity of the dark matter particles and $k_f$ is the symmetry factor  for the identical particles in the final states. $\overline{|\mathcal{M}|^2}$ is the spin averaged matrix element squared for annihilation and co-annhilation channels. For the numerical calculation we have taken all possible interference terms involved in the matrix element square calculation which are not shown in the Appendix~\ref{annihi}.

Equipped with the $\big< \sigma v \big>$ for different annihilation modes  we now examine the thermal relic abundance of DM particle. $\phi_{\rm{DM}}$ ($A/T_0$ for IDM/ITM) via Freeze-out mechanism \cite{Banerjee:2019luv,freezeout}. The evolution of the number density of DM is obtained by solving the Boltzmann equation \cite{Araki:2011hm}
\be
\frac{dn_{\phi_{\rm{DM}}}}{dt}+3 H n_{\phi_{\rm{DM}}} =-\big< \sigma v \big>(n^2_{\phi_{\rm{DM}}} - n^2_{\phi_{\rm{DM}},eq}),
\ee
where H is the Hubble parameter,  $n_{\phi_{\rm{DM}}}$, $n_{\phi_{\rm{DM}}, eq}$ and $\sigma$ are the number density of DM particle, the number density in thermal equilibrium and the total annihilation cross-section of $\phi_{\rm{DM}}$ respectively. All the particles in the $Z_2$-odd multiplets for both IDM/ITM will eventually contribute with $\big< \sigma v \big>$. Before the onset of freeze-out, the universe was hot and dense and as the universe expands, the temperature falls down. In this scenario the respective dark matter particles will not be able to find each other fast enough to maintain the equilibrium abundance. So when the equilibrium ends and the freeze-out starts, inert particles $T_0$ and $A$, can contribute in the relic density of DM through freeze-out mechanism \cite{freezeout}. Freeze-out  of $\phi_{DM}$ determines the DM relic abundance in  today's time which  gets constraints  from the WMAP and Planck experiments \cite{Planck} with the current value
\begin{eqnarray}
\rm \Omega_{DM} h^2=0.1199 \pm 0.0027, \label{Eq:6.1}
\end{eqnarray}
where $\rm h=0.67 \pm 0.012$ is the scaled current Hubble parameter in units of $\rm 100 km/s.Mpc$. Here, we use this value as upper bound on the contribution on dark matter production for the models IDM~\cite{Banerjee:2019luv} and ITM \cite{Tripletex}. 

Unlike IDM, the  mass splitting between dark matter $(T_0)$ and charged components $(T^{\pm})$ is much smaller for ITM, $\mathcal{O}(166)$ MeV. Thus the co-annihilation $T_0 T^{\pm} \rightarrow Z W^{\pm} $ contribution is larger as  compared to IDM.  Below we scan the parameter space for both IDM and ITM to find out the regions with correct DM relic as given in Eq.~\eqref{Eq:6.1}. 

\begin{figure}[H]
	\hspace*{-1.0cm}
	{\mbox{\subfigure[]{\includegraphics[width=0.5\linewidth,angle=-0]{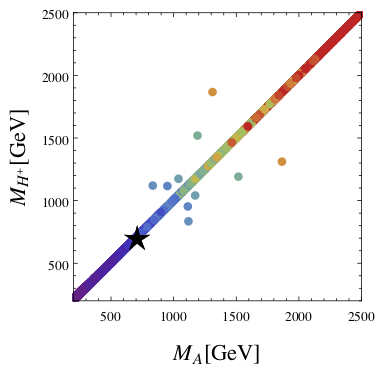}\label{f28}}
			\subfigure[]{\includegraphics[width=0.6\linewidth,angle=-0]{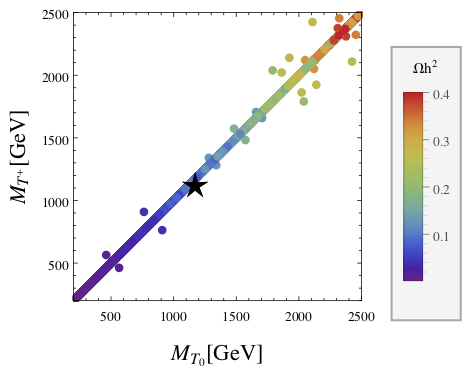}\label{f29}}}
		\caption{3D plot describing the variation of relic density with  dark matter mass and charged Higgs boson mass in GeV. \ref{f28}: IDM, \ref{f29}: ITM. The correct relic density corresponds to DM mass   $\gsim 700$ GeV in IDM and the mass splitting between DM mass and charged Higgs boson mass is order of $\mathcal{O}(1)$ GeV. In ITM scenario corresponds to DM mass of $\gsim 1176$ GeV with the mass gap being $\mathcal{O}(166)$ MeV at one-loop.}\label{fig10l}}
\end{figure}

For this scan we take the allowed parameter space from perturbativity and stability till Planck scale for the analysis of correct DM relic density by {\tt Micromegas 5.0.8}~\cite{Belanger:2008sj,Belanger:2006is,Belanger:2010pz}.  There we have taken contributions from all possible annhilation and co-annihilation channels and the interference effects therein.  Figure~\ref{fig10l} describes the variation of relic density with the masses of charged Higgs boson and DM ($A/T_0$ for IDM/ITM). The colour code of DM relic ($\Omega h^2$) is shown from blue to red for $0.0-0.4$ for both IDM and ITM respectively. The correct values of  $\Omega h^2 = 0.1199\pm 0.0027$ is specified by a star in both the cases. We can read from Figure~\ref{f28} that for IDM $M_{A} \gsim 700$ GeV corresponds to correct DM relic value. However, for ITM the correct relic value corresponds to $M_{T_0} \gsim 1176$ GeV as shown in Figure~\ref{f29}. The presence of one extra $Z_2$-odd scalar in IDM compared to ITM, results into higher the DM number density in IDM case and thus requires more annihilation or co-annihilation modes to obtain the correct relic compared to the  ITM case, leading to lower mass bound on DM mass for IDM.
Even for relatively heavier mass spectrum of  IDM corresponds mass gap of the order of $\mathcal{O}(1)$ GeV among the $Z_2$-odd particles. In comparison the  ITM scenario leads to even smaller mass gap  $\mathcal{O}(166)$ MeV coming from one-loop corrections, which leads to a dominant co-annihilation processes obtaining the correct relic as pointed out earlier. 

\section{Constrains from Direct Dark Matter experiments} \label{DMdirect}
In this section, we discuss the direct detection prospects of DM candidate for both IDM and ITM scenarios. Dark matter can be detected via elastic scattering with terrestrial detectors, the so-called direct detection method. From the particle physics point of view, the quantity that determines the direct detection rate is the dark matter-nucleon ($\rm{DM}-N$) scattering cross-section. In the IDM, the $\rm{DM}-N$ scattering process relevant for direct detection is Higgs-mediated. The tree-level spin-independent DM-nucleon interaction cross section, in IDM scenario \cite{Diaz:2015pyv,Garcia-Cely:2015ysa} is given by Eq.~\eqref{DD}
\be\label{DD}
 \rm \sigma_{SI}=\frac{\lambda_{345}^2 f_N^2}{4 \pi M_h^4}\frac{M_N^4}{(M_N+M_{A})^2},
 \ee
were $M_h$ is the mass of the SM-like Higgs boson, $\rm M_{A}$ is the mass of the DM candidate, $M_N$ is the nucleon mass that we took equal to the average of proton and neutron masses, $f_N$ is the nucleon form factor, taken equal to 0.3 for the subsequent analysis and $\lambda_{345}=\lambda_3 +\lambda_4 - 2\lambda_5$ with $\lambda_5 >0$, is the combined coupling  that is responsible for the scattering. we have used {\tt Micromegas 5.0.8}~\cite{Belanger:2008sj,Belanger:2006is,Belanger:2010pz} to calculate the direct spin-independent scattering cross-sections  and DM relic density for the parameter space and later compare with the experimental bounds from different direct detection experiments as discussed later.

In the case of ITM, the $T_0$ DM candidate can interact with nucleon by exchanging Higgs boson and the  DM-nucleon scattering cross section is given by \cite{Tripletex} Eq~\ref{DDt}
\be\label{DDt}
 \rm \sigma_{SI}=\frac{\lambda_{ht}^2 f_N^2}{4 \pi M_h^4}\frac{M_N^4}{(M_N+M_{T_0})^2},
  \ee
where the coupling constant $f_N$ is given by nuclear matrix elements and $\rm M_N=0.939$ GeV is nucleon mass which is average of the proton and neutron masses, $M_h$ is the SM-like Higgs boson mass, $M_{T_0}$ is the dark matter mass and $\lambda_{ht}$ is only responsible Higgs coupling here. 

 There are several experiments to detect DM particles directly through the elastic DM-nucleon scattering. The strong bounds on the DM-nucleon cross section are obtained from XENON100 \cite{Aprile:2012nq}, LUX \cite{Akerib:2013tjd} and XENON1T \cite{Aprile:2015uzo} experiments. The minimum upper limits on the spin independent cross sections are: 
  \bea
 \rm XENON100: \rm \sigma_{SI} \leq 2.0\times 10^{-45}\,cm^2\\\
\rm LUX: \rm \sigma_{SI} \leq 7.6 \times 10^{-46} \,cm^2\,\\
\rm XENON1T:\rm  \sigma_{SI} \leq 1.6\times 10^{-47} \,cm^2\,\\
\rm XENONnT: \rm \sigma_{SI} \leq 1.6 \times 10^{-48}\, cm^2.
 \eea

\begin{figure}
	{\mbox{\subfigure[]{\includegraphics[width=0.48\linewidth,angle=-0]{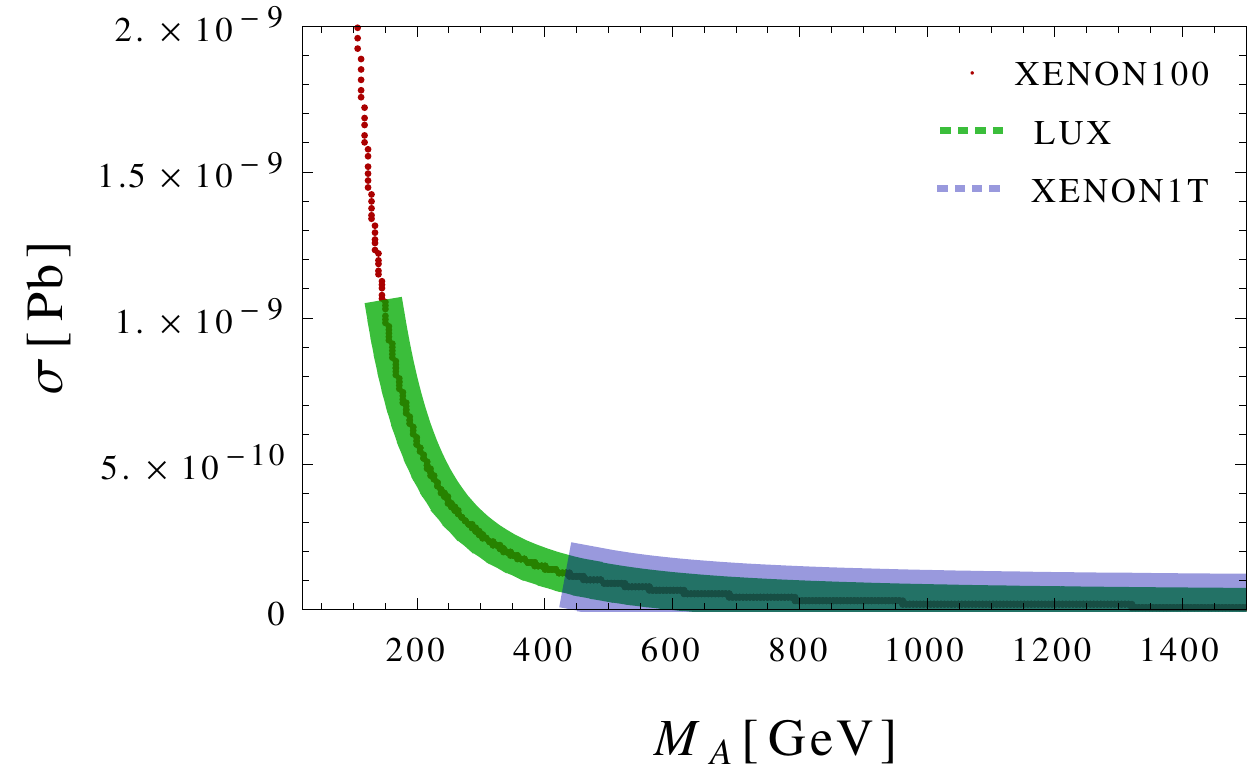}\label{f37}}
			\subfigure[]{\includegraphics[width=0.50\linewidth,angle=-0]{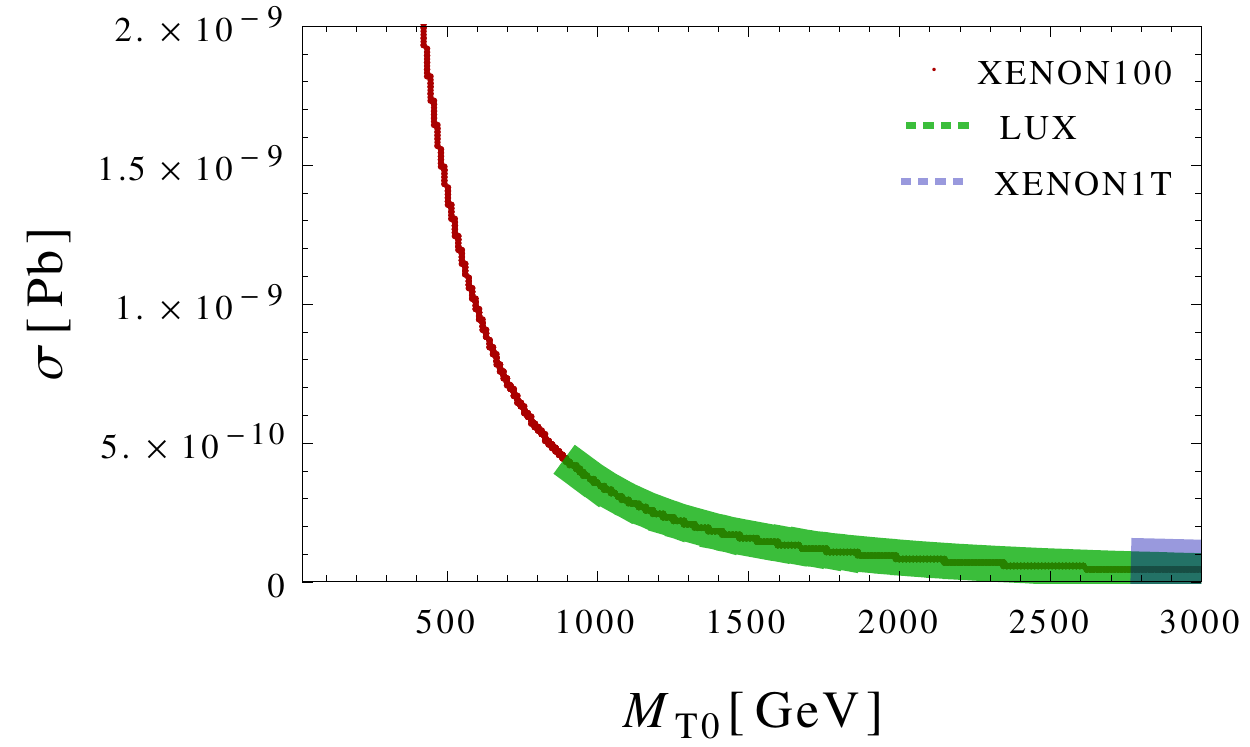}\label{f38}}}
		\caption{SI cross-section verses dark matter mass in GeV. Here we have shown XENON100, LUX and XENON1T data in red, green and blue regions respectively. Widths in green and blue regions are to make them transparent such that other bounds are visible.}\label{fig11l}}
\end{figure}

Figure~\ref{fig11l} describes the variation of spin independent $\rm(SI)$ DM-nucleon scattering cross-section with DM mass for both IDM and ITM. The red colour corresponds to the cross-section bound satisfied by XENON100 experiment \cite{Aprile:2012nq}, green colour satisfies the LUX experimental bound \cite{Akerib:2013tjd} and the blue colour corresponds to the experimental bound of XENON1T experiment \cite{Aprile:2015uzo} for both IDM and ITM. The cross-section varies with the DM mass and the Higgs quartic coupling $\lambda_{345}$ for IDM and $\lambda_{ht}$ for ITM. If the Higgs quartic coupling is chosen to be small enough $\lambda_{345}=0.01$ for IDM, the minimum DM mass satisfying the XENON1T bound is $420$ GeV \ref{f37}. Unfortunately this value of quartic coupling in ITM i.e. $\lambda_{ht}$=0.01 is not allowed by the vacuum stability. The enhancement in Higgs quartic coupling $\lambda_{345/ht}=0.2$ increases the lower bound of DM mass to $2770$ GeV  by XENON1T data\ref{f38}.

\begin{figure}[H]
	\begin{center}
		\includegraphics[width=0.8\linewidth,angle=-0]{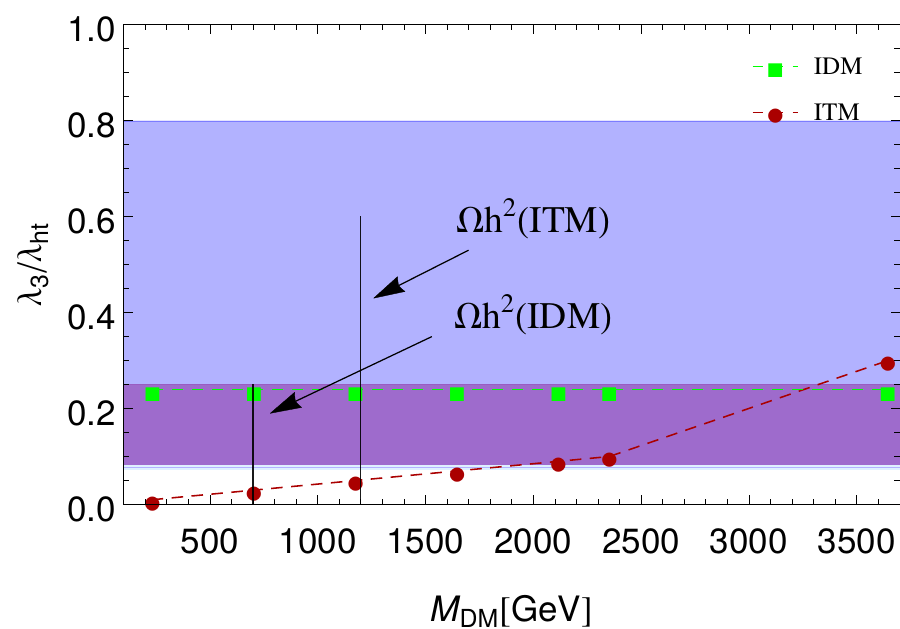}
		\caption{Variation of DM mass with Higgs quartic coupling $\lambda_3$ in IDM and ITM.  The light purple and blue color describe the allowed region by stability and perturbativity for  IDM and ITM respectively. The black vertical lines correspond to the relic density bound satisfied by DM mass $700, 1200$ GeV for IDM,  ITM respectively. The green and red coloured points describe the minimum value of $M_{\rm{DM}}$ for a given $\lambda_{345}/\lambda_{ht}$ for IDM and ITM respectively that satisfy the direct Dark matter constraint of XENON1T.}\label{fig12l}
	\end{center}
\end{figure}

The variation of DM mass with Higgs quartic coupling $\lambda_3$ in IDM and $\lambda_{ht}$ in ITM is depicted in Figure~\ref{fig12l}. The light purple and blue colour describe the allowed regions by stability and perturbativity till Planck scale for IDM and ITM respectively. The black vertical lines correspond to the relic density bound satisfied by DM mass $700, 1200$ GeV for IDM,  ITM respectively. The green and red colour points describe the minimum values of $M_{\rm{DM}}$ for a given $\lambda_{345}/\lambda_{ht}$ for IDM and ITM respectively that satisfy the direct Dark matter constraint of XENON1T \cite{Aprile:2015uzo}. In IDM the effective quartic coupling $\lambda_{345}$ allows to choose maximum allowed value of $\lambda_3$ satisfying the direct DM constraints, while in the case of ITM the minimum value of $M_{\rm{DM}}$ increases with increase in $\lambda_{ht}$.
\begin{figure}[H]
	\begin{center}
		\includegraphics[width=0.8\linewidth,angle=-0]{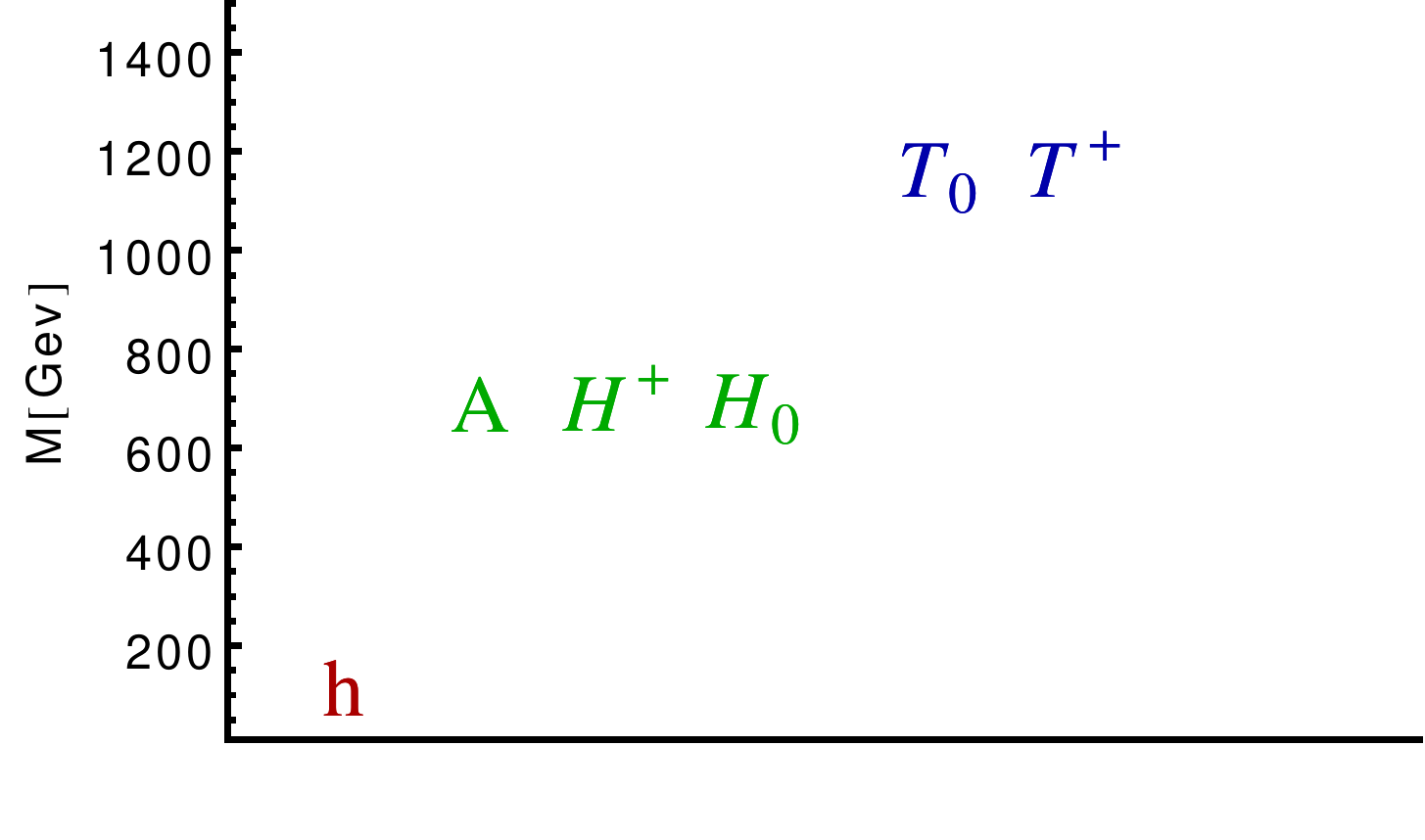}
		\caption{Physical mass eigenstates of the Higgs bosons in IDM and ITM corresponding to 
			lightest possible DM mass satisfying the correct DM relic. $h$ is the  $125.5$ GeV SM-like Higgs boson. Green colour corresponds to physical mass eigenstates in IDM. where $A$ is DM with $M_A=700.18$ GeV and  $M_H^{\pm}=702.364$ GeV, $M_{H_0}=708.314$ GeV respectively.  For ITM the lightest possible DM mass is $M_{T_{0}}=1176.00$ GeV along with almost degenrate charged Higgs mass $M_{T^{\pm}}=1176.16$ GeV represented in blue colour.}\label{fig13l}
	\end{center}
\end{figure}
Figure~\ref{fig13l} describes the mass spectrum for both IDM and ITM allowed by perturbativity and  vacuum stability till Planck scale, DM relic density and DM-nucleon scattering cross-section. The lightest allowed values for IDM  in the case are: $M_A=700.18$ GeV, $M_H^{\pm}=702.36$ GeV, $M_{H_0}=708.31$ GeV. The same reveals the lightest values for ITM are $M_{T0}=1176.00$ GeV and $M_{T^{\pm}}=1176.16$ GeV where as the SM-like Higgs stays with mass 125.5 GeV for both the cases. One more number of $Z_2$-odd field in IDM as compared to ITM which contributes to the number density of the dark matter. Thus IDM requires more annihilation cross-sections than ITM in getting the correct DM relic, which results in lower DM mass ($\sim 700$) GeV for IDM as compared to $\sim 1.2$ TeV for ITM.
\section{Constraints from H.E.S.S. and Fermi-Lat experiemtns}\label{HESS}
Indirect detection of dark matter is an interesting way to probe particle dark matter models. Among the few targets are Galactic centre and Dwarf Spheroidal Galaxies (dSphs), where dark matter annihilate or semi-annihilate into electron, positron, neutrinos, etc., yield gamma rays of different energies which are then observed by various telescopes. If the gamma-ray observation from different galactic sources are used as standard candles then any excess on the measured gamma-ray spectra can be used to probe the dark matter annihilation or co-annihilation channels.

The expected gamma-ray flux coming from the dark  matter annihilation for $\rm DM \, DM \to SM  \, SM$ can be written as

\be
\frac{d\phi_{\gamma}}{dE}=\frac{1}{8\pi m^2_{DM}}<\sigma v> \frac{dN_{\gamma}}{dE} J,
\ee
where $m_{DM}$ is the DM mass; which is $m_A(m_{T_0})$ for IDM (ITM), $<\sigma v> $ is the annihilation cross-section, $\frac{dN_{\gamma}}{dE}$ is the gamma-ray spectrum  and $J$  is the $J$-factor which takes into account all the astrophysical processes and is given by,

\be
J= \int_{r.o.i} d\Omega \int_{l.o.s}  \rho^2_{\rm DM}  dl,
\ee
where $\rho^2_{\rm DM}$ is the DM density  over the region of interest (r.o.i) and
the line of sight (l.o.s). In general  $J_i$ from different  dSphs have uncertainties and a combined analysis of 15 dSphs have been used \cite{FermiLAT}. Now for different choices of the final state annihilation channel, dark matter mass we can compute the gamma-ray  spectrum and compare with the experimental data to put bounds on those annihilation modes. Here for the datasets we compare  with two following experimental data sets to put bounds on $<\sigma v>$: 

\begin{itemize}
\item Fermi-LAT gamma-ray observations in the direction of dwarf spheroidal galaxies \cite{FermiLAT}; 
\item  H.E.S.S. gamma-ray observations in the direction of the Galactic Center \cite{HESS}.  
\end{itemize}
The Fermi-LAT satellite has measured over the years gamma-ray covering an energy range of 500 MeV to 500 GeV  and no excess has been reported in the direction of dSphs \cite{FermiLAT}. Thus stringent limits were imposed on the dark matter annihilation cross-sections for the standard annihilation channels.  On the other hand the High Energy Stereoscopic System (H.E.S.S.) gave us a new look at high energy gamma-rays from the Galactic Centre with current sensitivity of DM mass of 100 TeV \cite{HESS}. 
\begin{figure}[H]
\hspace*{-1cm}
		{\mbox{\subfigure[]{\includegraphics[width=0.55\linewidth,angle=-0]{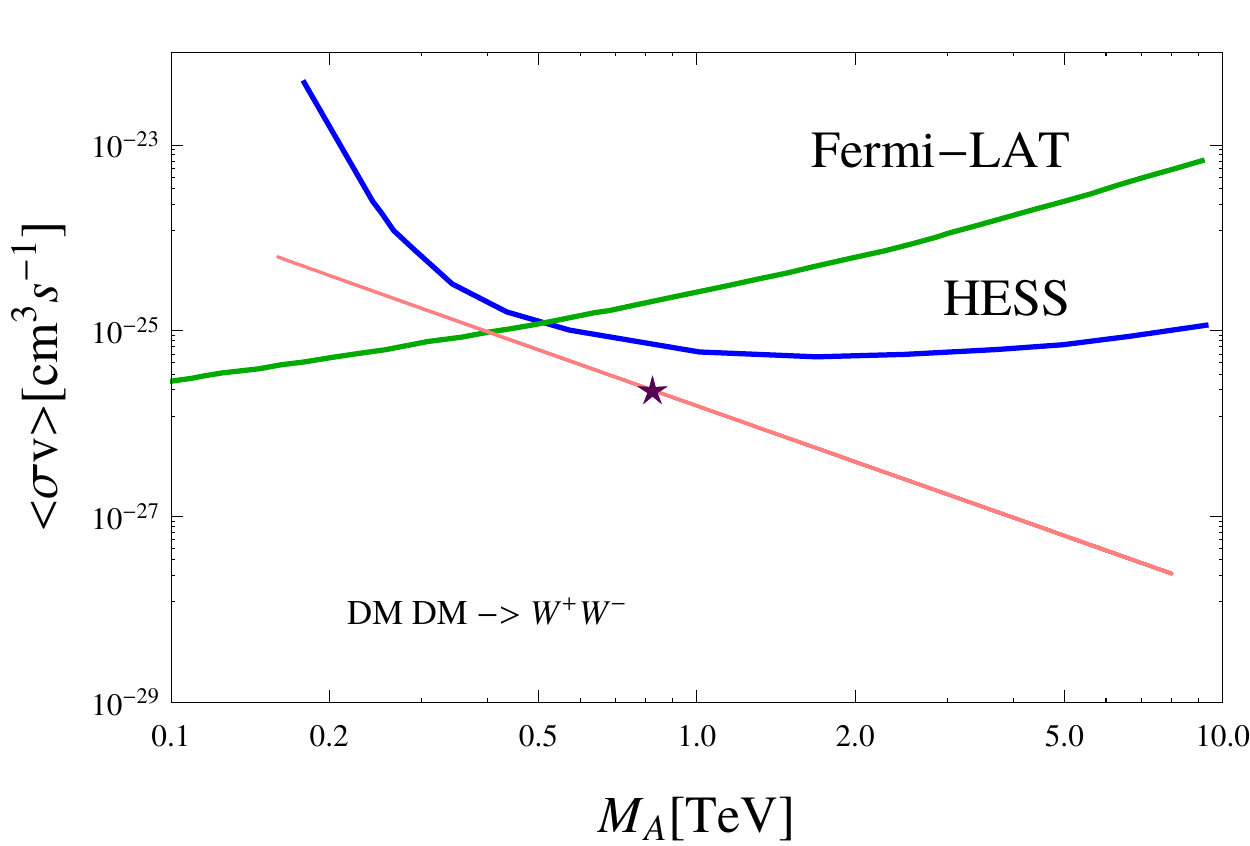}\label{f57}}
				\subfigure[]{\includegraphics[width=0.55\linewidth,angle=-0]{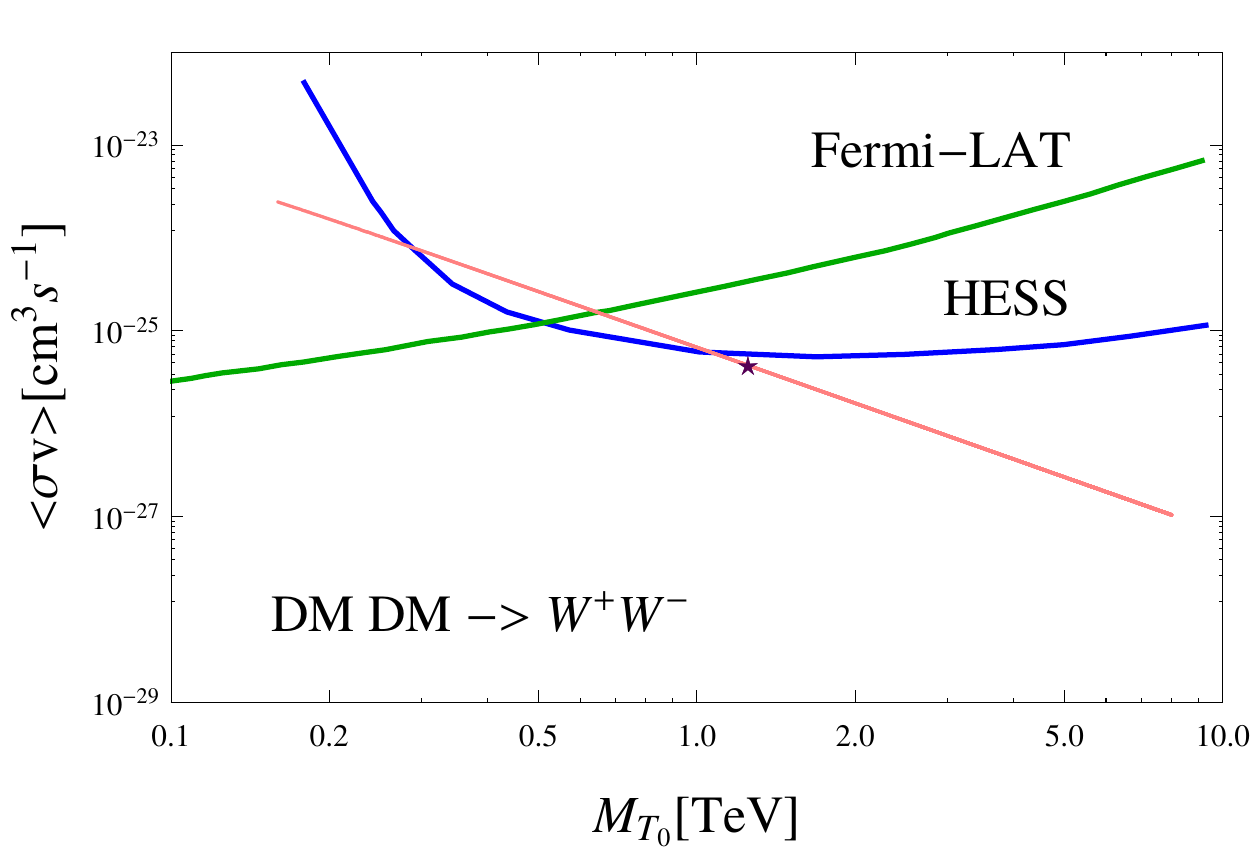}\label{f58}}}}
		\caption{$<\sigma v>$ in $W^\pm W^\mp$ mode  verses the DM mass as shown by pink lines in \ref{f57} for IDM and in \ref{f58} for ITM respectively. The Blue line corresponds to the H.E.S.S bounds \cite{HESS} and the green line corresponds to Fermi-LAT bounds \cite{FermiLAT} in  $W^\pm W^\mp$ mode. The start ($\star$) the points are chosen benchmark points as discussed in Table~\ref{decay}. }\label{FermiHESS}
\end{figure}

Since both the cases (IDM and ITM) the dark matter annihilate to $W^\pm W^\mp$ directly, the bounds on $<\sigma v>$ in $W^\pm W^\mp$ mode from H.E.S.S \cite{HESS} and Fermi-LAT \cite{FermiLAT} would be very evident. We impose such bounds on our parameter space as shown in Figure~\ref{FermiHESS} describes $<\sigma v>$ in $W^\pm W^\mp$ mode verses the DM mass by pink lines: Figure~\ref{f57} for IDM and Figure~\ref{f58} for ITM  respectively. The Blue line corresponds to the H.E.S.S bounds \cite{HESS} and the green line corresponds to Fermi-LAT bounds \cite{FermiLAT} in  $W^\pm W^\mp$ mode.  As expected due to triplet coupling to $W^\pm$ is larger (See Eq.~\ref{Tripeq}) in comparison with the doublets, the cross-section in $W^\pm W^\mp$ mode  is larger for a given mass. The start ($\star$) points are the chosen benchmark points as discussed in Table~\ref{decay} are allowed by both H.E.S.S \cite{HESS} and Fermi-LAT \cite{FermiLAT} data in $W^\pm W^\mp$ mode. In the context of IDM other indirect bounds are discussed in the literature \cite{hessIDM}.

\section{Dependence on the validity scale}\label{scale}
In this section we discuss how the parameter space depends on the validity scale of perturbativity and vacuum stability along with the relic and direct DM constraints. While implementing that we consider three different scales; namely the Planck scale ($10^{19}$ GeV), the GUT scale ($10^{15}$ GeV) and the $10^{4}$ GeV scale as the upper limit of the theory. It would be interesting to see how two different DM models differ in such different requirements. 
\subsection{Validity till Planck scale}
\begin{figure}[H]
	\hspace*{+0.5cm}
	{\mbox{\subfigure[]{\includegraphics[width=0.50\linewidth,angle=-0]{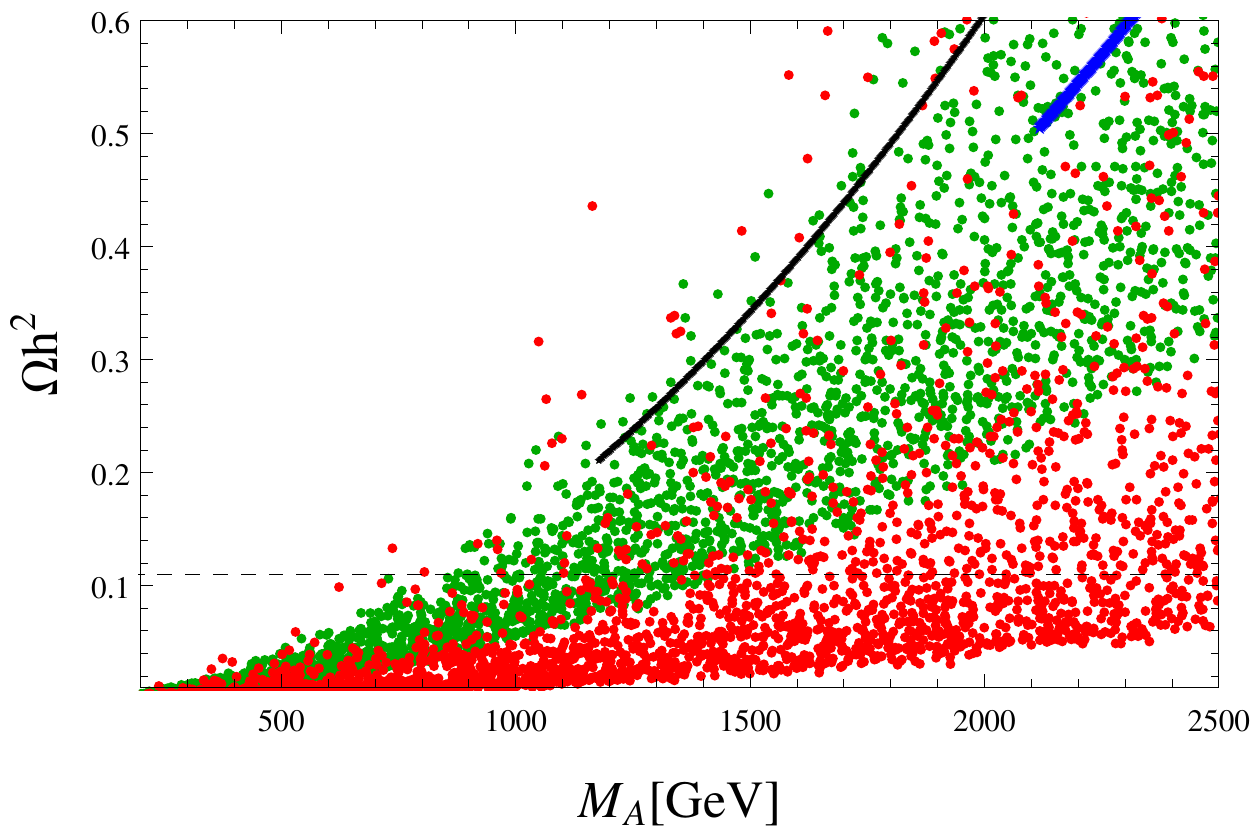}\label{f30}}
			\subfigure[]{\includegraphics[width=0.50\linewidth,angle=-0]{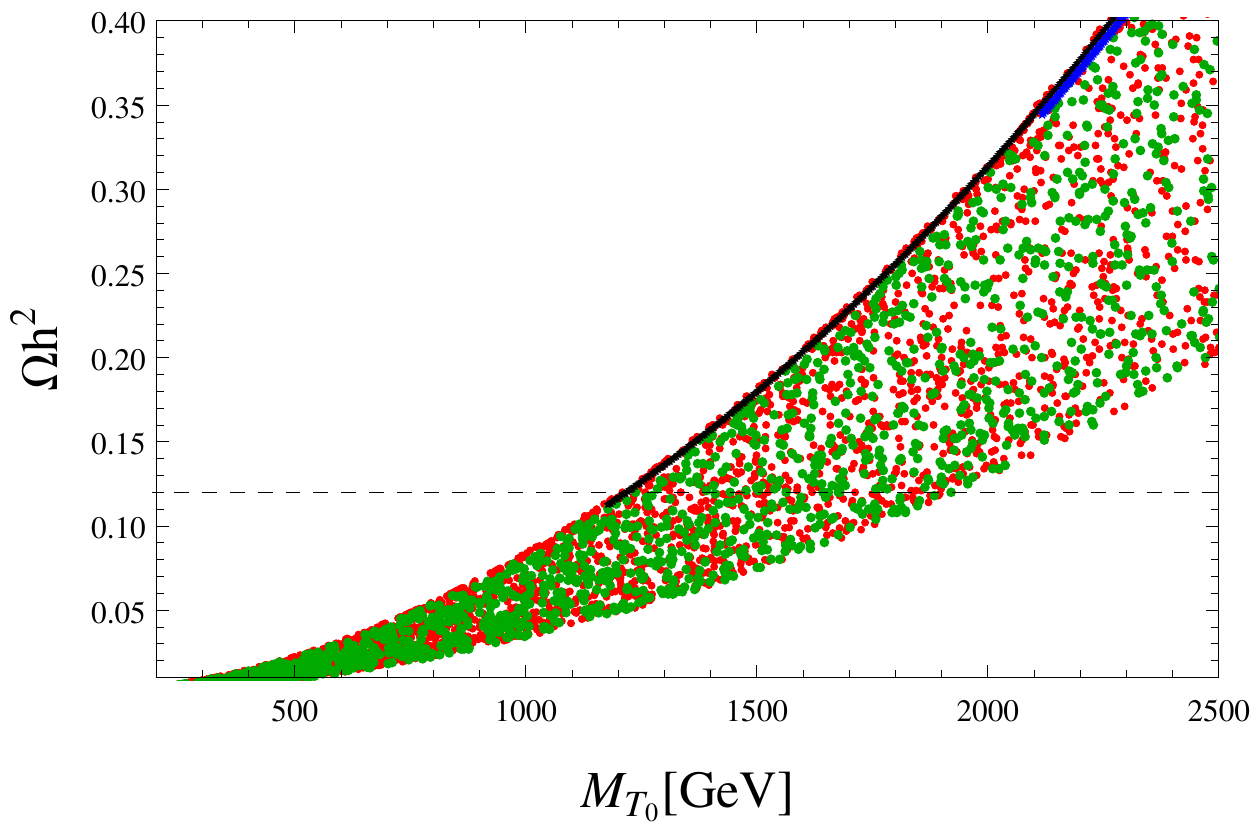}\label{f31}}}
		\caption{Relic density vs dark matter mass in GeV. \ref{f30}: Inert Higgs Doublet Model; \ref{f31}: Inert Higgs Triplet Model. Red color corresponds to the electroweak symmetry breaking allowed points, Green color  corresponds to the points which are allowed by both perturbativity and stability till Planck scale. The black  and the blue lines correspond to those points which are allowed by   direct detection cross-section bound of XENON1T for two different values of Higgs quartic coupling $\lambda_{345}=0.05, 0.09$ in IDM and $\lambda_{ht}=0.05,0.09$ in ITM.}\label{fig14l}}
\end{figure}

Here we consider that all the dimensionless couplings remain perturbative and the EW vacuum remains stable till Planck scale ($\mu \lesssim 10^{19}$ GeV). In Figure~\ref{fig14l} we present the parameter points in DM mass verses DM relic density for both IDM and ITM. The Red coloured points are allowed by the electroweak symmetry breaking. Among those points, the Green coloured points correspond to the points which are allowed by both perturbativity and stability till Planck scale ($\mu \lesssim 10^{19}$ GeV). The black and blue lines correspond to those points which are allowed by direct detection cross-section bound of XENON1T \cite{Aprile:2015uzo} for two different  benchmark scenarios chosen for IDM and ITM.  The benchmark points chosen for direct detection  are $\lambda_{345} =0.050$ $( \lambda_3=0.200, \lambda_4=0.100, \lambda_5=0.125)$ and $\lambda_{345}=0.09$ $(\lambda_3=0.200, \lambda_4=0.200, \lambda_5=0.155)$  for IDM as shown in Figure~\ref{f30} described by black and blue lines.  We see that the similar constraints for ITM are presented in  Figure~\ref{f31} for    $\lambda_{ht}=0.05$ and $\lambda_{ht}=0.09$ respectively. In the case of ITM, the  quartic coupling value $\lambda_{ht}=0.05$ is allowed by perturbativity till Planck scale but only to $\mu \lesssim 10^{9}$ GeV by vacuum stability, while $\lambda_{ht}=0.09$  is allowed by both till Planck scale. The dashed horizontal line defines the correct DM relic density as given in  Eq: \ref{Eq:6.1}.

\begin{figure}
	\hspace*{+0.5cm}
	{\mbox{\subfigure[]{\includegraphics[width=0.50\linewidth,angle=-0]{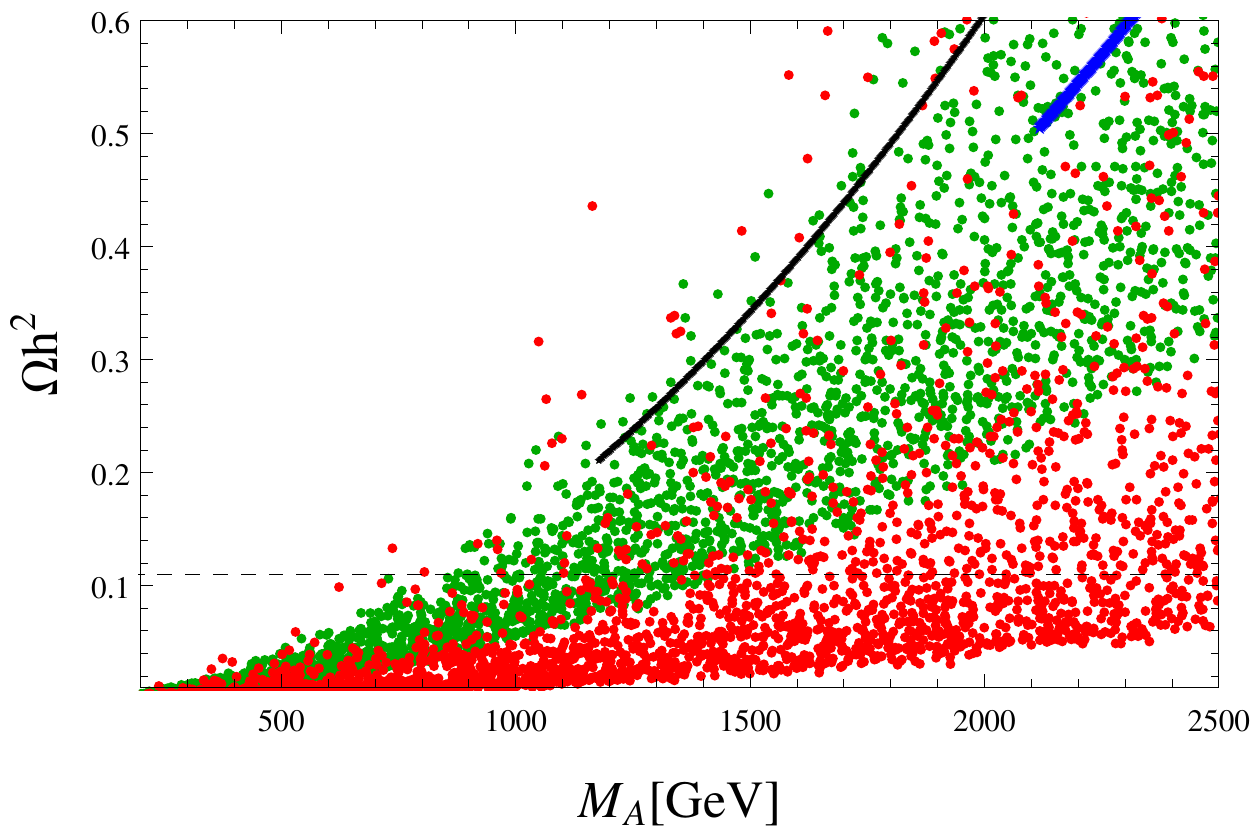}\label{f33}}
			\subfigure[]{\includegraphics[width=0.50\linewidth,angle=-0]{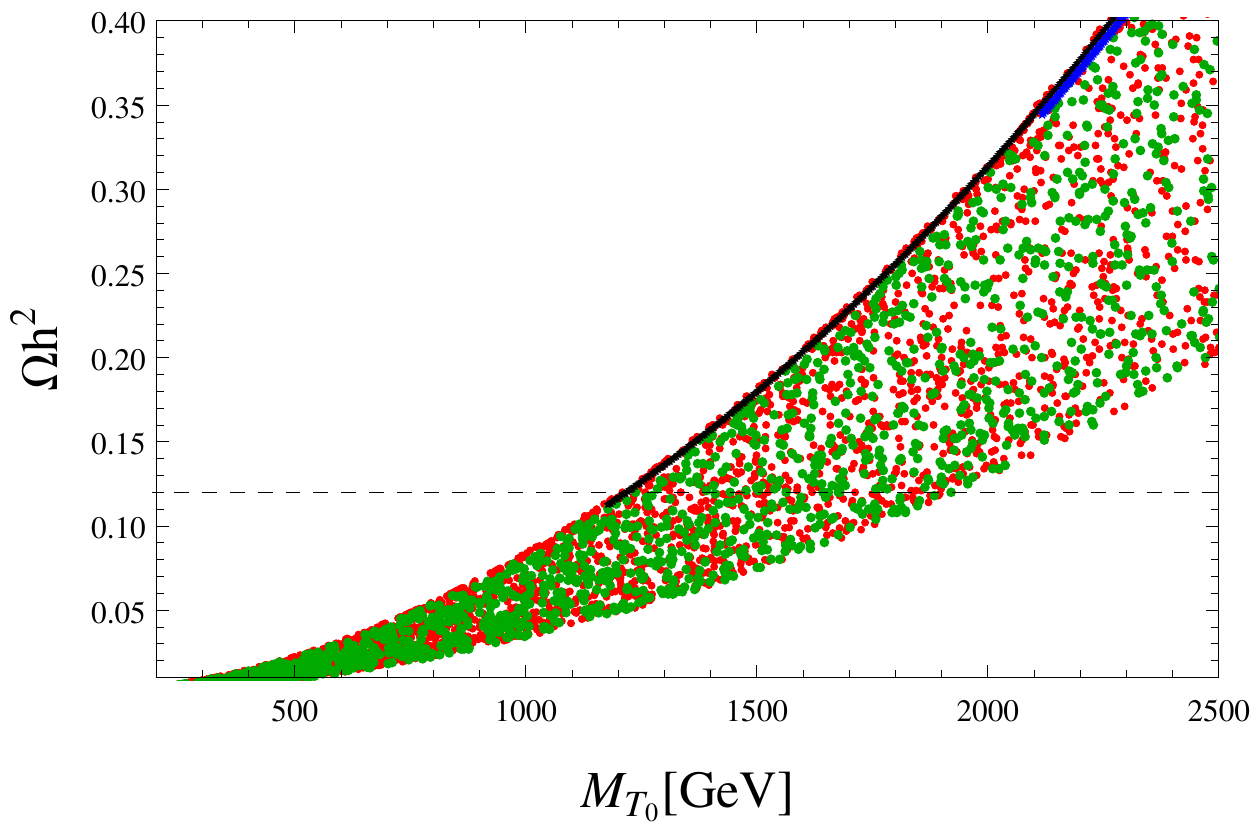}\label{f34}}}
		\caption{Relic density vs dark matter mass in GeV. \ref{f33}: Inert Higgs Doublet Model; \ref{f34}: Inert Higgs Triplet Model.Red color corresponds to the electroweak symmetry breaking allowed points, Green color  corresponds to the points which are allowed by both perturbativity and stability till Gut scale. The black  and blue lines correspond to those points which are allowed by   direct detection cross-section bound of XENON1T.}\label{fig15l}}
\end{figure}
\subsection{Validity till GUT scale}
Figure~\ref{fig15l} shows the DM mass verses relic density variation in IDM and ITM. Simialr to previous case here also green colour corresponds to the points which are allowed by both perturbativity and vacuum stability till GUT scale ($10^{15}$ GeV). For IDM and ITM, the allowed parameter space by both perturbativity and vacuum stability remain same as Planck scale. The black and blue lines again correspond to those points which are allowed by the direct detection cross-section bound of XENON1T \cite{Aprile:2015uzo}. The corresponding benchmark points are chosen $\lambda_{345}/\lambda_{ht}=0.05, 0.09$ for  IDM/ITM respectively as shown in  Figure~\ref{f33} and Figure~\ref{f34}. As discussed earlier for ITM, the EW vacuum is stable till $\mu \sim 10^{9}$ GeV for $\lambda_{ht}=0.05$.

x\subsection{Validity till $10^4$ GeV }
\begin{figure}[H]
	\hspace*{+0.5cm}
	{\mbox{\subfigure[]{\includegraphics[width=0.50\linewidth,angle=-0]{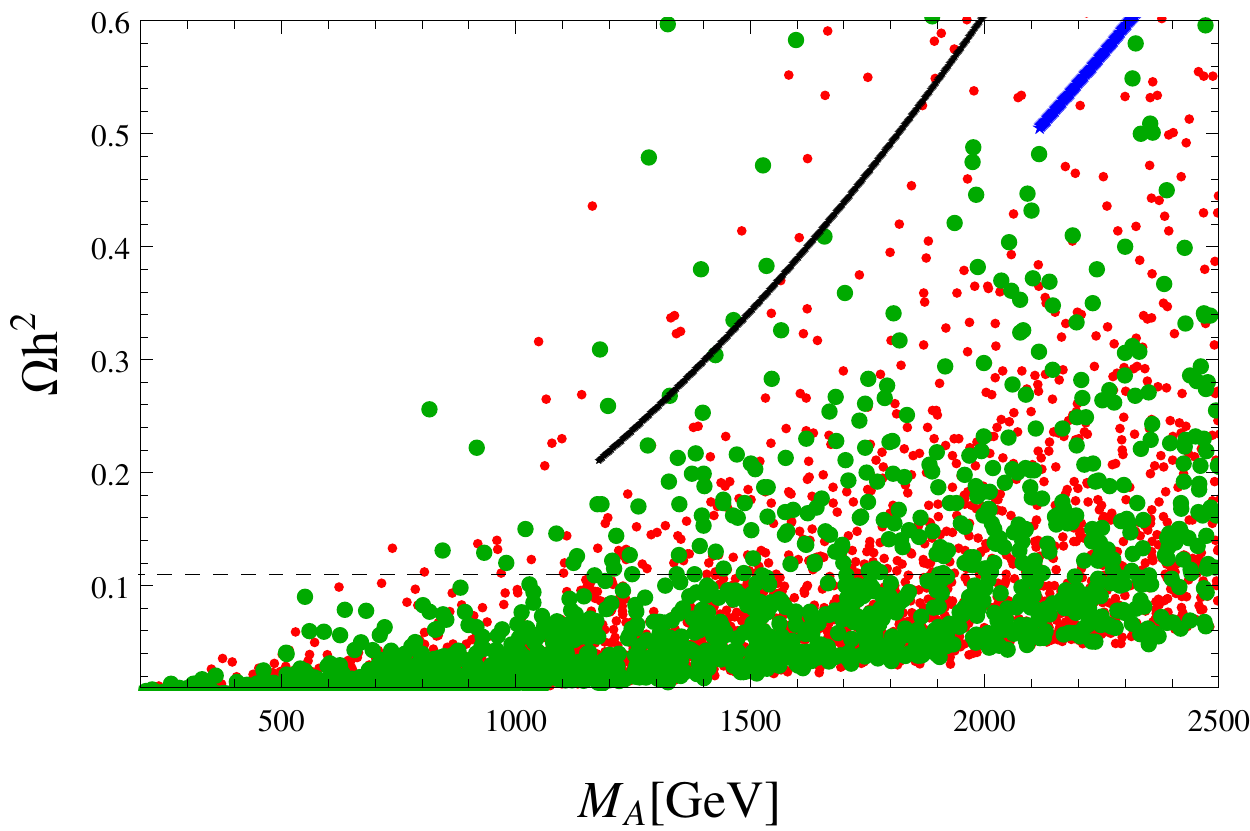}\label{f35}}
			\subfigure[]{\includegraphics[width=0.50\linewidth,angle=-0]{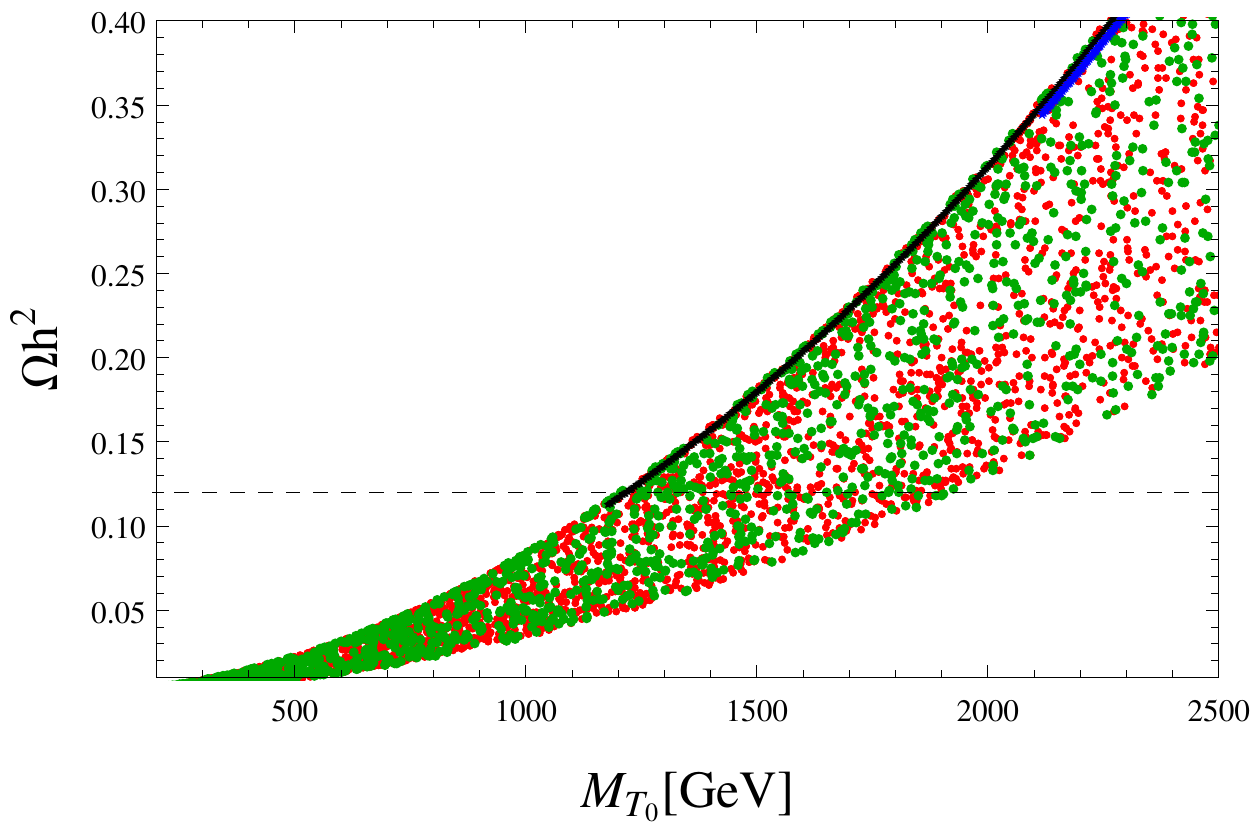}\label{f36}}}
		\caption{Relic density verses dark matter mass in GeV:  \ref{f35} Inert Higgs Doublet Model; \ref{f36} Inert Higgs Triplet Model. Red colour correspond to the electroweak symmetry breaking allowed points, Green colour  correspond to the points which are allowed by both perturbativity and stability till $10^{4}$ GeV scale. The black line corresponds to those points which are allowed by   direct detection cross-section bound of XENON1T.}\label{fig16l}}
\end{figure}

The above analysis is repeated for the benchmark points which are allowed by perturbativuty, vacuum stability, DM relic bound and direct detection cross-section bound till scale  $\mu \sim 10^{4}$ GeV as shown in Figure~\ref{fig16l}. In this scenario, green colour corresponds to points which are allowed by both perturbativity and vacuum stability till $10^{4}$ GeV scale. The allowed parameter space by vacuum stability and perturbativity increases for  both IDM and ITM as we see more green points as compared to Figure~\ref{fig14l} and Figure~\ref{fig15l}. The corresponding benchmark points are chosen $\lambda_{345}/\lambda_{ht}=0.05, 0.09$ for  IDM/ITM respectively as shown in  Figure~\ref{f35} and Figure~\ref{f36} and all the points are allowed by the perturbativity and vacuum stability constraints till $\mu \sim 10^{4}$ GeV.

\section{LHC Phenomenology}\label{pheno}
LHC is looking for the heavier states specially for the another Higgs bosons for both CP-even and CP-odd but so far no new resonances are found out and only cross-section bounds have been given by both CMS and ATLAS \cite{Aaboud:2017sjh,Sirunyan:2018zut}. In this article we consider the extension of SM with a inert $SU(2)$ doublet or inert $Y=0$ $SU(2)$ triplet. In both the cases the extra scalar gives rise to a lightest $Z_2$-odd particle which does not decay and  can contribute as missing energy in the collider~\cite{Kalinowski:2019cxe,Wan:2018eaz}.  

IDM has  one  pseudoscalar Higgs boson ($A$), one CP-even Higgs boson ($H_0$) and the charged Higgs boson $(H^{\pm})$ and all  are from the inert doublet $\Phi_2$, which is  $Z_2$ odd and their mass splittings are mostly $\lesssim M_W$ in allowed mass range, making a quasi-degenerate mass spectra.  Contary to IDM, ITM has only a CP-even real Higgs boson ($T_0$) and a charged Higgs boson ($T^\pm$). In this case their tree-level masses are identical unlike IDM case and only mass splitting of 166 MeV comes from  loop-corrections. 

In ITM the triplet does not take part in EWSB and so there is no mass mixing between the doublet and triplet which is very different from the supersymmetric triplet case \cite{TESSM,Bandyopadhyay:2014tha,TNSSM, Bandyopadhyay:2015tva} where such mixing occur from the superpotential. Moreover, $Y=0$ triplet nature does not allow it to couple to fermion in both SUSY and non-SUSY cases disparate from $Y=2$ triplet case of Type-II seesaw.  The normal $Y=0$ triplet which takes part in EWSB, breaks the custodial symmetry $(v_T \neq 0)$ which implies $g_{W^{\mp}-Z-H^{\pm}}\neq 0$ at tree-level. This  makes $\rho > 1$, which strongly  constrains  $v_T \lesssim 5$ GeV  \cite{PDG}. In case of ITM, we have $v_T=0$ as triplet stays in $Z_2$-odd, which certainly ceases the $g_{W^{\mp}-Z-H^{\pm}}$ coupling to exist.  Thus the charged Higgs boson decays to mono-lepton or di-jet plus $\slashed{E_{T}}$ via off-shell $W^{\pm}$ and DM unlike tri-lepton plus missing energy in case of triplets that gets vev and breaks custodial symmetry at tree-level \cite{Bandyopadhyay:2015oga,Bandyopadhyay:2014vma,Bandyopadhyay:2017klv,TavaresVelasco:2003ka}.
 
 Associated production of charged Higgs boson with another triplet neutral scalar in ITM scenario thus gives rise to mono-lepton or di-jet plus missing energy signature.  A pair of charged Higgs boson will give rise to di-lepton plus  missing energy \cite{Chiang:2020rcv, Bell:2020gug}. The signatures of ITM and IDM~\cite{Arhrib:2014pva,Lu:2019lok,Belyaev:2016lok} are very similar and the only difference is that in case of IDM we have additional neutral scalar (CP-even or CP-odd) which gives rise to distinguishing signature and thus can be separated from the ITM. Due to $Z_2$-odd, both inert Higgs bosons do not couple to fermions and their decay only happen via gauge mode on- or off-shell.

\begin{table}
	\begin{center}
		\renewcommand{\arraystretch}{1.4}
		\begin{tabular}{||c|c|c|c|c|c||}
			\hline\hline
			{\multirow{2}{*}{Model}} &
			{\multirow{2}{*}{Masses in GeV}}&
			{\multirow{2}{*}{Decay Modes}}&	{\multirow{2}{*}{BR in \% }}& {\multirow{2}{*}{Decay Width}}&{\multirow{2}{*}{Decay Length}} \\
			&&&& in GeV& in m\\
			\hline
			\multirow{5}{*}{IDM}	& \multirow{5}{*}{ $M_{A}= 898.48$}     & $H_{0} \rightarrow  A \bar{d} d$& $12.21 $& \multirow{5}{*}{$5.80 \times 10^{-7}$} &\multirow{5}{*}{$3.40 \times 10^{-10}$} \\
			& \multirow{6}{*}{$M_{H}^{\pm}= 902.69$} & $  H_{0} \rightarrow  A \bar{s} s$& $ 12.20 $&& \\
			&  \multirow{7}{*}{$M_{H_0}=911.88$} & $\displaystyle H_0\rightarrow \sum_{i=2,3}{A \bar{\nu_i}\nu_i }$ &$10.75 $&  &\\
			& & $H_{0} \rightarrow  A \bar{u} u$&$9.58 $& & \\
			& & $  H_{0} \rightarrow  A \bar{c} c$&$ 9.01 $& &\\
			\cline{3-6}
			& & $ H^{\pm} \rightarrow  A d \bar{u} $&$38.66 $& & \\
			& & $\displaystyle  H^{\pm}\rightarrow \sum_{i=1,2,3}{A \bar{\nu_i}\ell_i }$&$ 32.85 $&\multirow{2}{*}{$3.69 \times 10^{-9}$} &\multirow{2}{*}{$5.34 \times 10^{-8}$}\\
			& & $ H^{\pm} \rightarrow  A s \bar{c} $&$ 28.47 $& &\\
			\cline{1-6}
			\multirow{2}{*}{ITM}	& \multirow{1}{*}{$M_{T_0}= 1178.60$} &$ T^{\pm} \rightarrow  T_0 \bar{d} u $ & 72.72 & \multirow{2}{*}{$7.58\times 10^{-17}$ }&  \multirow{2}{*}{2.64}\\
			&\multirow{1}{*}{$ M_T^{\pm}=1178.76$} & $ T^{\pm} \rightarrow  T_0 \nu \ell^{\pm} $ & 24.30 & &\\
			\hline
		\end{tabular}
		\caption{Dominant 3-body decay modes and corresponding branching ratios, decay width and decay length for the $Z_2$-odd Higgs bosons for the benchmark points of IDM and ITM.  }\label{decay}
	\end{center}
\end{table}

In Table~\ref{decay} we present the benchmark points for the future collider study which are allowed by the vacuum stability, perturbativity bounds till Planck scale, dark matter relic and DM constraints. The heavy  Higgs boson and charged Higgs boson mass stay around 912 GeV and 903 GeV  respectively with the pseudoscalar  boson mass around 899 GeV. In this allowed mass range, the mass gap among the other heavier Higgs bosons are of the order of $\mathcal{O}(1)$ GeV, giving rise to naturally soft decay products for the associated Higgs productions. For the ITM case the mass splitting between $T^\pm$ and $T^0$ is $\sim 166$ MeV which comes from the loop correction.

Here the decays of $Z_2$ odd Higgs bosons ($H^\pm/H_0/T^\pm)$ are only possible via  three-body decays to quarks and leptons plus the DM particle via off-shell gauge boson due to insufficient phase space to decay into two on-shell gauge bosons. In these compressed  scenarios of IDM and ITM, the dominant decay modes for heavy Higgs boson(H) and charged Higgs boson $H^{\pm}/T^\pm$ are  $H_0 \rightarrow A Z^*$  and  $ H^{\pm}/T^\pm \rightarrow A/T_0\, W^{\pm *}$, with off-shell W/Z bosons. After integrating out gauge bosons, the decay width for  dominant $H_0 \rightarrow A f \overline{f}$ and  $H^{\pm}/T^\pm \rightarrow A/T_0 f \overline{f'}$ channels can be approximately given by \cite{3bdcy},
\bea
\Gamma(H_0 \rightarrow A f \overline{f})&=&\frac{1}{120\pi^3}\frac{g^4_2}{m_W^4}(\Delta m_0)^5 \sum_{i}N^i_c[(a^i_V)^2 + (a^i_{A})^2]\times \theta(\Delta m_0-2m_i), \nonumber \\
\Gamma(H^{\pm}/T^\pm \rightarrow A/T_0 f \overline{f'})&=&\frac{1}{120\pi^3}\frac{g^4_2}{m_W^4}(\Delta m^{\pm})^5 \sum_{j k}N^j_c[|c^{jk}_V|^2 +|c^{jk}_A|^2]\times \theta(\Delta m^{\pm}-m_j-m_{k}).\nonumber
\eea
where $N^i(j)_c$ is the colour factor of the SM fermions in the decay. The step function $\theta$ comes from the four-momentum conservation. The  electroweak couplings $a^i_V$ and $a^i_A$ are given by
\bea
a^i_V=\frac{1}{2}(T^3_i-2Q_is^2_W),  \qquad a^i_{A}=-\frac{1}{2}T^3_i,\nonumber
\eea
where i runs over all SM fermion, $\rm Q_i(T^3_i)$ is the charge (the third component of isospin) for the i-th fermion, and $\rm s_W$ stands for $\rm sin\theta_W$ with $\theta_W$ being the Weinberg angle. Similarly, the couplings $c^{jk}_V$  and $c^{jk}_A$ for lepton sectors can be represented as 
\bea
c^{jk}_V=-c^{jk}_A=\frac{1}{2 \sqrt{2}}\delta^{jk},\nonumber
\eea 
and for quark sectors
\bea
c^{jk}_V=-c^{jk}_A=\frac{1}{2 \sqrt{2}}V^{jk}_{CKM},\nonumber
\eea 
where j(k) runs over up-type (down-type) fermions and $V_{CKM}$ is the Cabbibo-Kobayashi-Maskawa matrix. Here $\Delta m_0$ and  $\Delta m^\pm$ are the mass splittings for $H_0-A$ and $H^\pm/T^\pm -A/T_0$ pairs respectively which can be crucial giving rise to displaced decays.  For ITM, $\Delta m^\pm \sim 166$ MeV, which comes from the loop correction thus always will give displaced charged Higgs decay. On the other hand, for IDM both $\Delta m_0$ and  $\Delta m^\pm$ have some tree-level contributions which can also lead to prompt decay like in our BP in Table~\ref{decay}.

 In Table~\ref{decay}  we also show the dominant three-body decay modes for the heavy CP-even  Higgs boson in IDM  with branching fractions of  BR$(H_0 \rightarrow A \bar{d}d) \sim 12.21 \%$ and BR$(H_0 \rightarrow A \bar{s}s) \sim 12.20 \%$ respectively with a total decay width  of $\sim 5.80 \times 10^{-7}$ GeV. This corresponds to decay length of $\sim 10^{-10}$ meter,  which essentially give rise to a prompt decay. The other subdominant decay modes are with BR$\displaystyle(H_0\rightarrow \sum_{i=2,3}{A \bar{\nu_i}\nu_i )}\sim 10.75\%$ and BR$\displaystyle(H_0\rightarrow A \bar{u}u )\sim 9.58\%$ respectively. For the charged Higgs the domiant modes are $A d \bar{u},  A \bar{\nu} \ell, \, A s \bar{c}$ with branching ratios $38.7\%, \, 32.9\%$ and $28.5\%$respectively. 

  Similarly lower panel of Table~\ref{decay} shows the benchmark point for the ITM scenario. Here the charged Higgs bosons and the triplet neutral scalar stay almost mass degenerate with n$M_{T_0}=$1178.60 GeV and $M_{T^{\pm}}=$1178.76 GeV respectively. Such spectrum only allows the three body decays with branching ratios of BR $(T^{\pm} \rightarrow  T_0 \bar{d} u )\sim 72.72 \% $  and BR$(T^{\pm} \rightarrow  T_0 \nu \ell^{\pm}) \sim 24.30 \%$ respectively. A very small decay width of $7.58 \times 10^{-17}$ GeV easily gives rise to $\mathcal{O}(2)$ meter displaced charged Higgs boson decay.\cite{Huitu:2010uc,BLscalar,Bandyopadhyay:2015iij,Bandyopadhyay:2014sma,Bandyopadhyay:2011qm,Bandyopadhyay:2010cu,Bandyopadhyay:2010wp}
\begin{table}
	\begin{center}
		\renewcommand{\arraystretch}{1.4}
		\begin{tabular}{||c|c|c|c|c|c||}
			\hline\hline
			\multirow{3}{*}{Energy}&
			\multicolumn{3}{|c|}{IDM} &
			\multicolumn{2}{|c||}{ITM}\\
				\cline{2-6}
			&
			$\sigma(H^{\pm}H^{\mp})$  &	 $\sigma(H^{\pm}H_{0})$& 	 $\sigma(H^{\pm}A)$	& 	 $\sigma(T^{\pm}T^{\mp})$  & 	 $\sigma(T^{\pm}T_{0})$\\
			&	in fb & in fb &in fb&  in fb & in fb  \\
			\cline{1-6}
			14 TeV & $1.88 \times 10^{-2}$ & $3.49 \times 10^{-2}$& $3.64 \times 10^{-2}$ & $3.07 \times 10^{-3}$& $6.82 \times 10^{-3}$\\
			\hline
			100 TeV	&	1.87  & 3.29 & 3.30 & $6.16 \times 10^{-1} $& $1.23$\\
			\hline
		\end{tabular}
		\caption{Production cross-section at LHC for 14 TeV and 100  TeV center of mass energy for the benchmark points in Table~\ref{decay}. }\label{cross}
	\end{center}
\end{table}

Next we focus on the production cross-sections of the chosen benchmark points at the LHC with centre of mass energy of 14, 100 TeV~\cite{Ilnicka:2015sra}. In Table~\ref{cross} present the cross-sections of various associated Higgs production modes at the LHC with centre of mass energy of 14 and 100 TeV. Here we used {\tt CalcHEP 3.7.5}~\cite{Belyaev:2012qa} for calculating the tree-level cross sections and decay branching fraction for the chosen benchmark points. For the cross-sections {\tt NNPDF 3.0 QED LO} \cite{pdf} is used as parton distribution function and $\sqrt{\hat{s}}$ is used as scale, where $s=E^2_{cm}$ is the known Mandelmstam variable. The associated Higgs productions include the production modes of $H^{\pm}H^{\mp}$, $H^{\pm}H_0$, $H^{\pm}A$ in IDM and $T^{\pm}T^{\mp}$, $T^{\pm}T_0$ in ITM as shown in Table~\ref{cross}. The charged Higgs pair production and associated productions  cross-sections at tree-level are $\sigma(H^{\pm}H^{\mp})=1.88 \times 10^{-2}$ fb, $\sigma(H^{\pm}H_{0})=3.49 \times 10^{-2}$ fb and $\sigma(H^{\pm}A)=3.64 \times 10^{-2}$ fb  respectively for IDM. Similar cross-sections for ITM are given by $\sigma(T^{\pm}T^{\mp})=3.07 \times 10^{-3}$ fb, $\sigma(T^{\pm}T_{0})=6.82 \times 10^{-3}$ fb respectively at the LHC with 14 TeV centre of mass energy. It is evident that the cross-sections are very low due to electro-weak nature of the process and around TeV mass of the particles. Nevertheless the situation improves  at 100 TeV with $\sigma(H^{\pm}H^{\mp})=1.87 $ fb, $\sigma(H^{\pm}H_{0})=3.29 $ fb,  $\sigma(H^{\pm}A)=3.30$ fb  for IDM and $\sigma(T^{\pm}T^{\mp})=6.16 \times 10^{-1}$ fb, $\sigma(T^{\pm}T_{0})=1.23 $ fb for ITM respectively. At 100 TeV LHC and with sufficiently large integrated luminosity studying the mono-lepton plus missing energy with prompt and displaced leptons one can distinguish such scenarios. IDM has one more massive mode compared to ITM which could also be instrumental  in distinguishing such scenarios as we demonstrate below.

Before going to further analysis here we describe the set up  and work flow of the collider simulation at the LHC.  For some BSM models have been extracted by writing the Lagrangian in SARAH \cite{Staub:2013tta} and then the corresponding CalCHEP \cite{Belyaev:2012qa} model files are also generated. We used CalcHEP to generate events in  {\tt lhe} format than can be read by PYTHIA6 \cite{Sjostrand:2006za}.  PYTHIA6 is used for parton and hadron-level simulation using the Fastjet-3.2.3 \cite{FastJet} with anti-kT algorithm. For the completeness of  this simulation we switch on the initial state radiation (ISR), final state radiation (DSR) and multiple interactions (MI). 
For this, the jet size have been selected to be R = 0.5, with the following cuts:
\begin{itemize}
	\item Calorimeter coverage: $|\eta| < 4.5$.
	\item Minimum transeverse momentum of each jet: $p_{T,min}^{jet} = 20.0$ GeV; jets are ordered in $p_{T}$.
	\item Jets are reconstructed out of only stable hadrons and no hard lepton.

	\item Selected leptons are hadronically clean, \textit{i.e,} hadronic activity within a cone of $\Delta R < 0.3$ around each lepton should be less than $15\%$ of the leptonic transeverse momentum, \textit{i.e.} $ p_{T}^{\mathrm{had}}< 0.15 p_{T}^{\text{lep}}$ within the cone.
	\item In order to make the leptons distinct from the jet, we put $\Delta R_{lj} > 0.4$ and $\Delta R_{ll} > 0.2$ to distinguish them from other leptons, where $\Delta R_{ij}=\sqrt{\Delta \eta^2_{ij}+ \Delta \phi^2_{ij}}$.
\end{itemize}

For the case of IDM the leptons that comes from the decays of the charged Higgs boson are prompt ones as can be read from Table~\ref{decay}. Whereas, the leptons coming from the charged Higgs boson in case of ITM are displaced ones by few mm to few m. For such displaced leptons we do not have any SM backgrounds. One common feature that the both scenarios posses is that due to very compressed spectrum the missing energy cancels between the two DM particles, one coming from the charged Higgs  decay and the other produced in association. The similar behaviour is also observed in charged Higgs pair production and generic to compressed spectrum scenarios found in supersymmetry \cite{PBASr} and Universal Extra Dimensions \cite{UED}. Nevertheless, for the IDM scenario due to a mass gap around 5-10 GeV among the charged Higgs and other other $Z_2$-odd neutral Higgs bosons, $p_T$  of the the leptons coming from the charged Higgs boson can be  around 20 GeV considering the boost effect at 100 TeV centre of mass energy. The important point  is to note that the leptons coming from SM  gauge bosons $W^\pm, \, Z$  would be relatively hard $\sim 40 $ GeV or more and the missing energy from the $W^\pm$ decays peaks around $50$ GeV. Drell-Yan (DY) processes via photon and $Z$ boson on/off-shell comes  always with two hard leptons in the final state. Process like $\gamma \, W^\pm$ can give rise to mono-lepton in the final states but always occupied the photon and relatively large missing energies. To eliminate this possible SM backgrounds for the IDM final sate we choose 

\be\label{fs1}
n_\ell =1 + n_j=0 +n_\gamma=0 + p^\ell_T <30 \, \rm{GeV} + \ptmiss \leq 30\, \rm{GeV}.
\ee
We present the numbers for hadronically quiet mono-lepton plus missing energy signatures as pointed out in Eq.~\eqref{fs1} in Table~\ref{SigIDM} at 100 TeV centre of mass energy at the LHC at  an integrated luminosity of 1000 fb$^{-1}$ for the benchmark point of IDM given in Table~\ref{decay}. The numbers at 1000fb$^{-1}$ of integrated luminosity suggests that around 18$\sigma$ signal significance is possible at the LHC with $E_{CM}=100$ TeV. However, due to lower cross-sections at 14 TeV even with 3000 fb$^{-1}$ of  integrated luminosity is not enough for a $5\sigma$ discovery and so the numbers are not presented here. As mentioned before, the situation improves a lot for ITM due to displaced leptonic signatures around mm to m range and the final state of $n_\ell \geq 1$ has no SM backgrounds as presented in Table~\ref{SigITM} at an integrated luminosity of 1000 fb$^{-1}$ at the LHC with $\rm E_{CM}=100$ TeV. Numbers suggests that around $12\sigma$ discovery is possible. In the context of other scenario mono-lepton signature at the LHC has been looked for \cite{tim}.


\begin{table}
	\begin{center}
		\renewcommand{\arraystretch}{1.4}
		\begin{tabular}{||c|c|c|c|c|c|c|c|c|c||}\hline
			\multirow{2}{*}{Signal}&$\rm E_{CM}$ &\multicolumn{3}{c||}{IDM}&\multicolumn{4}{c||}{Backgrounds}\\\cline{3-9}
			&in TeV&$H^\pm A$&$H^\pm H^\mp$& $H^\pm H_0$&$W^\pm W^\mp$&$W^\pm Z$&$ZZ$&DY\\\hline
			$n_\ell =1 + n_j=0 +n_\gamma=0 $&\multirow{2}{*}{100}&105.6&96.2&123.1&0.0&0.0&0.0&0.0\\	\cline{3-9}
				$+ p^\ell_T <30 \, \rm{GeV} + \ptmiss \leq 30\, \rm{GeV}$& &\multicolumn{3}{c||}{Total=324.9}&\multicolumn{4}{c||}{0.0}\\ 	\cline{2-9}
			\hline
		\end{tabular}
		\caption{$n_\ell=1 + 5\leq p^\ell_T \leq 20 $GeV signature at center of Mass energy of 100 TeV for the chosen BP of IDM at an integrated luminosity of 1000 fb$^{-1}$}\label{SigIDM}
	\end{center}
\end{table}


\begin{table}
	\begin{center}
		\renewcommand{\arraystretch}{1.4}
		\begin{tabular}{||c|c|c|c||}\hline
			Signal&$\rm E_{CM}$ &\multicolumn{2}{c||}{ITM}\\\hline
			$n_\ell \geq 1\,\,\&$ &\multirow{2}{*}{100}&89.4&59.0\\
			\cline{3-4}
				$1mm \leq d \leq 10$ m&&\multicolumn{2}{c||}{148.4}\\\hline
		\end{tabular}
		\caption{$n_\ell \geq 1 +\ptmiss <20 \, \rm{GeV}$ signature  with displacement $1mm \leq d \leq 10$ at centre of Mass energy of 100 TeV for the chosen BP of ITM at an integrated luminosity of 1000 fb$^{-1}$}\label{SigITM}
	\end{center}
\end{table}

\section{Conclusions} \label{concl}
In this article we consider two possible extensions of SM which give
 rise to a potential DM candidate and further extensions of which can address many other phenomenological issues \cite{2HDMpheno,Tripletex}. For this purpose $Z_2$-odd $SU(2)$ doublet extension, IDM and $Y=0$ $SU(2)$ triplet extension, ITM are analysed. The EWSB conditions in case of IDM give rise to extra CP-even($H_0$) and CP-odd($A$) Higgs bosons along with a charged Higgs boson $H^\pm$. Here lightest of the two neutral Higgs boson can be the DM candidate. However, for ITM there is only one CP-even($T_0$) neutral Higgs boson and one charged Higgs boson ($T^\pm$) that come from the $Z_2$ odd triplet multiplet. The EW mass gap among these $Z_2$-odd particles varies between $\mathcal{O}(M_W)$ to $\mathcal{O}(1)$ GeV in case of IDM at the tree-level.  In comparison the $Z_2$ odd particles in ITM are all mass degenerate at the tree-level and only $\mathcal{O}(1)$ MeV mass splitting comes from loop correction.
 
 After EWSB we checked the perturbative unitarity of all the dimensionless couplings for both IDM and ITM scenarios. Due to existence of large numbers of scalars IDM scenario gets perturbative bounds below Planck scale even with relatively smaller values of one of the Higgs quartic couplings at the EW scale i.e. $\lambda_i\simeq 0.1-0.2$. On the other hand, ITM scenario remains perturbative till Planck scale for higher values of  Higgs quartic coupling,   i.e. $\lambda_{t,ht} \lesssim 0.8$ and $\lambda_{ht}\lesssim 0.5,\, \lambda_t$=1.3. Similar to perturbativity, the stability of EW vacuum gives bounds on the parameter space by requiring that SM direction of the Higgs potential is stable and for SM such validity scale is $\mu \sim 10^{9-10}$ GeV \cite{Buttazzo:2013uya}.
 Introduction of the $Z_2$ scalar in both the cases i.e. IDM and ITM moves the region to greater stability. Thus models with right-handed neutrinos with large Yukawa can be in the stable region by the help of these scalars\cite{exwfermion}-\cite{Garg:2017iva}.
 
 After checking the perturbative unitarity and stability we move to calculate the DM relic abundance for both the scenarios. The dominant mode of annihilation for the both the cases are into $W^\pm W^\mp$ and co-annihilation is in association with the charged Higgs boson into $W^\pm Z$. However, due to presence of one extra $Z_2$ scalar in IDM compared to ITM, the DM number density is relatively on higher side than ITM. This requires more annihilation or co-annihilation to obtain the correct relic compared to the  ITM case, leading to lower mass bound on DM mass i.e. $m_{\rm{DM}} \gsim 700$ GeV in IDM compared to ITM, where it is $m_{\rm{DM}} \gsim 1176$ GeV. Later we also considered the direct-DM bounds from DM-nucleon scattering cross-section from XENON100, LUX and XENON1T \cite{Aprile:2012nq,Akerib:2013tjd,Aprile:2015uzo}. The corresponding indirect bounds on $<\sigma v>$ in $W^\pm W^\mp$ mode from H.E.S.S \cite{HESS} and Fermi-LAT \cite{FermiLAT} are also taken into account.

 At the end we studied their decay modes by calculating their decay branching fractions for the allowed benchmark points. We also estimate their production cross-sections for various associated Higgs-DM production modes at the LHC for the centre of mass energy of 14, 100 TeV respectively. Compressed spectrum for ITM will easily lead to displaced mono- or di-charged leptonic or displaced jet final states along with missing energy. Such displaced case however not so natural in case of IDM. Nevertheless, such inert scenarios can easily be distinguished from the normal Type-I 2HDM and $Y=0$ real scalar triplet, where both of them take part in EWSB as their decay products are not so restrictive.  Finally a PYTHIA levele signal-background analysis shows that the displaced lepton plus missing energy for  ITM and hadronically  quiet mono-leptonic signature for the IDM at the LHC can be viable modes to probe these scenarios. Since  14 TeV numbers  are not that significant owing TeV scale  phenomenology, we presented the numbers at 100 TeV at the LHC at 1000 fb$^{-1}$ of  integrated luminosity. However,  a  $5\sigma$ discovery is  expected  in $75-170$ fb$^{-1}$ luminosity at the LHC with $E_{\rm CM}=100$ TeV.

\section*{Acknowledgements} PB wants to thank SERB project (CRG/2018/004971) for the financial support towards this work. SJ thanks DST/INSPIRES/03/2018/001207 for the financial support towards finishing this work. SJ thanks Arjun Kumar for useful discussions in Higgs Triplet. PB and SJ thank  Eung Jin Chun, Marco Cirelli and Anirban Karan for useful discussions.
\appendix

\section{Two-loop $\beta$-functions for IDM} \label{betaf1}
\subsection{Scalar Quartic Couplings}
\footnotesize{
	\begingroup
	\allowdisplaybreaks
	\begin{align*}
	\beta_{\lambda_{1=h}} \ =  \ &
	\frac{1}{16\pi^2} \Bigg[	\frac{27}{200} g_{1}^{4} +\frac{9}{20} g_{1}^{2} g_{2}^{2} +\frac{9}{8} g_{2}^{4} -\frac{9}{5} g_{1}^{2} \lambda_1 -9 g_{2}^{2} \lambda_1 +24 \lambda_{1}^{2} +2 \lambda_{3}^{2} +2 \lambda_3 \lambda_4 +\lambda_{4}^{2}+4 \lambda_{5}^{2} \nonumber \\ 
	&\quad \quad +12 \lambda_1 \mbox{Tr}\Big({Y_d  Y_{d}^{\dagger}}\Big) +4 \lambda_1 \mbox{Tr}\Big({Y_e  Y_{e}^{\dagger}}\Big) +12 \lambda_1 \mbox{Tr}\Big({Y_u  Y_{u}^{\dagger}}\Big) -6 \mbox{Tr}\Big({Y_d  Y_{d}^{\dagger}  Y_d  Y_{d}^{\dagger}}\Big) -2 \mbox{Tr}\Big({Y_e  Y_{e}^{\dagger}  Y_e  Y_{e}^{\dagger}}\Big) \nonumber \\ 
	&\quad \quad -6 \mbox{Tr}\Big({Y_u  Y_{u}^{\dagger}  Y_u  Y_{u}^{\dagger}}\Big) \Bigg] \nonumber \\
	&+\frac{1}{(16\pi^2)^2}\Bigg[-\frac{3537}{2000} g_{1}^{6} -\frac{1719}{400} g_{1}^{4} g_{2}^{2} -\frac{303}{80} g_{1}^{2} g_{2}^{4} +\frac{291}{16} g_{2}^{6} +\frac{1953}{200} g_{1}^{4} \lambda_1 +\frac{117}{20} g_{1}^{2} g_{2}^{2} \lambda_1 -\frac{51}{8} g_{2}^{4} \lambda_1 +\frac{108}{5} g_{1}^{2} \lambda_{1}^{2} \nonumber \\ 
	&\quad \quad +108 g_{2}^{2} \lambda_{1}^{2} -312 \lambda_{1}^{3} +\frac{9}{10} g_{1}^{4} \lambda_3 +\frac{15}{2} g_{2}^{4} \lambda_3 +\frac{12}{5} g_{1}^{2} \lambda_{3}^{2} +12 g_{2}^{2} \lambda_{3}^{2} -20 \lambda_1 \lambda_{3}^{2} -8 \lambda_{3}^{3} +\frac{9}{20} g_{1}^{4} \lambda_4 \nonumber \\ 
	&\quad \quad +\frac{3}{2} g_{1}^{2} g_{2}^{2} \lambda_4 +\frac{15}{4} g_{2}^{4} \lambda_4 +\frac{12}{5} g_{1}^{2} \lambda_3 \lambda_4 +12 g_{2}^{2} \lambda_3 \lambda_4 -20 \lambda_1 \lambda_3 \lambda_4 -12 \lambda_{3}^{2} \lambda_4 +\frac{6}{5} g_{1}^{2} \lambda_{4}^{2} \nonumber \\ 
	& \quad \quad +3 g_{2}^{2} \lambda_{4}^{2} -12 \lambda_1 \lambda_{4}^{2} -16 \lambda_3 \lambda_{4}^{2} -6 \lambda_{4}^{3} -\frac{12}{5} g_{1}^{2} \lambda_{5}^{2} -56 \lambda_1 \lambda_{5}^{2} -80 \lambda_3 \lambda_{5}^{2} -88 \lambda_4 \lambda_{5}^{2} \nonumber \\ 
	&\quad \quad +\frac{1}{20} \Big(-5 \Big(64 \lambda_1 \Big(-5 g_{3}^{2}  + 9 \lambda_1 \Big) -90 g_{2}^{2} \lambda_1  + 9 g_{2}^{4} \Big) + 9 g_{1}^{4}  + g_{1}^{2} \Big(50 \lambda_1  + 54 g_{2}^{2} \Big)\Big)\mbox{Tr}\Big({Y_d  Y_{d}^{\dagger}}\Big) \nonumber \\ 
	&\quad \quad -\frac{3}{20} \Big(15 g_{1}^{4}  -2 g_{1}^{2} \Big(11 g_{2}^{2}  + 25 \lambda_1 \Big) + 5 \Big(-10 g_{2}^{2} \lambda_1  + 64 \lambda_{1}^{2}  + g_{2}^{4}\Big)\Big)\mbox{Tr}\Big({Y_e  Y_{e}^{\dagger}}\Big) -\frac{171}{100} g_{1}^{4} \mbox{Tr}\Big({Y_u  Y_{u}^{\dagger}}\Big) \nonumber \\ 
	&\quad \quad +\frac{63}{10} g_{1}^{2} g_{2}^{2} \mbox{Tr}\Big({Y_u  Y_{u}^{\dagger}}\Big) -\frac{9}{4} g_{2}^{4} \mbox{Tr}\Big({Y_u  Y_{u}^{\dagger}}\Big) +\frac{17}{2} g_{1}^{2} \lambda_1 \mbox{Tr}\Big({Y_u  Y_{u}^{\dagger}}\Big) +\frac{45}{2} g_{2}^{2} \lambda_1 \mbox{Tr}\Big({Y_u  Y_{u}^{\dagger}}\Big) \nonumber \\ 
	&\quad \quad +80 g_{3}^{2} \lambda_1 \mbox{Tr}\Big({Y_u  Y_{u}^{\dagger}}\Big) -144 \lambda_{1}^{2} \mbox{Tr}\Big({Y_u  Y_{u}^{\dagger}}\Big) +\frac{4}{5} g_{1}^{2} \mbox{Tr}\Big({Y_d  Y_{d}^{\dagger}  Y_d  Y_{d}^{\dagger}}\Big) -32 g_{3}^{2} \mbox{Tr}\Big({Y_d  Y_{d}^{\dagger}  Y_d  Y_{d}^{\dagger}}\Big) \nonumber \\ 
	&\quad \quad -3 \lambda_1 \mbox{Tr}\Big({Y_d  Y_{d}^{\dagger}  Y_d  Y_{d}^{\dagger}}\Big) -42 \lambda_1 \mbox{Tr}\Big({Y_d  Y_{d}^{\dagger}  Y_u  Y_{u}^{\dagger}}\Big) -\frac{12}{5} g_{1}^{2} \mbox{Tr}\Big({Y_e  Y_{e}^{\dagger}  Y_e  Y_{e}^{\dagger}}\Big) - \lambda_1 \mbox{Tr}\Big({Y_e  Y_{e}^{\dagger}  Y_e  Y_{e}^{\dagger}}\Big) \nonumber \\ 
	&\quad \quad -\frac{8}{5} g_{1}^{2} \mbox{Tr}\Big({Y_u  Y_{u}^{\dagger}  Y_u  Y_{u}^{\dagger}}\Big) -32 g_{3}^{2} \mbox{Tr}\Big({Y_u  Y_{u}^{\dagger}  Y_u  Y_{u}^{\dagger}}\Big) -3 \lambda_1 \mbox{Tr}\Big({Y_u  Y_{u}^{\dagger}  Y_u  Y_{u}^{\dagger}}\Big) +30 \mbox{Tr}\Big({Y_d  Y_{d}^{\dagger}  Y_d  Y_{d}^{\dagger}  Y_d  Y_{d}^{\dagger}}\Big) \nonumber \\ 
	&\quad \quad -6 \mbox{Tr}\Big({Y_d  Y_{d}^{\dagger}  Y_u  Y_{u}^{\dagger}  Y_d  Y_{d}^{\dagger}}\Big) -6 \mbox{Tr}\Big({Y_d  Y_{d}^{\dagger}  Y_u  Y_{u}^{\dagger}  Y_u  Y_{u}^{\dagger}}\Big) +10 \mbox{Tr}\Big({Y_e  Y_{e}^{\dagger}  Y_e  Y_{e}^{\dagger}  Y_e  Y_{e}^{\dagger}}\Big) +30 \mbox{Tr}\Big({Y_u  Y_{u}^{\dagger}  Y_u  Y_{u}^{\dagger}  Y_u  Y_{u}^{\dagger}}\Big)\Bigg] \, . \\
	\beta_{\lambda_2} \  = \ &
	\frac{1}{16\pi^2}\Bigg[
24 \lambda_{2}^{2}  + 2 \lambda_{3}^{2}  + 2 \lambda_3 \lambda_4  + 4 \lambda_{5}^{2}  -9 g_{2}^{2} \lambda_2  + \frac{27}{200} g_{1}^{4}  + \frac{9}{20} g_{1}^{2} \Big(-4 \lambda_2  + g_{2}^{2}\Big) + \frac{9}{8} g_{2}^{4}  + \lambda_{4}^{2}\Bigg] \nonumber \\
	&\quad \quad +\frac{1}{(16\pi^2)^2}\Bigg[-\frac{3537}{2000} g_{1}^{6} -\frac{1719}{400} g_{1}^{4} g_{2}^{2} -\frac{303}{80} g_{1}^{2} g_{2}^{4} +\frac{291}{16} g_{2}^{6} +\frac{1953}{200} g_{1}^{4} \lambda_2 +\frac{117}{20} g_{1}^{2} g_{2}^{2} \lambda_2 -\frac{51}{8} g_{2}^{4} \lambda_2  \nonumber \\ 
	&\quad \quad +108 g_{2}^{2} \lambda_{2}^{2} -312 \lambda_{2}^{3} +\frac{9}{10} g_{1}^{4} \lambda_3 +\frac{15}{2} g_{2}^{4} \lambda_3 +\frac{12}{5} g_{1}^{2} \lambda_{3}^{2} +12 g_{2}^{2} \lambda_{3}^{2} -20 \lambda_2 \lambda_{3}^{2} -8 \lambda_{3}^{3} +\frac{9}{20} g_{1}^{4} \lambda_4 \nonumber \\ 
	&\quad \quad +\frac{3}{2} g_{1}^{2} g_{2}^{2} \lambda_4 +\frac{15}{4} g_{2}^{4} \lambda_4 +\frac{12}{5} g_{1}^{2} \lambda_3 \lambda_4 +12 g_{2}^{2} \lambda_3 \lambda_4 -20 \lambda_2 \lambda_3 \lambda_4 -12 \lambda_{3}^{2} \lambda_4 +\frac{6}{5} g_{1}^{2} \lambda_{4}^{2} \nonumber \\ 
	&\quad \quad +3 g_{2}^{2} \lambda_{4}^{2} -12 \lambda_2 \lambda_{4}^{2} -16 \lambda_3 \lambda_{4}^{2} -6 \lambda_{4}^{3} -\frac{12}{5} g_{1}^{2} \lambda_{5}^{2} -56 \lambda_2 \lambda_{5}^{2} -80 \lambda_3 \lambda_{5}^{2} -88 \lambda_4 \lambda_{5}^{2}+\frac{108}{5} g_{1}^{2} \lambda_{2}^{2} \nonumber \\ 
	&\quad \quad -6 \Big(2 \lambda_{3}^{2}  + 2 \lambda_3 \lambda_4  + 4 \lambda_{5}^{2}  + \lambda_{4}^{2}\Big)\mbox{Tr}\Big({Y_d  Y_{d}^{\dagger}}\Big) -2 \Big(2 \lambda_{3}^{2}  + 2 \lambda_3 \lambda_4  + 4 \lambda_{5}^{2}  + \lambda_{4}^{2}\Big)\mbox{Tr}\Big({Y_e  Y_{e}^{\dagger}}\Big) \nonumber \\ 
	&\quad \quad -12 \lambda_{3}^{2} \mbox{Tr}\Big({Y_u  Y_{u}^{\dagger}}\Big) -12 \lambda_3 \lambda_4 \mbox{Tr}\Big({Y_u  Y_{u}^{\dagger}}\Big) -6 \lambda_{4}^{2} \mbox{Tr}\Big({Y_u  Y_{u}^{\dagger}}\Big) -24 \lambda_{5}^{2} \mbox{Tr}\Big({Y_u  Y_{u}^{\dagger}}\Big)\Bigg] \, .  \\
	\beta_{\lambda_3} \ =  \ &
	\frac{1}{16\pi^2}\Bigg[\frac{27}{100} g_{1}^{4} -\frac{9}{10} g_{1}^{2} g_{2}^{2} +\frac{9}{4} g_{2}^{4} -\frac{9}{5} g_{1}^{2} \lambda_3 -9 g_{2}^{2} \lambda_3 +12 \lambda_1 \lambda_3 +12 \lambda_2 \lambda_3 +4 \lambda_{3}^{2} +4 \lambda_1 \lambda_4 +4 \lambda_2 \lambda_4 +2 \lambda_{4}^{2} \nonumber \\ 
	&\quad \quad +8 \lambda_{5}^{2} +6 \lambda_3 \mbox{Tr}\Big({Y_d  Y_{d}^{\dagger}}\Big) +2 \lambda_3 \mbox{Tr}\Big({Y_e  Y_{e}^{\dagger}}\Big) +6 \lambda_3 \mbox{Tr}\Big({Y_u  Y_{u}^{\dagger}}\Big)\Bigg]\nonumber \\
	&+\frac{1}{(16\pi^2)^2}\Bigg[-\frac{3537}{1000} g_{1}^{6} +\frac{909}{200} g_{1}^{4} g_{2}^{2} +\frac{33}{40} g_{1}^{2} g_{2}^{4} +\frac{291}{8} g_{2}^{6} +\frac{27}{10} g_{1}^{4} \lambda_1 -3 g_{1}^{2} g_{2}^{2} \lambda_1 +\frac{45}{2} g_{2}^{4} \lambda_1 +\frac{27}{10} g_{1}^{4} \lambda_2 \nonumber \\ 
	&\quad \quad -3 g_{1}^{2} g_{2}^{2} \lambda_2 +\frac{45}{2} g_{2}^{4} \lambda_2 +\frac{1773}{200} g_{1}^{4} \lambda_3 +\frac{33}{20} g_{1}^{2} g_{2}^{2} \lambda_3 -\frac{111}{8} g_{2}^{4} \lambda_3 +\frac{72}{5} g_{1}^{2} \lambda_1 \lambda_3 +72 g_{2}^{2} \lambda_1 \lambda_3 \nonumber \\ 
	&\quad \quad -60 \lambda_{1}^{2} \lambda_3 +\frac{72}{5} g_{1}^{2} \lambda_2 \lambda_3 +72 g_{2}^{2} \lambda_2 \lambda_3 -60 \lambda_{2}^{2} \lambda_3 +\frac{6}{5} g_{1}^{2} \lambda_{3}^{2} +6 g_{2}^{2} \lambda_{3}^{2} -72 \lambda_1 \lambda_{3}^{2} -72 \lambda_2 \lambda_{3}^{2} \nonumber \\ 
	&\quad \quad -12 \lambda_{3}^{3} +\frac{9}{10} g_{1}^{4} \lambda_4 -\frac{9}{5} g_{1}^{2} g_{2}^{2} \lambda_4 +\frac{15}{2} g_{2}^{4} \lambda_4 +\frac{24}{5} g_{1}^{2} \lambda_1 \lambda_4 +36 g_{2}^{2} \lambda_1 \lambda_4 -16 \lambda_{1}^{2} \lambda_4 +\frac{24}{5} g_{1}^{2} \lambda_2 \lambda_4 \nonumber \\ 
	&\quad \quad +36 g_{2}^{2} \lambda_2 \lambda_4 -16 \lambda_{2}^{2} \lambda_4 -12 g_{2}^{2} \lambda_3 \lambda_4 -32 \lambda_1 \lambda_3 \lambda_4 -32 \lambda_2 \lambda_3 \lambda_4 -4 \lambda_{3}^{2} \lambda_4 -\frac{6}{5} g_{1}^{2} \lambda_{4}^{2} \nonumber \\ 
	&\quad \quad +6 g_{2}^{2} \lambda_{4}^{2} -28 \lambda_1 \lambda_{4}^{2} -28 \lambda_2 \lambda_{4}^{2} -16 \lambda_3 \lambda_{4}^{2} -12 \lambda_{4}^{3} +\frac{48}{5} g_{1}^{2} \lambda_{5}^{2} -144 \lambda_1 \lambda_{5}^{2} -144 \lambda_2 \lambda_{5}^{2} \nonumber \\ 
	&\quad \quad -72 \lambda_3 \lambda_{5}^{2} -176 \lambda_4 \lambda_{5}^{2}+\frac{1}{20} \Big(-5 \Big(-45 g_{2}^{2} \lambda_3  + 8 \Big(-20 g_{3}^{2} \lambda_3  + 3 \Big(2 \lambda_{3}^{2} + 4 \lambda_1 \Big(3 \lambda_3  + \lambda_4\Big)  \nonumber \\ 
	&\quad \quad  + 4 \lambda_{5}^{2}  + \lambda_{4}^{2}\Big)\Big) + 9 g_{2}^{4} \Big) + 9 g_{1}^{4}  + g_{1}^{2} \Big(25 \lambda_3  -54 g_{2}^{2} \Big)\Big)\mbox{Tr}\Big({Y_d  Y_{d}^{\dagger}}\Big)-\frac{1}{20} \Big(45 g_{1}^{4} \nonumber \\ 
	&\quad \quad   + 5 \Big(-15 g_{2}^{2} \lambda_3  + 3 g_{2}^{4}  + 8 \Big(2 \lambda_{3}^{2}  + 4 \lambda_1 \Big(3 \lambda_3  + \lambda_4\Big) + 4 \lambda_{5}^{2}  + \lambda_{4}^{2}\Big)\Big) + g_{1}^{2} \Big(66 g_{2}^{2}  -75 \lambda_3 \Big)\Big)\mbox{Tr}\Big({Y_e  Y_{e}^{\dagger}}\Big) \nonumber \\ 
	&\quad \quad -\frac{171}{100} g_{1}^{4} \mbox{Tr}\Big({Y_u  Y_{u}^{\dagger}}\Big) -\frac{63}{10} g_{1}^{2} g_{2}^{2} \mbox{Tr}\Big({Y_u  Y_{u}^{\dagger}}\Big) -\frac{9}{4} g_{2}^{4} \mbox{Tr}\Big({Y_u  Y_{u}^{\dagger}}\Big) +\frac{17}{4} g_{1}^{2} \lambda_3 \mbox{Tr}\Big({Y_u  Y_{u}^{\dagger}}\Big) \nonumber \\ 
	&\quad \quad +\frac{45}{4} g_{2}^{2} \lambda_3 \mbox{Tr}\Big({Y_u  Y_{u}^{\dagger}}\Big) +40 g_{3}^{2} \lambda_3 \mbox{Tr}\Big({Y_u  Y_{u}^{\dagger}}\Big) -72 \lambda_1 \lambda_3 \mbox{Tr}\Big({Y_u  Y_{u}^{\dagger}}\Big) -12 \lambda_{3}^{2} \mbox{Tr}\Big({Y_u  Y_{u}^{\dagger}}\Big) \nonumber \\ 
	&\quad \quad -24 \lambda_1 \lambda_4 \mbox{Tr}\Big({Y_u  Y_{u}^{\dagger}}\Big) -6 \lambda_{4}^{2} \mbox{Tr}\Big({Y_u  Y_{u}^{\dagger}}\Big) -24 \lambda_{5}^{2} \mbox{Tr}\Big({Y_u  Y_{u}^{\dagger}}\Big) -\frac{27}{2} \lambda_3 \mbox{Tr}\Big({Y_d  Y_{d}^{\dagger}  Y_d  Y_{d}^{\dagger}}\Big) \nonumber \\ 
	&\quad \quad -21 \lambda_3 \mbox{Tr}\Big({Y_d  Y_{d}^{\dagger}  Y_u  Y_{u}^{\dagger}}\Big) -24 \lambda_4 \mbox{Tr}\Big({Y_d  Y_{d}^{\dagger}  Y_u  Y_{u}^{\dagger}}\Big) -\frac{9}{2} \lambda_3 \mbox{Tr}\Big({Y_e  Y_{e}^{\dagger}  Y_e  Y_{e}^{\dagger}}\Big) -\frac{27}{2} \lambda_3 \mbox{Tr}\Big({Y_u  Y_{u}^{\dagger}  Y_u  Y_{u}^{\dagger}}\Big) \Bigg] \, . \\
	\beta_{\lambda_4} \ = \  &
	\frac{1}{16\pi^2}\Bigg[\frac{9}{5} g_{1}^{2} g_{2}^{2} -\frac{9}{5} g_{1}^{2} \lambda_4 -9 g_{2}^{2} \lambda_4 +4 \lambda_1 \lambda_4 +4 \lambda_2 \lambda_4 +8 \lambda_3 \lambda_4 +4 \lambda_{4}^{2} +32 \lambda_{5}^{2} +6 \lambda_4 \mbox{Tr}\Big({Y_d  Y_{d}^{\dagger}}\Big) \nonumber \\ 
	&\quad \quad +2 \lambda_4 \mbox{Tr}\Big({Y_e  Y_{e}^{\dagger}}\Big) +6 \lambda_4 \mbox{Tr}\Big({Y_u  Y_{u}^{\dagger}}\Big)\Bigg]\nonumber \\
	&+\frac{1}{(16\pi^2)^2} \Bigg[-\frac{657}{50} g_{1}^{4} g_{2}^{2} -\frac{42}{5} g_{1}^{2} g_{2}^{4} +6 g_{1}^{2} g_{2}^{2} \lambda_1 +6 g_{1}^{2} g_{2}^{2} \lambda_2 +\frac{6}{5} g_{1}^{2} g_{2}^{2} \lambda_3 +\frac{1413}{200} g_{1}^{4} \lambda_4 +\frac{153}{20} g_{1}^{2} g_{2}^{2} \lambda_4 \nonumber \\ 
	&\quad \quad -\frac{231}{8} g_{2}^{4} \lambda_4 +\frac{24}{5} g_{1}^{2} \lambda_1 \lambda_4 -28 \lambda_{1}^{2} \lambda_4 +\frac{24}{5} g_{1}^{2} \lambda_2 \lambda_4 -28 \lambda_{2}^{2} \lambda_4 +\frac{12}{5} g_{1}^{2} \lambda_3 \lambda_4 +36 g_{2}^{2} \lambda_3 \lambda_4 \nonumber \\ 
	&\quad \quad -80 \lambda_1 \lambda_3 \lambda_4 -80 \lambda_2 \lambda_3 \lambda_4 -28 \lambda_{3}^{2} \lambda_4 +\frac{24}{5} g_{1}^{2} \lambda_{4}^{2} +18 g_{2}^{2} \lambda_{4}^{2} -40 \lambda_1 \lambda_{4}^{2} -40 \lambda_2 \lambda_{4}^{2} -28 \lambda_3 \lambda_{4}^{2} \nonumber \\ 
	&\quad \quad +\frac{192}{5} g_{1}^{2} \lambda_{5}^{2} +216 g_{2}^{2} \lambda_{5}^{2} -192 \lambda_1 \lambda_{5}^{2} -192 \lambda_2 \lambda_{5}^{2} -192 \lambda_3 \lambda_{5}^{2} -104 \lambda_4 \lambda_{5}^{2} \nonumber \\ 
	&\quad \quad +\Big(4 \Big(10 g_{3}^{2} \lambda_4  -3 \Big(2 \lambda_1 \lambda_4  + 2 \lambda_3 \lambda_4  + 8 \lambda_{5}^{2}  + \lambda_{4}^{2}\Big)\Big) + \frac{45}{4} g_{2}^{2} \lambda_4  + g_{1}^{2} \Big(\frac{27}{5} g_{2}^{2}  + \frac{5}{4} \lambda_4 \Big)\Big)\mbox{Tr}\Big({Y_d  Y_{d}^{\dagger}}\Big) \nonumber \\ 
	&\quad \quad +\Big(-4 \Big(2 \lambda_1 \lambda_4  + 2 \lambda_3 \lambda_4  + 8 \lambda_{5}^{2}  + \lambda_{4}^{2}\Big) + \frac{15}{4} g_{2}^{2} \lambda_4  + \frac{3}{20} g_{1}^{2} \Big(25 \lambda_4  + 44 g_{2}^{2} \Big)\Big)\mbox{Tr}\Big({Y_e  Y_{e}^{\dagger}}\Big) \nonumber \\ 
	&\quad \quad +\frac{63}{5} g_{1}^{2} g_{2}^{2} \mbox{Tr}\Big({Y_u  Y_{u}^{\dagger}}\Big) +\frac{17}{4} g_{1}^{2} \lambda_4 \mbox{Tr}\Big({Y_u  Y_{u}^{\dagger}}\Big) +\frac{45}{4} g_{2}^{2} \lambda_4 \mbox{Tr}\Big({Y_u  Y_{u}^{\dagger}}\Big) +40 g_{3}^{2} \lambda_4 \mbox{Tr}\Big({Y_u  Y_{u}^{\dagger}}\Big) \nonumber \\ 
	&\quad \quad -24 \lambda_1 \lambda_4 \mbox{Tr}\Big({Y_u  Y_{u}^{\dagger}}\Big) -24 \lambda_3 \lambda_4 \mbox{Tr}\Big({Y_u  Y_{u}^{\dagger}}\Big) -12 \lambda_{4}^{2} \mbox{Tr}\Big({Y_u  Y_{u}^{\dagger}}\Big) -96 \lambda_{5}^{2} \mbox{Tr}\Big({Y_u  Y_{u}^{\dagger}}\Big) \nonumber \\ 
	&\quad \quad -\frac{27}{2} \lambda_4 \mbox{Tr}\Big({Y_d  Y_{d}^{\dagger}  Y_d  Y_{d}^{\dagger}}\Big) +27 \lambda_4 \mbox{Tr}\Big({Y_d  Y_{d}^{\dagger}  Y_u  Y_{u}^{\dagger}}\Big) -\frac{9}{2} \lambda_4 \mbox{Tr}\Big({Y_e  Y_{e}^{\dagger}  Y_e  Y_{e}^{\dagger}}\Big) -\frac{27}{2} \lambda_4 \mbox{Tr}\Big({Y_u  Y_{u}^{\dagger}  Y_u  Y_{u}^{\dagger}}\Big)\Bigg] \, . \\
	\beta_{\lambda_5} \ = \ &
	\frac{1}{16\pi^2}\Bigg[-\frac{9}{5} g_{1}^{2} \lambda_5 -9 g_{2}^{2} \lambda_5 +4 \lambda_1 \lambda_5 +4 \lambda_2 \lambda_5 +8 \lambda_3 \lambda_5 +12 \lambda_4 \lambda_5 +6 \lambda_5 \mbox{Tr}\Big({Y_d  Y_{d}^{\dagger}}\Big) +2 \lambda_5 \mbox{Tr}\Big({Y_e  Y_{e}^{\dagger}}\Big) \nonumber \\ 
	&\quad \quad +6 \lambda_5 \mbox{Tr}\Big({Y_u  Y_{u}^{\dagger}}\Big)\Bigg] 
	\nonumber \\
	& +\frac{1}{(16\pi^2)^2}\Bigg[\frac{1413}{200} g_{1}^{4} \lambda_5 +\frac{57}{20} g_{1}^{2} g_{2}^{2} \lambda_5 -\frac{231}{8} g_{2}^{4} \lambda_5 -\frac{12}{5} g_{1}^{2} \lambda_1 \lambda_5 -28 \lambda_{1}^{2} \lambda_5 -\frac{12}{5} g_{1}^{2} \lambda_2 \lambda_5 -28 \lambda_{2}^{2} \lambda_5 \nonumber \\ 
	&\quad \quad +\frac{48}{5} g_{1}^{2} \lambda_3 \lambda_5 +36 g_{2}^{2} \lambda_3 \lambda_5 -80 \lambda_1 \lambda_3 \lambda_5 -80 \lambda_2 \lambda_3 \lambda_5 -28 \lambda_{3}^{2} \lambda_5 +\frac{72}{5} g_{1}^{2} \lambda_4 \lambda_5 +72 g_{2}^{2} \lambda_4 \lambda_5 \nonumber \\ 
	&\quad \quad -88 \lambda_1 \lambda_4 \lambda_5 -88 \lambda_2 \lambda_4 \lambda_5 -76 \lambda_3 \lambda_4 \lambda_5 -32 \lambda_{4}^{2} \lambda_5 +24 \lambda_{5}^{3} \nonumber \\ 
	&\quad \quad +\frac{1}{4} \Big(16 \Big(10 g_{3}^{2}  -6 \lambda_1  -6 \lambda_3  -9 \lambda_4 \Big) + 45 g_{2}^{2}  + 5 g_{1}^{2} \Big)\lambda_5 \mbox{Tr}\Big({Y_d  Y_{d}^{\dagger}}\Big) \nonumber \\ 
	&\quad \quad +\frac{1}{4} \Big(15 g_{1}^{2}  + 15 g_{2}^{2}  -16 \Big(2 \lambda_1  + 2 \lambda_3  + 3 \lambda_4 \Big)\Big)\lambda_5 \mbox{Tr}\Big({Y_e  Y_{e}^{\dagger}}\Big) +\frac{17}{4} g_{1}^{2} \lambda_5 \mbox{Tr}\Big({Y_u  Y_{u}^{\dagger}}\Big) +\frac{45}{4} g_{2}^{2} \lambda_5 \mbox{Tr}\Big({Y_u  Y_{u}^{\dagger}}\Big) \nonumber \\ 
	&\quad \quad +40 g_{3}^{2} \lambda_5 \mbox{Tr}\Big({Y_u  Y_{u}^{\dagger}}\Big) -24 \lambda_1 \lambda_5 \mbox{Tr}\Big({Y_u  Y_{u}^{\dagger}}\Big) -24 \lambda_3 \lambda_5 \mbox{Tr}\Big({Y_u  Y_{u}^{\dagger}}\Big) -36 \lambda_4 \lambda_5 \mbox{Tr}\Big({Y_u  Y_{u}^{\dagger}}\Big) \nonumber \\ 
	&\quad \quad -\frac{3}{2} \lambda_5 \mbox{Tr}\Big({Y_d  Y_{d}^{\dagger}  Y_d  Y_{d}^{\dagger}}\Big) +3 \lambda_5 \mbox{Tr}\Big({Y_d  Y_{d}^{\dagger}  Y_u  Y_{u}^{\dagger}}\Big) -\frac{1}{2} \lambda_5 \mbox{Tr}\Big({Y_e  Y_{e}^{\dagger}  Y_e  Y_{e}^{\dagger}}\Big) -\frac{3}{2} \lambda_5 \mbox{Tr}\Big({Y_u  Y_{u}^{\dagger}  Y_u  Y_{u}^{\dagger}}\Big)\Bigg] \, .
	\end{align*}
	\endgroup

\subsection{Gauge Couplings }
\begingroup
\allowdisplaybreaks
\begin{align*}
\beta_{g_1} \ = \ & 
\frac{1}{16\pi^2}\Bigg[\frac{21}{5} g_{1}^{3}\Bigg]
+\frac{1}{(16\pi^2)^2}\Bigg[\frac{1}{50} g_{1}^{3} \Big(180 g_{2}^{2}  + 208 g_{1}^{2}  -25 \mbox{Tr}\Big({Y_d  Y_{d}^{\dagger}}\Big)  + 440 g_{3}^{2}  -75 \mbox{Tr}\Big({Y_e  Y_{e}^{\dagger}}\Big)  -85 \mbox{Tr}\Big({Y_u  Y_{u}^{\dagger}}\Big) \Big)\Bigg] \, . \\
\beta_{g_2}  \ =  \ &
\frac{1}{16\pi^2}\Bigg[-3 g_{2}^{3}\Bigg]+\frac{1}{(16\pi^2)^2}\Bigg[
\frac{1}{10} g_{2}^{3} \Big(120 g_{3}^{2}  + 12 g_{1}^{2}  -15 \mbox{Tr}\Big({Y_d  Y_{d}^{\dagger}}\Big)  -15 \mbox{Tr}\Big({Y_u  Y_{u}^{\dagger}}\Big)  -5 \mbox{Tr}\Big({Y_e  Y_{e}^{\dagger}}\Big)  + 80 g_{2}^{2} \Big)\Bigg] \, .  \\
\beta_{g_3} \ = \ &  
\frac{1}{16\pi^2}\Bigg[-7 g_{3}^{3}\Bigg]+
\frac{1}{(16\pi^2)^2}\Bigg[-\frac{1}{10} g_{3}^{3} \Big(-11 g_{1}^{2}  + 20 \mbox{Tr}\Big({Y_d  Y_{d}^{\dagger}}\Big)  + 20 \mbox{Tr}\Big({Y_u  Y_{u}^{\dagger}}\Big)  + 260 g_{3}^{2}  -45 g_{2}^{2} \Big)\Bigg]  \, .
\end{align*}
\endgroup

\subsection{Yukawa Coupling}
\begingroup
\allowdisplaybreaks
\begin{align*}
\beta_{Y_u} \  = \ & 
\frac{1}{16\pi^2}\Bigg[-\frac{3}{2} \Big(- {Y_u  Y_{u}^{\dagger}  Y_u}  + {Y_d  Y_{d}^{\dagger}  Y_u}\Big)\nonumber \\ 
&\quad \quad +Y_u \Big(3 \mbox{Tr}\Big({Y_d  Y_{d}^{\dagger}}\Big)  + 3 \mbox{Tr}\Big({Y_u  Y_{u}^{\dagger}}\Big)  -8 g_{3}^{2}  -\frac{17}{20} g_{1}^{2}  -\frac{9}{4} g_{2}^{2}  + \mbox{Tr}\Big({Y_e  Y_{e}^{\dagger}}\Big)\Big)\Bigg]\nonumber \\
&+
\frac{1}{(16\pi^2)^2}\Bigg[\frac{1}{80} \Big(20 \Big(11 {Y_d  Y_{d}^{\dagger}  Y_d  Y_{d}^{\dagger}  Y_u}  -4 {Y_d  Y_{d}^{\dagger}  Y_u  Y_{u}^{\dagger}  Y_u}  + 6 {Y_u  Y_{u}^{\dagger}  Y_u  Y_{u}^{\dagger}  Y_u}  - {Y_u  Y_{u}^{\dagger}  Y_d  Y_{d}^{\dagger}  Y_u} \Big)\nonumber \\ 
&\quad \quad +{Y_u  Y_{u}^{\dagger}  Y_u} \Big(1280 g_{3}^{2}  -180 \mbox{Tr}\Big({Y_e  Y_{e}^{\dagger}}\Big)  + 223 g_{1}^{2}  -540 \mbox{Tr}\Big({Y_d  Y_{d}^{\dagger}}\Big)  -540 \mbox{Tr}\Big({Y_u  Y_{u}^{\dagger}}\Big)  + 675 g_{2}^{2}  -960 \lambda_1 \Big)\nonumber \\ 
&\quad \quad +{Y_d  Y_{d}^{\dagger}  Y_u} \Big(100 \mbox{Tr}\Big({Y_e  Y_{e}^{\dagger}}\Big)  -1280 g_{3}^{2}  + 300 \mbox{Tr}\Big({Y_d  Y_{d}^{\dagger}}\Big)  + 300 \mbox{Tr}\Big({Y_u  Y_{u}^{\dagger}}\Big)  -43 g_{1}^{2}  + 45 g_{2}^{2} \Big)\Big)\nonumber \\ 
&\quad \quad +Y_u \Big(\frac{1267}{600} g_{1}^{4} -\frac{9}{20} g_{1}^{2} g_{2}^{2} -\frac{21}{4} g_{2}^{4} +\frac{19}{15} g_{1}^{2} g_{3}^{2} +9 g_{2}^{2} g_{3}^{2} -108 g_{3}^{4} +6 \lambda_{1}^{2} +\lambda_{3}^{2}+\lambda_3 \lambda_4 +\lambda_{4}^{2}+6 \lambda_{5}^{2} \nonumber \\ 
&\quad \quad +\frac{5}{8} \Big(32 g_{3}^{2}  + 9 g_{2}^{2}  + g_{1}^{2}\Big)\mbox{Tr}\Big({Y_d  Y_{d}^{\dagger}}\Big) +\frac{15}{8} \Big(g_{1}^{2} + g_{2}^{2}\Big)\mbox{Tr}\Big({Y_e  Y_{e}^{\dagger}}\Big) +\frac{17}{8} g_{1}^{2} \mbox{Tr}\Big({Y_u  Y_{u}^{\dagger}}\Big) +\frac{45}{8} g_{2}^{2} \mbox{Tr}\Big({Y_u  Y_{u}^{\dagger}}\Big) \nonumber \\ 
&\quad \quad +20 g_{3}^{2} \mbox{Tr}\Big({Y_u  Y_{u}^{\dagger}}\Big) -\frac{27}{4} \mbox{Tr}\Big({Y_d  Y_{d}^{\dagger}  Y_d  Y_{d}^{\dagger}}\Big) +\frac{3}{2} \mbox{Tr}\Big({Y_d  Y_{d}^{\dagger}  Y_u  Y_{u}^{\dagger}}\Big) -\frac{9}{4} \mbox{Tr}\Big({Y_e  Y_{e}^{\dagger}  Y_e  Y_{e}^{\dagger}}\Big) -\frac{27}{4} \mbox{Tr}\Big({Y_u  Y_{u}^{\dagger}  Y_u  Y_{u}^{\dagger}}\Big) \Big)\Bigg] \, .
\end{align*}
\endgroup

\section{Two-loop $\beta$-functions for ITM} \label{betaf2}
\subsection{Scalar Quartic Couplings}
\footnotesize{
	\begingroup
	\allowdisplaybreaks
	\begin{align*}
	\beta_{\lambda=\lambda_{h}} \ =  \ &
	\frac{1}{16\pi^2} \Bigg[	\frac{27}{200} g_{1}^{4} +\frac{9}{20} g_{1}^{2} g_{2}^{2} +\frac{9}{8} g_{2}^{4} -\frac{9}{5} g_{1}^{2} \lambda -9 g_{2}^{2} \lambda +24 \lambda^{2} +8 \lambda_{HT}^{2} +12 \lambda \mbox{Tr}\Big({Y_d  Y_{d}^{\dagger}}\Big) +4 \lambda \mbox{Tr}\Big({Y_e  Y_{e}^{\dagger}}\Big) \nonumber \\ 
	&\quad \quad +12 \lambda \mbox{Tr}\Big({Y_u  Y_{u}^{\dagger}}\Big) -6 \mbox{Tr}\Big({Y_d  Y_{d}^{\dagger}  Y_d  Y_{d}^{\dagger}}\Big) -2 \mbox{Tr}\Big({Y_e  Y_{e}^{\dagger}  Y_e  Y_{e}^{\dagger}}\Big) -6 \mbox{Tr}\Big({Y_u  Y_{u}^{\dagger}  Y_u  Y_{u}^{\dagger}}\Big) \Bigg] \nonumber \\
	& +\frac{1}{(16\pi^2)^2}\Bigg[-\frac{3411}{2000} g_{1}^{6} -\frac{1677}{400} g_{1}^{4} g_{2}^{2} -\frac{317}{80} g_{1}^{2} g_{2}^{4} +\frac{277}{16} g_{2}^{6} +\frac{1887}{200} g_{1}^{4} \lambda +\frac{117}{20} g_{1}^{2} g_{2}^{2} \lambda -\frac{29}{8} g_{2}^{4} \lambda  \nonumber \\ 
	&\quad \quad +\frac{108}{5} g_{1}^{2} \lambda^{2} +108 g_{2}^{2} \lambda^{2} -312 \lambda^{3} +10 g_{2}^{4} \lambda_{HT} +32 g_{2}^{2} \lambda_{HT}^{2} -80 \lambda \lambda_{HT}^{2} -128 \lambda_{HT}^{3} \nonumber \\ 
	&\quad \quad +\frac{1}{20} \Big(-5 \Big(64 \lambda \Big(-5 g_{3}^{2}  + 9 \lambda \Big) -90 g_{2}^{2} \lambda  + 9 g_{2}^{4} \Big) + 9 g_{1}^{4}  + g_{1}^{2} \Big(50 \lambda  + 54 g_{2}^{2} \Big)\Big)\mbox{Tr}\Big({Y_d  Y_{d}^{\dagger}}\Big) \nonumber \\ 
	&\quad \quad -\frac{3}{20} \Big(15 g_{1}^{4}  -2 g_{1}^{2} \Big(11 g_{2}^{2}  + 25 \lambda \Big) + 5 \Big(-10 g_{2}^{2} \lambda  + 64 \lambda^{2}  + g_{2}^{4}\Big)\Big)\mbox{Tr}\Big({Y_e  Y_{e}^{\dagger}}\Big) -\frac{171}{100} g_{1}^{4} \mbox{Tr}\Big({Y_u  Y_{u}^{\dagger}}\Big) \nonumber \\ 
	&\quad \quad +\frac{63}{10} g_{1}^{2} g_{2}^{2} \mbox{Tr}\Big({Y_u  Y_{u}^{\dagger}}\Big) -\frac{9}{4} g_{2}^{4} \mbox{Tr}\Big({Y_u  Y_{u}^{\dagger}}\Big) +\frac{17}{2} g_{1}^{2} \lambda \mbox{Tr}\Big({Y_u  Y_{u}^{\dagger}}\Big) +\frac{45}{2} g_{2}^{2} \lambda \mbox{Tr}\Big({Y_u  Y_{u}^{\dagger}}\Big) \nonumber \\ 
	&\quad \quad +80 g_{3}^{2} \lambda \mbox{Tr}\Big({Y_u  Y_{u}^{\dagger}}\Big) -144 \lambda^{2} \mbox{Tr}\Big({Y_u  Y_{u}^{\dagger}}\Big) +\frac{4}{5} g_{1}^{2} \mbox{Tr}\Big({Y_d  Y_{d}^{\dagger}  Y_d  Y_{d}^{\dagger}}\Big) -32 g_{3}^{2} \mbox{Tr}\Big({Y_d  Y_{d}^{\dagger}  Y_d  Y_{d}^{\dagger}}\Big) \nonumber \\ 
	&\quad \quad -3 \lambda \mbox{Tr}\Big({Y_d  Y_{d}^{\dagger}  Y_d  Y_{d}^{\dagger}}\Big) -42 \lambda \mbox{Tr}\Big({Y_d  Y_{u}^{\dagger}  Y_u  Y_{d}^{\dagger}}\Big) -\frac{12}{5} g_{1}^{2} \mbox{Tr}\Big({Y_e  Y_{e}^{\dagger}  Y_e  Y_{e}^{\dagger}}\Big) - \lambda \mbox{Tr}\Big({Y_e  Y_{e}^{\dagger}  Y_e  Y_{e}^{\dagger}}\Big) \nonumber \\ 
	&\quad \quad -\frac{8}{5} g_{1}^{2} \mbox{Tr}\Big({Y_u  Y_{u}^{\dagger}  Y_u  Y_{u}^{\dagger}}\Big) -32 g_{3}^{2} \mbox{Tr}\Big({Y_u  Y_{u}^{\dagger}  Y_u  Y_{u}^{\dagger}}\Big) -3 \lambda \mbox{Tr}\Big({Y_u  Y_{u}^{\dagger}  Y_u  Y_{u}^{\dagger}}\Big) +30 \mbox{Tr}\Big({Y_d  Y_{d}^{\dagger}  Y_d  Y_{d}^{\dagger}  Y_d  Y_{d}^{\dagger}}\Big) \nonumber \\ 
	&\quad \quad -12 \mbox{Tr}\Big({Y_d  Y_{d}^{\dagger}  Y_d  Y_{u}^{\dagger}  Y_u  Y_{d}^{\dagger}}\Big) +6 \mbox{Tr}\Big({Y_d  Y_{u}^{\dagger}  Y_u  Y_{d}^{\dagger}  Y_d  Y_{d}^{\dagger}}\Big) -6 \mbox{Tr}\Big({Y_d  Y_{u}^{\dagger}  Y_u  Y_{u}^{\dagger}  Y_u  Y_{d}^{\dagger}}\Big) \nonumber \\ 
	&\quad \quad +10 \mbox{Tr}\Big({Y_e  Y_{e}^{\dagger}  Y_e  Y_{e}^{\dagger}  Y_e  Y_{e}^{\dagger}}\Big) +30 \mbox{Tr}\Big({Y_u  Y_{u}^{\dagger}  Y_u  Y_{u}^{\dagger}  Y_u  Y_{u}^{\dagger}}\Big)\Bigg] \, . \\
	\beta_{\lambda_T} \  = \ &
	\frac{1}{16\pi^2}\Bigg[
-24 g_{2}^{2} \lambda_T  + 88 \lambda_{T}^{2}  + 8 \lambda_{HT}^{2}  + \frac{3}{2} g_{2}^{4}\Bigg] \nonumber \\
	& +\frac{1}{(16\pi^2)^2}\Bigg[-\frac{68}{3} g_{2}^{6} +10 g_{2}^{4} \lambda_{HT} +\frac{48}{5} g_{1}^{2} \lambda_{HT}^{2} +48 g_{2}^{2} \lambda_{HT}^{2} -128 \lambda_{HT}^{3} +\frac{94}{3} g_{2}^{4} \lambda_T -320 \lambda_{HT}^{2} \lambda_T +640 g_{2}^{2} \lambda_{T}^{2}\nonumber \\ 
	&\quad \quad  -4416 \lambda_{T}^{3} -48 \lambda_{HT}^{2} \mbox{Tr}\Big({Y_d  Y_{d}^{\dagger}}\Big) -16 \lambda_{HT}^{2} \mbox{Tr}\Big({Y_e  Y_{e}^{\dagger}}\Big) -48 \lambda_{HT}^{2} \mbox{Tr}\Big({Y_u  Y_{u}^{\dagger}}\Big)\Bigg] \, .  \\
	\beta_{\lambda_{HT}} \ =  \ &
	\frac{1}{16\pi^2}\Bigg[\frac{3}{4} g_{2}^{4} -\frac{9}{10} g_{1}^{2} \lambda_{HT} -\frac{33}{2} g_{2}^{2} \lambda_{HT} +12 \lambda \lambda_{HT} +16 \lambda_{HT}^{2} +24 \lambda_{HT} \lambda_T +6 \lambda_{HT} \mbox{Tr}\Big({Y_d  Y_{d}^{\dagger}}\Big) +2 \lambda_{HT} \mbox{Tr}\Big({Y_e  Y_{e}^{\dagger}}\Big) \nonumber \\ 
	&\quad \quad +6 \lambda_{HT} \mbox{Tr}\Big({Y_u  Y_{u}^{\dagger}}\Big)\Bigg]\nonumber \\
	&+\frac{1}{(16\pi^2)^2}\Bigg[-\frac{9}{16} g_{1}^{2} g_{2}^{4} +\frac{329}{48} g_{2}^{6} +\frac{15}{2} g_{2}^{4} \lambda +\frac{1671}{400} g_{1}^{4} \lambda_{HT} +\frac{9}{8} g_{1}^{2} g_{2}^{2} \lambda_{HT} -\frac{1087}{48} g_{2}^{4} \lambda_{HT} +\frac{72}{5} g_{1}^{2} \lambda \lambda_{HT}  \nonumber \\ 
	&\quad \quad +72 g_{2}^{2} \lambda \lambda_{HT}-60 \lambda^{2} \lambda_{HT} +\frac{12}{5} g_{1}^{2} \lambda_{HT}^{2} +44 g_{2}^{2} \lambda_{HT}^{2} -288 \lambda \lambda_{HT}^{2} -168 \lambda_{HT}^{3} +20 g_{2}^{4} \lambda_T +144 g_{2}^{2} \lambda_{HT} \lambda_T  \nonumber \\ 
	&\quad \quad -576 \lambda_{HT}^{2} \lambda_T-544 \lambda_{HT} \lambda_{T}^{2} -\frac{1}{4} \Big(3 g_{2}^{4}  -45 g_{2}^{2} \lambda_{HT}  + \lambda_{HT} \Big(-160 g_{3}^{2}  + 192 \lambda_{HT}  + 288 \lambda  -5 g_{1}^{2} \Big)\Big)\mbox{Tr}\Big({Y_d  Y_{d}^{\dagger}}\Big) \nonumber \\ 
	&\quad \quad -\frac{1}{4} \Big(-15 g_{2}^{2} \lambda_{HT}  + \lambda_{HT} \Big(-15 g_{1}^{2}  + 64 \lambda_{HT}  + 96 \lambda \Big) + g_{2}^{4}\Big)\mbox{Tr}\Big({Y_e  Y_{e}^{\dagger}}\Big) -\frac{3}{4} g_{2}^{4} \mbox{Tr}\Big({Y_u  Y_{u}^{\dagger}}\Big)  \nonumber \\ 
	&\quad \quad +\frac{17}{4} g_{1}^{2} \lambda_{HT} \mbox{Tr}\Big({Y_u  Y_{u}^{\dagger}}\Big)+\frac{45}{4} g_{2}^{2} \lambda_{HT} \mbox{Tr}\Big({Y_u  Y_{u}^{\dagger}}\Big) +40 g_{3}^{2} \lambda_{HT} \mbox{Tr}\Big({Y_u  Y_{u}^{\dagger}}\Big) -72 \lambda \lambda_{HT} \mbox{Tr}\Big({Y_u  Y_{u}^{\dagger}}\Big) \nonumber \\ 
	&\quad \quad  -48 \lambda_{HT}^{2} \mbox{Tr}\Big({Y_u  Y_{u}^{\dagger}}\Big)-\frac{27}{2} \lambda_{HT} \mbox{Tr}\Big({Y_d  Y_{d}^{\dagger}  Y_d  Y_{d}^{\dagger}}\Big) -21 \lambda_{HT} \mbox{Tr}\Big({Y_d  Y_{u}^{\dagger}  Y_u  Y_{d}^{\dagger}}\Big) -\frac{9}{2} \lambda_{HT} \mbox{Tr}\Big({Y_e  Y_{e}^{\dagger}  Y_e  Y_{e}^{\dagger}}\Big) \nonumber
	\\
	&\quad \quad  -\frac{27}{2} \lambda_{HT} \mbox{Tr}\Big({Y_u  Y_{u}^{\dagger}  Y_u  Y_{u}^{\dagger}}\Big) \Bigg] \, . \\
	\end{align*}
	\endgroup
	
	\subsection{Gauge Couplings }
	\begingroup
	\allowdisplaybreaks
	\begin{align*}
	\beta_{g_1} \ = \ & 
	\frac{1}{16\pi^2}\Bigg[\frac{41}{10} g_{1}^{3}\Bigg]
	+\frac{1}{(16\pi^2)^2}\Bigg[\frac{1}{50} g_{1}^{3} \Big(135 g_{2}^{2}  + 199 g_{1}^{2}  -25 \mbox{Tr}\Big({Y_d  Y_{d}^{\dagger}}\Big)  + 440 g_{3}^{2}  -75 \mbox{Tr}\Big({Y_e  Y_{e}^{\dagger}}\Big)  -85 \mbox{Tr}\Big({Y_u  Y_{u}^{\dagger}}\Big) \Big)\Bigg] \, . \\
	\beta_{g_2}  \ =  \ &
	\frac{1}{16\pi^2}\Bigg[-\frac{17}{6} g_{2}^{3}\Bigg]+\frac{1}{(16\pi^2)^2}\Bigg[
\frac{1}{30} g_{2}^{3} \Big(-15 \mbox{Tr}\Big({Y_e  Y_{e}^{\dagger}}\Big)  + 27 g_{1}^{2}  + 360 g_{3}^{2}  + 455 g_{2}^{2}  -45 \mbox{Tr}\Big({Y_d  Y_{d}^{\dagger}}\Big)  -45 \mbox{Tr}\Big({Y_u  Y_{u}^{\dagger}}\Big) \Big)\Bigg] \, .  \\
	\beta_{g_3} \ = \ &  
	\frac{1}{16\pi^2}\Bigg[-7 g_{3}^{3}\Bigg]+
	\frac{1}{(16\pi^2)^2}\Bigg[-\frac{1}{10} g_{3}^{3} \Big(-11 g_{1}^{2}  + 20 \mbox{Tr}\Big({Y_d  Y_{d}^{\dagger}}\Big)  + 20 \mbox{Tr}\Big({Y_u  Y_{u}^{\dagger}}\Big)  + 260 g_{3}^{2}  -45 g_{2}^{2} \Big)\Bigg]  \, .
	\end{align*}
	\endgroup
	
	\subsection{Yukawa Coupling}
	\begingroup
	\allowdisplaybreaks
	\begin{align*}
	\beta_{Y_u} \  = \ & 
	\frac{1}{16\pi^2}\Bigg[-\frac{3}{2} \Big(- {Y_u  Y_{u}^{\dagger}  Y_u}  + {Y_u  Y_{d}^{\dagger}  Y_d}\Big)\nonumber \\ 
	&+Y_u \Big(3 \mbox{Tr}\Big({Y_d  Y_{d}^{\dagger}}\Big)  + 3 \mbox{Tr}\Big({Y_u  Y_{u}^{\dagger}}\Big)  -8 g_{3}^{2}  -\frac{17}{20} g_{1}^{2}  -\frac{9}{4} g_{2}^{2}  + \mbox{Tr}\Big({Y_e  Y_{e}^{\dagger}}\Big)\Big)\Bigg]\nonumber \\
	&+
	\frac{1}{(16\pi^2)^2}\Bigg[\frac{1}{80} \Big(20 \Big(11 {Y_u  Y_{d}^{\dagger}  Y_d  Y_{d}^{\dagger}  Y_d}  -4 {Y_u  Y_{u}^{\dagger}  Y_u  Y_{d}^{\dagger}  Y_d}  + 6 {Y_u  Y_{u}^{\dagger}  Y_u  Y_{u}^{\dagger}  Y_u}  - {Y_u  Y_{d}^{\dagger}  Y_d  Y_{u}^{\dagger}  Y_u} \Big)\nonumber \\ 
	&\quad \quad +{Y_u  Y_{u}^{\dagger}  Y_u} \Big(1280 g_{3}^{2}  -180 \mbox{Tr}\Big({Y_e  Y_{e}^{\dagger}}\Big)  + 223 g_{1}^{2}  -540 \mbox{Tr}\Big({Y_d  Y_{d}^{\dagger}}\Big)  -540 \mbox{Tr}\Big({Y_u  Y_{u}^{\dagger}}\Big)  + 675 g_{2}^{2}  -960 \lambda \Big)\nonumber \\ 
	&\quad \quad +{Y_u  Y_{d}^{\dagger}  Y_d} \Big(100 \mbox{Tr}\Big({Y_e  Y_{e}^{\dagger}}\Big)  -1280 g_{3}^{2}  + 300 \mbox{Tr}\Big({Y_d  Y_{d}^{\dagger}}\Big)  + 300 \mbox{Tr}\Big({Y_u  Y_{u}^{\dagger}}\Big)  -43 g_{1}^{2}  + 45 g_{2}^{2} \Big)\Big)\nonumber \\ 
	&\quad \quad +Y_u \Big(\frac{1187}{600} g_{1}^{4} -\frac{9}{20} g_{1}^{2} g_{2}^{2} -\frac{19}{4} g_{2}^{4} +\frac{19}{15} g_{1}^{2} g_{3}^{2} +9 g_{2}^{2} g_{3}^{2} -108 g_{3}^{4} +6 \lambda^{2} +4 \lambda_{HT}^{2} \nonumber \\ 
	&\quad \quad +\frac{5}{8} \Big(32 g_{3}^{2}  + 9 g_{2}^{2}  + g_{1}^{2}\Big)\mbox{Tr}\Big({Y_d  Y_{d}^{\dagger}}\Big) +\frac{15}{8} \Big(g_{1}^{2} + g_{2}^{2}\Big)\mbox{Tr}\Big({Y_e  Y_{e}^{\dagger}}\Big) +\frac{17}{8} g_{1}^{2} \mbox{Tr}\Big({Y_u  Y_{u}^{\dagger}}\Big) +\frac{45}{8} g_{2}^{2} \mbox{Tr}\Big({Y_u  Y_{u}^{\dagger}}\Big) \nonumber \\ 
	&\quad \quad +20 g_{3}^{2} \mbox{Tr}\Big({Y_u  Y_{u}^{\dagger}}\Big) -\frac{27}{4} \mbox{Tr}\Big({Y_d  Y_{d}^{\dagger}  Y_d  Y_{d}^{\dagger}}\Big) +\frac{3}{2} \mbox{Tr}\Big({Y_d  Y_{u}^{\dagger}  Y_u  Y_{d}^{\dagger}}\Big) -\frac{9}{4} \mbox{Tr}\Big({Y_e  Y_{e}^{\dagger}  Y_e  Y_{e}^{\dagger}}\Big) -\frac{27}{4} \mbox{Tr}\Big({Y_u  Y_{u}^{\dagger}  Y_u  Y_{u}^{\dagger}}\Big) \Big)\Bigg] \, .
	\end{align*}
	\endgroup
\section{Dominant Annihilation cross-section for IDM and ITM} \label{annihi}
Here we provide the total amplitude squared for the dominant annihiliation process $DM DM \rightarrow W^+W^-/ZZ$ and co-annhilation $H^{\pm}/T^{\pm}+A/T_0 \rightarrow Z + W^{\pm}$ for IDM and ITM. The relevant Feynman diagrams are shown in Figure\ref{fig8l} and Figure\ref{fig9l}. We denote by $M_c$ the amplitude for the direct annihilation diagram and by $M_s$ the Higgs mediated diagram. $M_{u,t}$ correspond to the $H^{+}/T^{+}$ mediated diagrams. In the following, $p1$ and $p2$ denotes the 4-momentum of the annihilating $A/T_0$, $p3$ and $p4$ are the momentum of the 2 gauge bosons in the final-state and $\theta_W$ is the Weinberg angle.
\subsection{Process 1: $\rm~A(p1)+~A(p2) \rightarrow W^+(p3) + W^-(p4)$}
\begingroup
\allowdisplaybreaks
\begin{align*}
|\mathcal{M}_s|^2=&\frac{g^4_2v^4(\lambda_3+\lambda_4-2\lambda_5)^2}{8 M^4_{W^+} D_s}\Big[(p1.p2)^2+2(p1.p2)M^2_{A}-2(p1.p2)M^2_{W^+}\nonumber \\
&+M^4_{A}-2M^2_{A}M^2_{W^+}+3M^4_{W^+}\Big], \nonumber \\
D_s=& [(-p1-p2)^2-M^2_h]^2. \nonumber \\
|\mathcal{M}_t|^2= & \frac{g^4_2}{M^4_{W^+} D_t}\Big[(p1.p3)^4-2(p1.p3)^2M^2_{A}M^2_{W^+}+M^4_{A}M^4_{W^+}+61
\Big], \nonumber \\
D_t=& [(-p1+p3)^2-M^2_{H^+}]^2. \nonumber \\
|\mathcal{M}_u|^2= & \frac{{g^4_2}}{{M^4_{W^+}D_u}}\Big[(p1.p3)^4-4(p1.p3)^3(p1.p2)-4(p1.p3)^3M^2_{A}+6(p1.p3)^2(p1.p2)^2 \nonumber \\
&+12(p1.p3)^2(p1.p2)M^2_{A}  +6(p1.p3)^2M^4_{A}-2(p1.p3)^2M^2_{A}M^2_{W^+} \nonumber \\
&-4(p1.p3)(p1.p2)^3-12(p1.p3)(p1.p2)^2M^2_{A} -12(p1.p3)(p1.p2)M^4_{A} \nonumber \\
&+4(p1.p3)(p1.p2)M^2_{A}M^2_{W^+}-4(p1.p3)M^6_{A}+4(p1.p3)M^4_{A}M^2_{W^+} \nonumber \\
&+(p1.p2)^4+4(p1.p2)^3M^2_{A}+6(p1.p2)^2M^4_{A}-2(p1.p2)^2M^2_{A}M^2_{W^+} \nonumber \\
&+4(p1.p2)M^6_{A}-4(p1.p2)M^4_{A}M^2_{W^+}+M^8_{A}-2M^6_{A}M^2_{W^+}+M^4_{A}M^4_{W^+}\Big], \nonumber \\
D_u=& [(-p1+p4)^2-M^2_{H^+}]^2. \nonumber \\
|\mathcal{M}_c|^2= & \frac{g^4_2}{4M^4_{W^+} }\Big[(p1.p2)^2+2(p1.p2)M^2_{A}-2(p1.p2)M^2_{W^+}+M^4_{A}-2M^2_{A}M^2_{W^+}+3M^4_{W^+}\Big]. \nonumber 
\end{align*}
\subsection{Process 2: $\rm~A(p1)+~A(p2) \rightarrow Z(p3) + Z(p4)$}
\begin{align*}
|\mathcal{M}_s|^2=&\frac{g^4_2v^4(\lambda_3+\lambda_4-2\lambda_5)^2}{8 M^4_Z D_s}\Big[(p1.p2)^2+2(p1.p2)M^2_{A}-2(p1.p2)M^2_Z\nonumber \\
&+M^4_{A}-2M^2_{A}M^2_Z+3M^4_Z\Big], \nonumber \\
D_s=& [(-p1-p2)^2-M^2_h]^2. \nonumber \\
|\mathcal{M}_t|^2= & \frac{(-g_2Cos\theta_W-g_1Sin\theta_W)}{M^4_Z D_t}\Big[(p1.p3)^4-2(p1.p3)^2M^2_{A}M^2_Z+M^4_{A}M^4_Z\Big], \nonumber \\
D_t=& [(-p1+p3)^2-M^2_{H_0}]^2. \nonumber \\
|\mathcal{M}_c|^2= & \frac{(g_2Cos\theta_W+g_1Sin\theta_W)}{8M^4_Z }\Big[(p1.p2)^2+2(p1.p2)M^2_{A}-2(p1.p2)M^2_Z\nonumber \\
&+M^4_{A}-2M^2_{A}M^2_Z+3M^4_Z\Big]. \nonumber 
\end{align*}
\subsection{Process 3: $\rm~H^{\pm}(p1)+~A(p2) \rightarrow Z(p3) + W^{\pm}(p4)$}
\begin{align*}
|\mathcal{M}_s|^2=&\frac{g_2^4Cos\theta^2_W}{8M^6_{W^+}M^2_ZD_s}\Big[M_{W^+}^{12}-4 M_A^2 M_{W^+}^{10}+8 M_Z^2 M_{W^+}^{10}-12 M_{H^+}^2
M_{W^+}^{10}+8 \text{p1}.\text{p3} M_{W^+}^{10}+6 M_A^4
M_{W^+}^8\nonumber\\
&-18 M_Z^4 M_{W^+}^8+30 M_{H^+}^4 M_{W^+}^8-8
(\text{p1}.\text{p2})^2 M_{W^+}^8+16 (\text{p1}.\text{p3})^2
M_{W^+}^8-8 \text{p1}.\text{p3} M_A^2 M_{W^+}^8\nonumber\\
&-14 M_A^2 M_Z^2
M_{W^+}^8-40 \text{p1}.\text{p2} M_Z^2 M_{W^+}^8+72
\text{p1}.\text{p3} M_Z^2 M_{W^+}^8+20 M_A^2 M_{H^+}^2
M_{W^+}^8-10 M_Z^2 M_{H^+}^2 M_{W^+}^8\nonumber\\
&+48 \text{p1}.\text{p2}
M_{H^+}^2 M_{W^+}^8-40 \text{p1}.\text{p3} M_{H^+}^2 M_{W^+}^8-48
\text{p1}.\text{p2} \text{p1}.\text{p3} M_{W^+}^8-4 M_A^6
M_{W^+}^6+8 M_Z^6 M_{W^+}^6\nonumber\\
&-28 M_{H^+}^6 M_{W^+}^6-8
\text{p1}.\text{p3} M_A^4 M_{W^+}^6+38 M_A^2 M_Z^4 M_{W^+}^6+32
\text{p1}.\text{p2} M_Z^4 M_{W^+}^6-72 \text{p1}.\text{p3} M_Z^4
M_{W^+}^6\nonumber\\
&-28 M_A^2 M_{H^+}^4 M_{W^+}^6-10 M_Z^2 M_{H^+}^4
M_{W^+}^6-96 \text{p1}.\text{p2} M_{H^+}^4 M_{W^+}^6+56
\text{p1}.\text{p3} M_{H^+}^4 M_{W^+}^6\nonumber\\
&-64 \text{p1}.\text{p2}
(\text{p1}.\text{p3})^2 M_{W^+}^6+16 (\text{p1}.\text{p2})^2
M_A^2 M_{W^+}^6-32 (\text{p1}.\text{p3})^2 M_A^2 M_{W^+}^6+32
\text{p1}.\text{p2} \text{p1}.\text{p3} M_A^2 M_{W^+}^6\nonumber\\
&+2 M_A^4
M_Z^2 M_{W^+}^6-32 (\text{p1}.\text{p2})^2 M_Z^2 M_{W^+}^6+160
(\text{p1}.\text{p3})^2 M_Z^2 M_{W^+}^6+44 \text{p1}.\text{p2}
M_A^2 M_Z^2 M_{W^+}^6\nonumber\\
&-72 \text{p1}.\text{p3} M_A^2 M_Z^2
M_{W^+}^6-160 \text{p1}.\text{p2} \text{p1}.\text{p3} M_Z^2
M_{W^+}^6-4 M_A^4 M_{H^+}^2 M_{W^+}^6+10 M_Z^4 M_{H^+}^2
M_{W^+}^6\nonumber\\
&-80 (\text{p1}.\text{p2})^2 M_{H^+}^2 M_{W^+}^6-32
(\text{p1}.\text{p3})^2 M_{H^+}^2 M_{W^+}^6-32
\text{p1}.\text{p2} M_A^2 M_{H^+}^2 M_{W^+}^6+48
\text{p1}.\text{p3} M_A^2 M_{H^+}^2 M_{W^+}^6\nonumber\\
&+104 M_A^2 M_Z^2
M_{H^+}^2 M_{W^+}^6+20 \text{p1}.\text{p2} M_Z^2 M_{H^+}^2
M_{W^+}^6-88 \text{p1}.\text{p3} M_Z^2 M_{H^+}^2 M_{W^+}^6+160
\text{p1}.\text{p2} \text{p1}.\text{p3} M_{H^+}^2 M_{W^+}^6\nonumber\\
&+96
(\text{p1}.\text{p2})^2 \text{p1}.\text{p3} M_{W^+}^6+M_A^8
M_{W^+}^4+M_Z^8 M_{W^+}^4+9 M_{H^+}^8 M_{W^+}^4+8
\text{p1}.\text{p3} M_A^6 M_{W^+}^4-18 M_A^2 M_Z^6 M_{W^+}^4\nonumber\\
&+8
\text{p1}.\text{p2} M_Z^6 M_{W^+}^4-8 \text{p1}.\text{p3} M_Z^6
M_{W^+}^4+12 M_A^2 M_{H^+}^6 M_{W^+}^4+18 M_Z^2 M_{H^+}^6
M_{W^+}^4+48 \text{p1}.\text{p2} M_{H^+}^6 M_{W^+}^4\nonumber\\
&-24
\text{p1}.\text{p3} M_{H^+}^6 M_{W^+}^4+16
(\text{p1}.\text{p2})^4 M_{W^+}^4-8 (\text{p1}.\text{p2})^2 M_A^4
M_{W^+}^4+16 (\text{p1}.\text{p3})^2 M_A^4 M_{W^+}^4\nonumber\\
&+16
\text{p1}.\text{p2} \text{p1}.\text{p3} M_A^4 M_{W^+}^4-21 M_A^4
M_Z^4 M_{W^+}^4-40 (\text{p1}.\text{p2})^2 M_Z^4 M_{W^+}^4+16
(\text{p1}.\text{p3})^2 M_Z^4 M_{W^+}^4\nonumber\\
&-40 \text{p1}.\text{p2}
M_A^2 M_Z^4 M_{W^+}^4+72 \text{p1}.\text{p3} M_A^2 M_Z^4
M_{W^+}^4+16 \text{p1}.\text{p2} \text{p1}.\text{p3} M_Z^4
M_{W^+}^4-2 M_A^4 M_{H^+}^4 M_{W^+}^4\nonumber\\
&+11 M_Z^4 M_{H^+}^4
M_{W^+}^4+88 (\text{p1}.\text{p2})^2 M_{H^+}^4 M_{W^+}^4+16
(\text{p1}.\text{p3})^2 M_{H^+}^4 M_{W^+}^4+32
\text{p1}.\text{p2} M_A^2 M_{H^+}^4 M_{W^+}^4\nonumber\\
&-40
\text{p1}.\text{p3} M_A^2 M_{H^+}^4 M_{W^+}^4-22 M_A^2 M_Z^2
M_{H^+}^4 M_{W^+}^4+48 \text{p1}.\text{p2} M_Z^2 M_{H^+}^4
M_{W^+}^4+8 \text{p1}.\text{p3} M_Z^2 M_{H^+}^4 M_{W^+}^4\nonumber\\
&-112
\text{p1}.\text{p2} \text{p1}.\text{p3} M_{H^+}^4 M_{W^+}^4+64
(\text{p1}.\text{p2})^2 (\text{p1}.\text{p3})^2 M_{W^+}^4+64
\text{p1}.\text{p2} (\text{p1}.\text{p3})^2 M_A^2 M_{W^+}^4\nonumber\\
&-32
(\text{p1}.\text{p2})^2 \text{p1}.\text{p3} M_A^2 M_{W^+}^4+6
M_A^6 M_Z^2 M_{W^+}^4+8 \text{p1}.\text{p3} M_A^4 M_Z^2
M_{W^+}^4+32 (\text{p1}.\text{p2})^3 M_Z^2 M_{W^+}^4\nonumber\\
&-64
\text{p1}.\text{p2} (\text{p1}.\text{p3})^2 M_Z^2 M_{W^+}^4-8
(\text{p1}.\text{p2})^2 M_A^2 M_Z^2 M_{W^+}^4-32
(\text{p1}.\text{p3})^2 M_A^2 M_Z^2 M_{W^+}^4+32
\text{p1}.\text{p2} \text{p1}.\text{p3} M_A^2 M_Z^2 M_{W^+}^4\nonumber\\
&+32
(\text{p1}.\text{p2})^2 \text{p1}.\text{p3} M_Z^2 M_{W^+}^4-4
M_A^6 M_{H^+}^2 M_{W^+}^4+10 M_Z^6 M_{H^+}^2 M_{W^+}^4-16
\text{p1}.\text{p2} M_A^4 M_{H^+}^2 M_{W^+}^4\nonumber\\
&-8
\text{p1}.\text{p3} M_A^4 M_{H^+}^2 M_{W^+}^4+34 M_A^2 M_Z^4
M_{H^+}^2 M_{W^+}^4+24 \text{p1}.\text{p2} M_Z^4 M_{H^+}^2
M_{W^+}^4-56 \text{p1}.\text{p3} M_Z^4 M_{H^+}^2 M_{W^+}^4\nonumber\\
&+64
(\text{p1}.\text{p2})^3 M_{H^+}^2 M_{W^+}^4+64
\text{p1}.\text{p2} (\text{p1}.\text{p3})^2 M_{H^+}^2
M_{W^+}^4+16 (\text{p1}.\text{p2})^2 M_A^2 M_{H^+}^2 M_{W^+}^4\nonumber\\
&+32
(\text{p1}.\text{p3})^2 M_A^2 M_{H^+}^2 M_{W^+}^4-96
\text{p1}.\text{p2} \text{p1}.\text{p3} M_A^2 M_{H^+}^2
M_{W^+}^4-34 M_A^4 M_Z^2 M_{H^+}^2 M_{W^+}^4\nonumber\\
&+40
(\text{p1}.\text{p2})^2 M_Z^2 M_{H^+}^2 M_{W^+}^4-32
(\text{p1}.\text{p3})^2 M_Z^2 M_{H^+}^2 M_{W^+}^4-80
\text{p1}.\text{p2} M_A^2 M_Z^2 M_{H^+}^2 M_{W^+}^4\nonumber\\
&+16
\text{p1}.\text{p3} M_A^2 M_Z^2 M_{H^+}^2 M_{W^+}^4+32
\text{p1}.\text{p2} \text{p1}.\text{p3} M_Z^2 M_{H^+}^2
M_{W^+}^4-160 (\text{p1}.\text{p2})^2 \text{p1}.\text{p3}
M_{H^+}^2 M_{W^+}^4\nonumber\\
&-64 (\text{p1}.\text{p2})^3
\text{p1}.\text{p3} M_{W^+}^4-2 M_A^2 M_Z^8 M_{W^+}^2-6 M_Z^2
M_{H^+}^8 M_{W^+}^2+12 M_A^4 M_Z^6 M_{W^+}^2-4
\text{p1}.\text{p2} M_A^2 M_Z^6 M_{W^+}^2\nonumber\\
&+8 \text{p1}.\text{p3}
M_A^2 M_Z^6 M_{W^+}^2-4 M_Z^4 M_{H^+}^6 M_{W^+}^2-4 M_A^2 M_Z^2
M_{H^+}^6 M_{W^+}^2-28 \text{p1}.\text{p2} M_Z^2 M_{H^+}^6
M_{W^+}^2\nonumber\\
&+8 \text{p1}.\text{p3} M_Z^2 M_{H^+}^6 M_{W^+}^2+4
\text{p1}.\text{p2} M_A^4 M_Z^4 M_{W^+}^2+8
(\text{p1}.\text{p2})^2 M_A^2 M_Z^4 M_{W^+}^2+16 M_Z^6 M_{H^+}^4
M_{W^+}^2\nonumber\\
&-12 \text{p1}.\text{p2} M_Z^4 M_{H^+}^4 M_{W^+}^2+8
M_A^4 M_Z^2 M_{H^+}^4 M_{W^+}^2-40 (\text{p1}.\text{p2})^2 M_Z^2
M_{H^+}^4 M_{W^+}^2+4 \text{p1}.\text{p2} M_A^2 M_Z^2 M_{H^+}^4
M_{W^+}^2\nonumber\\
&+8 \text{p1}.\text{p3} M_A^2 M_Z^2 M_{H^+}^4
M_{W^+}^2+32 \text{p1}.\text{p2} \text{p1}.\text{p3} M_Z^2
M_{H^+}^4 M_{W^+}^2-2 M_A^8 M_Z^2 M_{W^+}^2-4 \text{p1}.\text{p2}
M_A^6 M_Z^2 M_{W^+}^2\nonumber\\
&-8 \text{p1}.\text{p3} M_A^6 M_Z^2
M_{W^+}^2+8 (\text{p1}.\text{p2})^2 M_A^4 M_Z^2 M_{W^+}^2-32
\text{p1}.\text{p2} \text{p1}.\text{p3} M_A^4 M_Z^2 M_{W^+}^2+16
(\text{p1}.\text{p2})^3 M_A^2 M_Z^2 M_{W^+}^2\nonumber\\
&-32
(\text{p1}.\text{p2})^2 \text{p1}.\text{p3} M_A^2 M_Z^2
M_{W^+}^2+2 M_Z^8 M_{H^+}^2 M_{W^+}^2-28 M_A^2 M_Z^6 M_{H^+}^2
M_{W^+}^2+4 \text{p1}.\text{p2} M_Z^6 M_{H^+}^2 M_{W^+}^2\nonumber\\
&-8
\text{p1}.\text{p3} M_Z^6 M_{H^+}^2 M_{W^+}^2+4 M_A^4 M_Z^4
M_{H^+}^2 M_{W^+}^2-8 (\text{p1}.\text{p2})^2 M_Z^4 M_{H^+}^2
M_{W^+}^2+8 \text{p1}.\text{p2} M_A^2 M_Z^4 M_{H^+}^2 M_{W^+}^2\nonumber\\
&+4
M_A^6 M_Z^2 M_{H^+}^2 M_{W^+}^2+28 \text{p1}.\text{p2} M_A^4
M_Z^2 M_{H^+}^2 M_{W^+}^2-8 \text{p1}.\text{p3} M_A^4 M_Z^2
M_{H^+}^2 M_{W^+}^2-16 (\text{p1}.\text{p2})^3 M_Z^2 M_{H^+}^2
M_{W^+}^2\nonumber\\
&+32 (\text{p1}.\text{p2})^2 M_A^2 M_Z^2 M_{H^+}^2
M_{W^+}^2+32 (\text{p1}.\text{p2})^2 \text{p1}.\text{p3} M_Z^2
M_{H^+}^2 M_{W^+}^2+M_A^4 M_Z^8+M_Z^4 M_{H^+}^8-2 M_A^6 M_Z^6\nonumber\\
&-4
\text{p1}.\text{p2} M_A^4 M_Z^6-2 M_Z^6 M_{H^+}^6+4
\text{p1}.\text{p2} M_Z^4 M_{H^+}^6+M_A^8 M_Z^4+4
\text{p1}.\text{p2} M_A^6 M_Z^4+4 (\text{p1}.\text{p2})^2 M_A^4
M_Z^4+M_Z^8 M_{H^+}^4\nonumber\\
&+2 M_A^2 M_Z^6 M_{H^+}^4-4
\text{p1}.\text{p2} M_Z^6 M_{H^+}^4-2 M_A^4 M_Z^4 M_{H^+}^4+4
(\text{p1}.\text{p2})^2 M_Z^4 M_{H^+}^4-4 \text{p1}.\text{p2}
M_A^2 M_Z^4 M_{H^+}^4-2 M_A^2 M_Z^8 M_{H^+}^2\nonumber\\
&+2 M_A^4 M_Z^6
M_{H^+}^2+8 \text{p1}.\text{p2} M_A^2 M_Z^6 M_{H^+}^2-4
\text{p1}.\text{p2} M_A^4 M_Z^4 M_{H^+}^2-8
(\text{p1}.\text{p2})^2 M_A^2 M_Z^4 M_{H^+}^2\Big],\nonumber\\
D_s=&[(-p1-p2)^2-M_{W^+}^2]^2.\nonumber\\
|\mathcal{M}_t|^2= &\frac{g^2_2(-g_2Cos\theta_W+g_1Sin\theta_W)^2}{4M^2_{W^+}M^2_ZD_t}\Big[-4 M_A^2 M_{H^+}^2 M_Z^2 \text{p1}.\text{p3}-2 M_A^2 M_{H^+}^2
(\text{p1}.\text{p3})^2+2 M_A^2 M_{H^+}^2 M_{W^+}^2 M_Z^2\nonumber\\
&+2 M_A^2
M_{H^+}^2 M_Z^4+2 M_A^2 M_{H^+}^4 M_Z^2-M_A^4 M_{H^+}^2 M_Z^2-2
M_A^2 M_{W^+}^2 (\text{p1}.\text{p3})^2-2 M_A^2 M_Z^2
(\text{p1}.\text{p3})^2\nonumber\\
&+M_A^4 (\text{p1}.\text{p3})^2+4 M_A^2
(\text{p1}.\text{p3})^3-4 M_{H^+}^2 M_{W^+}^2 M_Z^2
\text{p1}.\text{p3}-2 M_{H^+}^2 M_{W^+}^2
(\text{p1}.\text{p3})^2+4 M_{H^+}^2 M_Z^4 \text{p1}.\text{p3}\nonumber\\
&+4
M_{H^+}^4 M_Z^2 \text{p1}.\text{p3}-2 M_{H^+}^2 M_Z^2
(\text{p1}.\text{p3})^2+M_{H^+}^4 (\text{p1}.\text{p3})^2-4
M_{H^+}^2 (\text{p1}.\text{p3})^3+2 M_{H^+}^2 M_{W^+}^2
M_Z^4\nonumber\\
&-M_{H^+}^2 M_{W^+}^4 M_Z^2+2 M_{H^+}^4 M_{W^+}^2
M_Z^2-M_{H^+}^2 M_Z^6-2 M_{H^+}^4 M_Z^4-M_{H^+}^6 M_Z^2-2
M_{W^+}^2 M_Z^2 (\text{p1}.\text{p3})^2\nonumber\\
&+M_{W^+}^4
(\text{p1}.\text{p3})^2+4 M_{W^+}^2 (\text{p1}.\text{p3})^3+M_Z^4
(\text{p1}.\text{p3})^2-4 M_Z^2 (\text{p1}.\text{p3})^3+4
(\text{p1}.\text{p3})^4\Big], \nonumber \\
D_t=&[(-p1+p3)^2-M_{H^+}^2]^2.\nonumber\\
|\mathcal{M}_u|^2= & \frac{g^2_2(g_2Cos\theta_W+g_1Sin\theta_W)^2}{4M^2_{W^+}M^2_ZD_u}\Big[M_{H^+}^8+2 M_A^2 M_{H^+}^6+2 M_Z^2 M_{H^+}^6-3 M_{W^+}^2
M_{H^+}^6+6 \text{p1}.\text{p2} M_{H^+}^6 \nonumber \\
&-6 \text{p1}.\text{p3}
M_{H^+}^6+M_A^4 M_{H^+}^4+M_Z^4 M_{H^+}^4+3 M_{W^+}^4
M_{H^+}^4+13 (\text{p1}.\text{p2})^2 M_{H^+}^4+13
(\text{p1}.\text{p3})^2 M_{H^+}^4\nonumber \\
&+8 \text{p1}.\text{p2} M_A^2
M_{H^+}^4-8 \text{p1}.\text{p3} M_A^2 M_{H^+}^4-2 M_A^2 M_Z^2
M_{H^+}^4+8 \text{p1}.\text{p2} M_Z^2 M_{H^+}^4-8
\text{p1}.\text{p3} M_Z^2 M_{H^+}^4\nonumber \\
&-4 M_A^2 M_{W^+}^2 M_{H^+}^4-4
M_Z^2 M_{W^+}^2 M_{H^+}^4-12 \text{p1}.\text{p2} M_{W^+}^2
M_{H^+}^4+12 \text{p1}.\text{p3} M_{W^+}^2 M_{H^+}^4\nonumber \\
&-26
\text{p1}.\text{p2} \text{p1}.\text{p3} M_{H^+}^4-M_{W^+}^6
M_{H^+}^2+2 \text{p1}.\text{p2} M_A^4 M_{H^+}^2-2
\text{p1}.\text{p3} M_A^4 M_{H^+}^2+2 \text{p1}.\text{p2} M_Z^4
M_{H^+}^2\nonumber\\
&-2 \text{p1}.\text{p3} M_Z^4 M_{H^+}^2+2 M_A^2 M_{W^+}^4
M_{H^+}^2+2 M_Z^2 M_{W^+}^4 M_{H^+}^2+6 \text{p1}.\text{p2}
M_{W^+}^4 M_{H^+}^2-6 \text{p1}.\text{p3} M_{W^+}^4 M_{H^+}^2\nonumber\\
&+12
(\text{p1}.\text{p2})^3 M_{H^+}^2-12 (\text{p1}.\text{p3})^3
M_{H^+}^2+36 \text{p1}.\text{p2} (\text{p1}.\text{p3})^2
M_{H^+}^2+10 (\text{p1}.\text{p2})^2 M_A^2 M_{H^+}^2\nonumber\\
&+10
(\text{p1}.\text{p3})^2 M_A^2 M_{H^+}^2-20 \text{p1}.\text{p2}
\text{p1}.\text{p3} M_A^2 M_{H^+}^2+10 (\text{p1}.\text{p2})^2
M_Z^2 M_{H^+}^2+10 (\text{p1}.\text{p3})^2 M_Z^2 M_{H^+}^2\nonumber\\
&-4
\text{p1}.\text{p2} M_A^2 M_Z^2 M_{H^+}^2+4 \text{p1}.\text{p3}
M_A^2 M_Z^2 M_{H^+}^2-20 \text{p1}.\text{p2} \text{p1}.\text{p3}
M_Z^2 M_{H^+}^2-M_A^4 M_{W^+}^2 M_{H^+}^2\nonumber\\
&-M_Z^4 M_{W^+}^2
M_{H^+}^2-14 (\text{p1}.\text{p2})^2 M_{W^+}^2 M_{H^+}^2-14
(\text{p1}.\text{p3})^2 M_{W^+}^2 M_{H^+}^2-8 \text{p1}.\text{p2}
M_A^2 M_{W^+}^2 M_{H^+}^2\nonumber\\
&+8 \text{p1}.\text{p3} M_A^2 M_{W^+}^2
M_{H^+}^2+2 M_A^2 M_Z^2 M_{W^+}^2 M_{H^+}^2-8 \text{p1}.\text{p2}
M_Z^2 M_{W^+}^2 M_{H^+}^2+8 \text{p1}.\text{p3} M_Z^2 M_{W^+}^2
M_{H^+}^2\nonumber\\
&+28 \text{p1}.\text{p2} \text{p1}.\text{p3} M_{W^+}^2
M_{H^+}^2-36 (\text{p1}.\text{p2})^2 \text{p1}.\text{p3}
M_{H^+}^2+4 (\text{p1}.\text{p2})^4+4
(\text{p1}.\text{p3})^4+(\text{p1}.\text{p2})^2
M_A^4\nonumber\\
&+(\text{p1}.\text{p3})^2 M_A^4-2 \text{p1}.\text{p2}
\text{p1}.\text{p3} M_A^4+(\text{p1}.\text{p2})^2
M_Z^4+(\text{p1}.\text{p3})^2 M_Z^4-2 \text{p1}.\text{p2}
\text{p1}.\text{p3} M_Z^4+(\text{p1}.\text{p2})^2
M_{W^+}^4\nonumber\\
&+(\text{p1}.\text{p3})^2 M_{W^+}^4-2 \text{p1}.\text{p2}
\text{p1}.\text{p3} M_{W^+}^4-16 \text{p1}.\text{p2}
(\text{p1}.\text{p3})^3+24 (\text{p1}.\text{p2})^2
(\text{p1}.\text{p3})^2+4 (\text{p1}.\text{p2})^3 M_A^2\nonumber\\
&-4
(\text{p1}.\text{p3})^3 M_A^2+12 \text{p1}.\text{p2}
(\text{p1}.\text{p3})^2 M_A^2-12 (\text{p1}.\text{p2})^2
\text{p1}.\text{p3} M_A^2+4 (\text{p1}.\text{p2})^3 M_Z^2-4
(\text{p1}.\text{p3})^3 M_Z^2\nonumber\\
&+12 \text{p1}.\text{p2}
(\text{p1}.\text{p3})^2 M_Z^2-2 (\text{p1}.\text{p2})^2 M_A^2
M_Z^2-2 (\text{p1}.\text{p3})^2 M_A^2 M_Z^2+4 \text{p1}.\text{p2}
\text{p1}.\text{p3} M_A^2 M_Z^2\nonumber\\
&-12 (\text{p1}.\text{p2})^2
\text{p1}.\text{p3} M_Z^2-4 (\text{p1}.\text{p2})^3 M_{W^+}^2+4
(\text{p1}.\text{p3})^3 M_{W^+}^2-12 \text{p1}.\text{p2}
(\text{p1}.\text{p3})^2 M_{W^+}^2\nonumber\\
&-2 (\text{p1}.\text{p2})^2 M_A^2
M_{W^+}^2-2 (\text{p1}.\text{p3})^2 M_A^2 M_{W^+}^2+4
\text{p1}.\text{p2} \text{p1}.\text{p3} M_A^2 M_{W^+}^2-2
(\text{p1}.\text{p2})^2 M_Z^2 M_{W^+}^2\nonumber\\
&-2 (\text{p1}.\text{p3})^2
M_Z^2 M_{W^+}^2+4 \text{p1}.\text{p2} \text{p1}.\text{p3} M_Z^2
M_{W^+}^2+12 (\text{p1}.\text{p2})^2 \text{p1}.\text{p3}
M_{W^+}^2-16 (\text{p1}.\text{p2})^3 \text{p1}.\text{p3}\Big], \nonumber \\
D_u=&[(-p1+p4)^2-M^2_{H_0}]^2.\nonumber\\
|\mathcal{M}_c|^2= &\frac{g^2_1g^2_2Sin\theta^2_W}{8M^2_{W^+}M^2_Z}\Big[2 M_A^2 M_{H^+}^2+4 M_A^2 \text{p1}.\text{p2}-2 M_A^2 M_{W^+}^2-2 M_A^2 M_Z^2+M_A^4+4 M_{H^+}^2 \text{p1}.\text{p2}\nonumber\\
&-2 M_{H^+}^2
M_{W^+}^2-2 M_{H^+}^2 M_Z^2+M_{H^+}^4-4 M_{W^+}^2 \text{p1}.\text{p2}-4 M_Z^2 \text{p1}.\text{p2}+10 M_{W^+}^2
M_Z^2+M_{W^+}^4\nonumber\\
&+M_Z^4+4 (\text{p1}.\text{p2})^2\Big].\nonumber
\end{align*}
\subsection{Process 1: $\rm~T_0(p1)+~T_0(p2) \rightarrow W^+(p3) + W^-(p4)$}
\begin{align*}
|\mathcal{M}_s|^2 = & \frac{g^4_2v^4\lambda_{ht}^2}{4 M^4_{W^+} D_s}\Big[(p1.p2)^2+2(p1.p2)M^2_{T_0}-2(p1.p2)M^2_{W^+}+M^4_{T_0}-2M^2_{T_0}M^2_{W^+}+3M^4_{W^+}\Big], \nonumber \\
D_s=&  [(-p1-p2)^2-M^2_h]^2. \nonumber \\
|\mathcal{M}_t|^2= & \frac{g^4_2}{M^4_{W^+} D_t}\Big[(p1.p3)^4-2(p1.p3)^2M^2_{T_0}M^2_{W^+}+M^4_{T_0}M^4_{W^+}\Big], \nonumber \\
D_t=& [(-p1+p3)^2-M^2_{T^+}]^2. \nonumber \\
|\mathcal{M}_u|^2= & \frac{{g^4_2}}{{M^4_{W^+}D_u}}\Big[(p1.p3)^4-4(p1.p3)^3(p1.p2)-4(p1.p3)^3M^2_{T_0}+6(p1.p3)^2(p1.p2)^2 \nonumber \\
&+12(p1.p3)^2(p1.p2)M^2_{T_0}  +6(p1.p3)^2M^4_{T_0}-2(p1.p3)^2M^2_{T_0}M^2_{W^+} \nonumber \\
&-4(p1.p3)(p1.p2)^3-12(p1.p3)(p1.p2)^2M^2_{T_0} -12(p1.p3)(p1.p2)M^4_{T_0} \nonumber \\
&+4(p1.p3)(p1.p2)M^2_{T_0}M^2_{W^+}-4(p1.p3)M^6_{T_0}+4(p1.p3)M^4_{T_0}M^2_{W^+} \nonumber \\
&+(p1.p2)^4+4(p1.p2)^3M^2_{T_0}+6(p1.p2)^2M^4_{T_0}-2(p1.p2)^2M^2_{T_0}M^2_{W^+} \nonumber \\
&+4(p1.p2)M^6_{T_0}-4(p1.p2)M^4_{T_0}M^2_{W^+}+M^8_{T_0}-2M^6_{T_0}M^2_{W^+}+M^4_{T_0}M^4_{W^+}\Big], \nonumber \\
D_u=& [(-p1+p4)^2-M^2_{T^+}]^2. \nonumber \\
|\mathcal{M}_c|^2= & \frac{g^4_2}{M^4_{W^+} }\Big[(p1.p2)^2+2(p1.p2)M^2_{T_0}-2(p1.p2)M^2_{W^+}+M^4_{T_0}-2M^2_{T_0}M^2_{W^+}+3M^4_{W^+}\Big]. \nonumber 
\end{align*}
\subsection{Process 2: $\rm~T_0(p1)+~T_0(p2) \rightarrow Z(p3) + Z(p4)$}
\begin{align*}
|\mathcal{M}_s|^2 =  &\frac{(g_2Cos\theta_W+g_1Sin\theta_W)^4v^4\lambda_{ht}^2}{8 M^4_Z D_s}\Big[(p1.p2)^2+2(p1.p2)M^2_{T_0}-2(p1.p2)M^2_Z\nonumber\\
&+M^4_{T_0}-2M^2_{T_0}M^2_Z+3M^4_Z\Big], \nonumber \\
D_s=&  [(-p1-p2)^2-M^2_h]^2. \nonumber
\end{align*}
\subsection{Process 3: $\rm~T^{\pm}(p1)+~T_0(p2) \rightarrow Z(p3) + W^{\pm}(p4)$}
\begin{align*}
|\mathcal{M}_s|^2 =  &\frac{g^4_2Cos\theta^2_W}{8M_{W^+}^{6}M_Z^2D_s}\Big[M_{W^+}^{12}+8 M_Z^2 M_{W^+}^{10}-4 M_{T_0}^2 M_{W^+}^{10}-12 M_{T^+}^2 M_{W^+}^{10}+8 \text{p1}.\text{p3} M_{W^+}^{10}\nonumber\\
&-18 M_Z^4
M_{W^+}^8+6 M_{T_0}^4 M_{W^+}^8+30 M_{T^+}^4 M_{W^+}^8-8 (\text{p1}.\text{p2})^2 M_{W^+}^8+16 (\text{p1}.\text{p3})^2
M_{W^+}^8\nonumber\\
&-40 \text{p1}.\text{p2} M_Z^2 M_{W^+}^8+72 \text{p1}.\text{p3} M_Z^2 M_{W^+}^8-14 M_Z^2 M_{T_0}^2 M_{W^+}^8-8
\text{p1}.\text{p3} M_{T_0}^2 M_{W^+}^8\nonumber\\
&-10 M_Z^2 M_{T^+}^2 M_{W^+}^8+20 M_{T_0}^2 M_{T^+}^2 M_{W^+}^8+48 \text{p1}.\text{p2}
M_{T^+}^2 M_{W^+}^8-40 \text{p1}.\text{p3} M_{T^+}^2 M_{W^+}^8\nonumber\\
&-48 \text{p1}.\text{p2} \text{p1}.\text{p3} M_{W^+}^8+8 M_Z^6
M_{W^+}^6-4 M_{T_0}^6 M_{W^+}^6-28 M_{T^+}^6 M_{W^+}^6+32 \text{p1}.\text{p2} M_Z^4 M_{W^+}^6\nonumber\\
&-72 \text{p1}.\text{p3} M_Z^4
M_{W^+}^6+2 M_Z^2 M_{T_0}^4 M_{W^+}^6-8 \text{p1}.\text{p3} M_{T_0}^4 M_{W^+}^6-10 M_Z^2 M_{T^+}^4 M_{W^+}^6\nonumber\\
&-28 M_{T_0}^2
M_{T^+}^4 M_{W^+}^6-96 \text{p1}.\text{p2} M_{T^+}^4 M_{W^+}^6+56 \text{p1}.\text{p3} M_{T^+}^4 M_{W^+}^6-64
\text{p1}.\text{p2} (\text{p1}.\text{p3})^2 M_{W^+}^6\nonumber\\
&-32 (\text{p1}.\text{p2})^2 M_Z^2 M_{W^+}^6+160 (\text{p1}.\text{p3})^2
M_Z^2 M_{W^+}^6-160 \text{p1}.\text{p2} \text{p1}.\text{p3} M_Z^2 M_{W^+}^6+38 M_Z^4 M_{T_0}^2 M_{W^+}^6\nonumber\\
&+16
(\text{p1}.\text{p2})^2 M_{T_0}^2 M_{W^+}^6-32 (\text{p1}.\text{p3})^2 M_{T_0}^2 M_{W^+}^6+44 \text{p1}.\text{p2} M_Z^2
M_{T_0}^2 M_{W^+}^6-72 \text{p1}.\text{p3} M_Z^2 M_{T_0}^2 M_{W^+}^6\nonumber\\
&+32 \text{p1}.\text{p2} \text{p1}.\text{p3} M_{T_0}^2
M_{W^+}^6+10 M_Z^4 M_{T^+}^2 M_{W^+}^6-4 M_{T_0}^4 M_{T^+}^2 M_{W^+}^6-80 (\text{p1}.\text{p2})^2 M_{T^+}^2 M_{W^+}^6\nonumber\\
&-32
(\text{p1}.\text{p3})^2 M_{T^+}^2 M_{W^+}^6+20 \text{p1}.\text{p2} M_Z^2 M_{T^+}^2 M_{W^+}^6-88 \text{p1}.\text{p3} M_Z^2
M_{T^+}^2 M_{W^+}^6+104 M_Z^2 M_{T_0}^2 M_{T^+}^2 M_{W^+}^6\nonumber\\
&-32 \text{p1}.\text{p2} M_{T_0}^2 M_{T^+}^2 M_{W^+}^6+48
\text{p1}.\text{p3} M_{T_0}^2 M_{T^+}^2 M_{W^+}^6+160 \text{p1}.\text{p2} \text{p1}.\text{p3} M_{T^+}^2 M_{W^+}^6\nonumber\\
&+96
(\text{p1}.\text{p2})^2 \text{p1}.\text{p3} M_{W^+}^6+M_Z^8 M_{W^+}^4+M_{T_0}^8 M_{W^+}^4+9 M_{T^+}^8 M_{W^+}^4+8
\text{p1}.\text{p2} M_Z^6 M_{W^+}^4\nonumber\\
&-8 \text{p1}.\text{p3} M_Z^6 M_{W^+}^4+6 M_Z^2 M_{T_0}^6 M_{W^+}^4+8 \text{p1}.\text{p3}
M_{T_0}^6 M_{W^+}^4+18 M_Z^2 M_{T^+}^6 M_{W^+}^4+12 M_{T_0}^2 M_{T^+}^6 M_{W^+}^4\nonumber\\
&+48 \text{p1}.\text{p2} M_{T^+}^6
M_{W^+}^4-24 \text{p1}.\text{p3} M_{T^+}^6 M_{W^+}^4+16 (\text{p1}.\text{p2})^4 M_{W^+}^4-40 (\text{p1}.\text{p2})^2 M_Z^4
M_{W^+}^4\nonumber\\
&+16 (\text{p1}.\text{p3})^2 M_Z^4 M_{W^+}^4+16 \text{p1}.\text{p2} \text{p1}.\text{p3} M_Z^4 M_{W^+}^4-21 M_Z^4
M_{T_0}^4 M_{W^+}^4-8 (\text{p1}.\text{p2})^2 M_{T_0}^4 M_{W^+}^4\nonumber\\
&+16 (\text{p1}.\text{p3})^2 M_{T_0}^4 M_{W^+}^4+8
\text{p1}.\text{p3} M_Z^2 M_{T_0}^4 M_{W^+}^4+16 \text{p1}.\text{p2} \text{p1}.\text{p3} M_{T_0}^4 M_{W^+}^4+11 M_Z^4
M_{T^+}^4 M_{W^+}^4\nonumber\\
&-2 M_{T_0}^4 M_{T^+}^4 M_{W^+}^4+88 (\text{p1}.\text{p2})^2 M_{T^+}^4 M_{W^+}^4+16 (\text{p1}.\text{p3})^2
M_{T^+}^4 M_{W^+}^4+48 \text{p1}.\text{p2} M_Z^2 M_{T^+}^4 M_{W^+}^4\nonumber\\
&+8 \text{p1}.\text{p3} M_Z^2 M_{T^+}^4 M_{W^+}^4-22 M_Z^2
M_{T_0}^2 M_{T^+}^4 M_{W^+}^4+32 \text{p1}.\text{p2} M_{T_0}^2 M_{T^+}^4 M_{W^+}^4-40 \text{p1}.\text{p3} M_{T_0}^2 M_{T^+}^4
M_{W^+}^4\nonumber\\
&-112 \text{p1}.\text{p2} \text{p1}.\text{p3} M_{T^+}^4 M_{W^+}^4+64 (\text{p1}.\text{p2})^2 (\text{p1}.\text{p3})^2
M_{W^+}^4+32 (\text{p1}.\text{p2})^3 M_Z^2 M_{W^+}^4-64 \text{p1}.\text{p2} (\text{p1}.\text{p3})^2 M_Z^2 M_{W^+}^4\nonumber\\
&+32
(\text{p1}.\text{p2})^2 \text{p1}.\text{p3} M_Z^2 M_{W^+}^4-18 M_Z^6 M_{T_0}^2 M_{W^+}^4-40 \text{p1}.\text{p2} M_Z^4
M_{T_0}^2 M_{W^+}^4+72 \text{p1}.\text{p3} M_Z^4 M_{T_0}^2 M_{W^+}^4\nonumber\\
&+64 \text{p1}.\text{p2} (\text{p1}.\text{p3})^2 M_{T_0}^2
M_{W^+}^4-8 (\text{p1}.\text{p2})^2 M_Z^2 M_{T_0}^2 M_{W^+}^4-32 (\text{p1}.\text{p3})^2 M_Z^2 M_{T_0}^2 M_{W^+}^4\nonumber\\
&+32
\text{p1}.\text{p2} \text{p1}.\text{p3} M_Z^2 M_{T_0}^2 M_{W^+}^4-32 (\text{p1}.\text{p2})^2 \text{p1}.\text{p3} M_{T_0}^2
M_{W^+}^4+10 M_Z^6 M_{T^+}^2 M_{W^+}^4-4 M_{T_0}^6 M_{T^+}^2 M_{W^+}^4\nonumber\\
&+24 \text{p1}.\text{p2} M_Z^4 M_{T^+}^2 M_{W^+}^4-56
\text{p1}.\text{p3} M_Z^4 M_{T^+}^2 M_{W^+}^4-34 M_Z^2 M_{T_0}^4 M_{T^+}^2 M_{W^+}^4-16 \text{p1}.\text{p2} M_{T_0}^4
M_{T^+}^2 M_{W^+}^4\nonumber\\
&-8 \text{p1}.\text{p3} M_{T_0}^4 M_{T^+}^2 M_{W^+}^4+64 (\text{p1}.\text{p2})^3 M_{T^+}^2 M_{W^+}^4+64
\text{p1}.\text{p2} (\text{p1}.\text{p3})^2 M_{T^+}^2 M_{W^+}^4+40 (\text{p1}.\text{p2})^2 M_Z^2 M_{T^+}^2 M_{W^+}^4\nonumber\\
&-32
(\text{p1}.\text{p3})^2 M_Z^2 M_{T^+}^2 M_{W^+}^4+32 \text{p1}.\text{p2} \text{p1}.\text{p3} M_Z^2 M_{T^+}^2 M_{W^+}^4+34
M_Z^4 M_{T_0}^2 M_{T^+}^2 M_{W^+}^4\nonumber\\
&+16 (\text{p1}.\text{p2})^2 M_{T_0}^2 M_{T^+}^2 M_{W^+}^4+32 (\text{p1}.\text{p3})^2
M_{T_0}^2 M_{T^+}^2 M_{W^+}^4-80 \text{p1}.\text{p2} M_Z^2 M_{T_0}^2 M_{T^+}^2 M_{W^+}^4\nonumber\\
&+16 \text{p1}.\text{p3} M_Z^2
M_{T_0}^2 M_{T^+}^2 M_{W^+}^4-96 \text{p1}.\text{p2} \text{p1}.\text{p3} M_{T_0}^2 M_{T^+}^2 M_{W^+}^4-160
(\text{p1}.\text{p2})^2 \text{p1}.\text{p3} M_{T^+}^2 M_{W^+}^4\nonumber\\
&-64 (\text{p1}.\text{p2})^3 \text{p1}.\text{p3} M_{W^+}^4-2
M_Z^2 M_{T_0}^8 M_{W^+}^2-6 M_Z^2 M_{T^+}^8 M_{W^+}^2-4 \text{p1}.\text{p2} M_Z^2 M_{T_0}^6 M_{W^+}^2\nonumber\\
&-8 \text{p1}.\text{p3}
M_Z^2 M_{T_0}^6 M_{W^+}^2-4 M_Z^4 M_{T^+}^6 M_{W^+}^2-28 \text{p1}.\text{p2} M_Z^2 M_{T^+}^6 M_{W^+}^2+8 \text{p1}.\text{p3}
M_Z^2 M_{T^+}^6 M_{W^+}^2\nonumber\\
&-4 M_Z^2 M_{T_0}^2 M_{T^+}^6 M_{W^+}^2+12 M_Z^6 M_{T_0}^4 M_{W^+}^2+4 \text{p1}.\text{p2} M_Z^4
M_{T_0}^4 M_{W^+}^2+8 (\text{p1}.\text{p2})^2 M_Z^2 M_{T_0}^4 M_{W^+}^2\nonumber\\
&-32 \text{p1}.\text{p2} \text{p1}.\text{p3} M_Z^2
M_{T_0}^4 M_{W^+}^2+16 M_Z^6 M_{T^+}^4 M_{W^+}^2-12 \text{p1}.\text{p2} M_Z^4 M_{T^+}^4 M_{W^+}^2+8 M_Z^2 M_{T_0}^4 M_{T^+}^4
M_{W^+}^2\nonumber\\
&-40 (\text{p1}.\text{p2})^2 M_Z^2 M_{T^+}^4 M_{W^+}^2+32 \text{p1}.\text{p2} \text{p1}.\text{p3} M_Z^2 M_{T^+}^4
M_{W^+}^2+4 \text{p1}.\text{p2} M_Z^2 M_{T_0}^2 M_{T^+}^4 M_{W^+}^2\nonumber\\
&+8 \text{p1}.\text{p3} M_Z^2 M_{T_0}^2 M_{T^+}^4
M_{W^+}^2-2 M_Z^8 M_{T_0}^2 M_{W^+}^2-4 \text{p1}.\text{p2} M_Z^6 M_{T_0}^2 M_{W^+}^2+8 \text{p1}.\text{p3} M_Z^6 M_{T_0}^2
M_{W^+}^2\nonumber\\
&+8 (\text{p1}.\text{p2})^2 M_Z^4 M_{T_0}^2 M_{W^+}^2+16 (\text{p1}.\text{p2})^3 M_Z^2 M_{T_0}^2 M_{W^+}^2-32
(\text{p1}.\text{p2})^2 \text{p1}.\text{p3} M_Z^2 M_{T_0}^2 M_{W^+}^2+2 M_Z^8 M_{T^+}^2 M_{W^+}^2\nonumber\\
&+4 \text{p1}.\text{p2} M_Z^6
M_{T^+}^2 M_{W^+}^2-8 \text{p1}.\text{p3} M_Z^6 M_{T^+}^2 M_{W^+}^2+4 M_Z^2 M_{T_0}^6 M_{T^+}^2 M_{W^+}^2-8
(\text{p1}.\text{p2})^2 M_Z^4 M_{T^+}^2 M_{W^+}^2\nonumber\\
&+4 M_Z^4 M_{T_0}^4 M_{T^+}^2 M_{W^+}^2+28 \text{p1}.\text{p2} M_Z^2 M_{T_0}^4
M_{T^+}^2 M_{W^+}^2-8 \text{p1}.\text{p3} M_Z^2 M_{T_0}^4 M_{T^+}^2 M_{W^+}^2\nonumber\\
&-16 (\text{p1}.\text{p2})^3 M_Z^2 M_{T^+}^2
M_{W^+}^2+32 (\text{p1}.\text{p2})^2 \text{p1}.\text{p3} M_Z^2 M_{T^+}^2 M_{W^+}^2-28 M_Z^6 M_{T_0}^2 M_{T^+}^2 M_{W^+}^2\nonumber\\
&+8
\text{p1}.\text{p2} M_Z^4 M_{T_0}^2 M_{T^+}^2 M_{W^+}^2+32 (\text{p1}.\text{p2})^2 M_Z^2 M_{T_0}^2 M_{T^+}^2 M_{W^+}^2+M_Z^4
M_{T_0}^8+M_Z^4 M_{T^+}^8\nonumber\\
&-2 M_Z^6 M_{T_0}^6+4 \text{p1}.\text{p2} M_Z^4 M_{T_0}^6-2 M_Z^6 M_{T^+}^6+4 \text{p1}.\text{p2}
M_Z^4 M_{T^+}^6+M_Z^8 M_{T_0}^4-4 \text{p1}.\text{p2} M_Z^6 M_{T_0}^4\nonumber\\
&+4 (\text{p1}.\text{p2})^2 M_Z^4 M_{T_0}^4+M_Z^8
M_{T^+}^4-4 \text{p1}.\text{p2} M_Z^6 M_{T^+}^4+4 (\text{p1}.\text{p2})^2 M_Z^4 M_{T^+}^4-2 M_Z^4 M_{T_0}^4 M_{T^+}^4+2 M_Z^6
M_{T_0}^2 M_{T^+}^4\nonumber\\
&-4 \text{p1}.\text{p2} M_Z^4 M_{T_0}^2 M_{T^+}^4+2 M_Z^6 M_{T_0}^4 M_{T^+}^2-4 \text{p1}.\text{p2} M_Z^4
M_{T_0}^4 M_{T^+}^2-2 M_Z^8 M_{T_0}^2 M_{T^+}^2+8 \text{p1}.\text{p2} M_Z^6 M_{T_0}^2 M_{T^+}^2\nonumber\\
&-8 (\text{p1}.\text{p2})^2
M_Z^4 M_{T_0}^2 M_{T^+}^2\Big], \nonumber\\
D_s=&[(-p1-p2)^2-M_{W^+}^2]^2.\nonumber\\
|\mathcal{M}_t|^2 =  &\frac{g^4_2Cos\theta^2_W}{4M_{W^+}^2 M_Z^2D_t}\Big[-4 M_{T^+}^2 M_{W^+}^2 M_Z^2 \text{p1}.\text{p3}-2 M_{T^+}^2 M_{W^+}^2 (\text{p1}.\text{p3})^2+4 M_{T^+}^2 M_Z^4
\text{p1}.\text{p3}\nonumber\\
&+4 M_{T^+}^4 M_Z^2 \text{p1}.\text{p3}-2 M_{T^+}^2 M_Z^2 (\text{p1}.\text{p3})^2-4 M_{T_0}^2 M_{T^+}^2
M_Z^2 \text{p1}.\text{p3}+M_{T^+}^4 (\text{p1}.\text{p3})^2\nonumber\\
&-4 M_{T^+}^2 (\text{p1}.\text{p3})^3-2 M_{T_0}^2 M_{T^+}^2
(\text{p1}.\text{p3})^2-2 M_{T_0}^2 M_{W^+}^2 (\text{p1}.\text{p3})^2-2 M_{T_0}^2 M_Z^2 (\text{p1}.\text{p3})^2\nonumber\\
&+M_{T_0}^4
(\text{p1}.\text{p3})^2+4 M_{T_0}^2 (\text{p1}.\text{p3})^3-2 M_{W^+}^2 M_Z^2 (\text{p1}.\text{p3})^2+M_{W^+}^4
(\text{p1}.\text{p3})^2+4 M_{W^+}^2 (\text{p1}.\text{p3})^3\nonumber\\
&+M_Z^4 (\text{p1}.\text{p3})^2-4 M_Z^2 (\text{p1}.\text{p3})^3+2
M_{T^+}^2 M_{W^+}^2 M_Z^4-M_{T^+}^2 M_{W^+}^4 M_Z^2+2 M_{T^+}^4 M_{W^+}^2 M_Z^2
\nonumber\\
&+2 M_{T_0}^2 M_{T^+}^2 M_{W^+}^2
M_Z^2-M_{T^+}^2 M_Z^6-2 M_{T^+}^4 M_Z^4+2 M_{T_0}^2 M_{T^+}^2 M_Z^4-M_{T^+}^6 M_Z^2\nonumber\\
&+2 M_{T_0}^2 M_{T^+}^4 M_Z^2-M_{T_0}^4
M_{T^+}^2 M_Z^2+4 (\text{p1}.\text{p3})^4\Big], \nonumber\\
D_t=&[(-p1+p3)^2-M_{T^+}^2]^2. \nonumber\\
|\mathcal{M}_c|^2 =  &\frac{g^4_2Cos\theta^2_W}{8M_{W^+}^2M_Z^2}\Big[4 M_{T^+}^2 \text{p1}.\text{p2}+4 M_{T_0}^2 \text{p1}.\text{p2}-4 M_{W^+}^2 \text{p1}.\text{p2}-4 M_Z^2 \text{p1}.\text{p2}-2
M_{T^+}^2 M_{W^+}^2\nonumber\\
&-2 M_{T^+}^2 M_Z^2+M_{T^+}^4+2 M_{T_0}^2 M_{T^+}^2-2 M_{T_0}^2 M_{W^+}^2-2 M_{T_0}^2 M_Z^2+M_{T_0}^4+10
M_{W^+}^2 M_Z^2+M_{W^+}^4+M_Z^4\nonumber\\
&+4 (\text{p1}.\text{p2})^2\Big].\nonumber
\end{align*}
\endgroup
The interefernce terms are also taken into account which are not given here. The cross section $\sigma(DM DM \rightarrow W^+W^-/ZZ)$ can then be obtained, from the total amplitude, in the usual way.

\end{document}